\documentclass[a4paper,11pt]{article}
\pdfoutput=1
\usepackage{amssymb}
\usepackage{amsmath}
\usepackage{braket}
\usepackage{mathtools}
\usepackage[usenames,dvipsnames,svgnames,table]{xcolor}
\usepackage[utf8]{inputenc}
\usepackage{color}
\usepackage{cite}
\usepackage{subcaption}
\usepackage{lmodern}
\usepackage{footnote}
\usepackage[normalem]{ulem}
\usepackage{glossaries-extra}
\setabbreviationstyle[acronym]{long-short}
\glssetcategoryattribute{acronym}{nohyperfirst}{true}

\usepackage{slashed}
\usepackage[pdftex]{graphicx}
\usepackage{multirow}
\usepackage{jheppub}
\usepackage{here}
\usepackage{hyperref}
\usepackage{verbatim}
\usepackage[justification=centering,singlelinecheck=false]{caption}
\usepackage{cleveref}
\usepackage{listings}
\usepackage{fancyvrb}
\usepackage{dsfont}
\usepackage{nicefrac,xfrac}
\usepackage{svg}

\Crefname{equation}{eq.}{eqs.}
\Crefname{section}{section}{sections}
\Crefname{figure}{figure}{figures}
\Crefname{appendix}{appendix}{appendices}

\newcommand{\cB}{\mathcal{B}}

\newcommand{\cD}[0]{\mathcal D}

\newcommand{\cK}[0]{\mathcal K}

\newcommand{\cM}[0]{\mathcal M}
\newcommand{\cO}[0]{\mathcal O}

\newcommand{\cY}[0]{\mathcal Y}

\newcommand{\wt}[0]{\widetilde}

\newcommand{\df}[0]{\mathrm{df}}

\newcommand{\Kdf}[0]{{\cK_{\df,3}}}

\newcommand{\PV}[0]{{\mathrm{PV}}}
\newcommand{\Mdf}[0]{\mathcal{M}_{\df,3}}

\newcommand{\kdf}{\mathcal{K}_{\text{df},3} }

\newcommand{\bm}[0]{\boldsymbol}

\newcommand{\XL}[3]{\boldsymbol{\mathcal X}_{[kab]}^{(#1#2#3)\dagger}}
\newcommand{\XR}[3]{\boldsymbol{\mathcal X}_{[kab]}^{(#1#2#3)}}
\newcommand{\YL}[3]{\boldsymbol{\mathcal Y}^{[kab] \dagger}_{(#1#2#3)}}
\newcommand{\YR}[3]{\boldsymbol{\mathcal Y}^{[kab]}_{(#1#2#3)}}
\newcommand{\CL}[0]{\boldsymbol{\mathcal C}^\dagger}
\newcommand{\CR}[0]{\boldsymbol{\mathcal C}}

\newcommand{\Luscher}[0]{Luscher:1986n2,Luscher:1991n1}
\newcommand{\KSS}[0]{Kim:2005gf}
\newcommand{\RaulSpin}[0]{Briceno:2015csa}

\newcommand{\HSQCa}[0]{Hansen:2014eka}
\newcommand{\HSQCb}[0]{Hansen:2015zga}
\newcommand{\BHSQC}[0]{Briceno:2017tce}
\newcommand{\BHSnum}[0]{Briceno:2018mlh}
\newcommand{\dwave}[0]{Blanton:2019igq}
\newcommand{\largera}[0]{Romero-Lopez:2019qrt}
\newcommand{\isospin}[0]{Hansen:2020zhy}
\newcommand{\implement}[0]{Blanton:2021eyf}

\newcommand{\altSetUps}[0]{Briceno:2014oea,Lee:2017igf}

\newcommand{\BSQC}[0]{Blanton:2020gha}
\newcommand{\BSequiv}[0]{Blanton:2020jnm}
\newcommand{\BSnondegen}[0]{Blanton:2020gmf}
\newcommand{\BStwoplusone}[0]{Blanton:2021mih}

\newcommand{\MD}[0]{Mai:2017bge}

\newcommand{\ThreeQCDNumerics}[0]{%
Detmold:2011kw,
Mai:2018djl,
Horz:2019rrn,
Blanton:2019vdk,
Mai:2019fba,
Culver:2019vvu,
Fischer:2020jzp,
Hansen:2020otl,
NPLQCD:2020ozd,
Alexandru:2020xqf,
Brett:2021wyd,
Blanton:2021llb,
Mai:2021nul,
Garofalo:2022pux,
Draper:2023boj}

\newcommand{\ThreeBody}[0]{%
Detmold:2008gh,
Beane:2007qr,
Briceno:2012rv,
Polejaeva:2012ut,
Hansen:2014eka,
Hansen:2015zga,
Briceno:2017tce,
Hammer:2017uqm,
Konig:2017krd,
Hammer:2017kms,
Mai:2017bge,
Briceno:2018mlh,
Briceno:2018aml,
Blanton:2019igq,
Pang:2019dfe,
Jackura:2019bmu,
Briceno:2019muc,
Romero-Lopez:2019qrt,
Hansen:2020zhy,
Blanton:2020gha,
Blanton:2020jnm,
Pang:2020pkl,
Romero-Lopez:2020rdq,
Blanton:2020gmf,
Muller:2020vtt,
Blanton:2021mih,
Muller:2021uur,
Blanton:2021eyf,
Jackura:2022gib}

\newacronym{CMF}{CMF}{center-of-momentum frame}

\DeclareFixedFont{\ttb}{T1}{txtt}{bx}{n}{9}
\DeclareFixedFont{\ttm}{T1}{txtt}{m}{n}{9}

\title{Incorporating $DD\pi$ effects and left-hand cuts in lattice QCD studies of the $T_{cc}(3875)^+$}

\author[a]{Maxwell T. Hansen}
\affiliation[a]{School of Physics and Astronomy, University of Edinburgh, Edinburgh EH9 3JZ, UK}

\author[b]{\!, Fernando Romero-L\'opez}
\affiliation[b]{CTP, Massachusetts Institute of Technology, Cambridge, MA 02139, USA}

\author[c]{\!, and~Stephen~R.~Sharpe}
\affiliation[c]{Physics Department, University of Washington, Seattle, WA 98195-1560, USA}

\emailAdd{maxwell.hansen@ed.ac.uk}
\emailAdd{fernando@mit.edu}
\emailAdd{srsharpe@uw.edu}

\abstract{We generalize the relativistic field-theoretic three-particle finite-volume scattering formalism to describe generic $DD\pi$ systems in the charm $C=2$ sector. This includes the isospin-0 channel, in which the recently discovered doubly-charmed tetraquark $T_{cc}(3875)^+$ is expected to manifest as a pole in the $DD \pi \to DD \pi$ scattering amplitude. The formalism presented here can also be applied to lattice QCD settings in which the $D^*$ is bound and, in particular, remains valid below the left-hand cut in $D D^*$ scattering, thus resolving an issue in previous analyses of lattice-determined finite-volume energies.}

\allowdisplaybreaks

\preprint{MIT-CTP/5667}
\begin{document}

\maketitle
\flushbottom
\clearpage

\section{Introduction}
\label{sec:intro}

A plethora of exotic hadrons have been discovered in the last two decades~\cite{Ali:2017jda,Olsen:2017bmm,Karliner:2017qhf,Guo:2017jvc,Liu:2019zoy,Brambilla:2019esw,Gershon:2022xnn,Lebed:2022vfu}, and it is of fundamental interest to provide first-principles predictions of their properties from quantum chromodynamics (QCD). While lattice QCD is in principle well suited for such a task, in practice its applicability is limited by the large number of open decay modes for many of these exotic resonances. A gold-plated exotic in this context is the recently discovered doubly-charmed tetraquark $T_{\rm cc}(3875)^+$~\cite{LHCb:2021vvq,LHCb:2021auc}, with quantum numbers ${(I)J^P=(0)1^+}$, which decays only into the three-hadron mode $DD\pi$. Given the recent progress in three-hadron spectroscopy from lattice QCD~\cite{\ThreeQCDNumerics}, future studies of the $T_{\rm cc}(3875)^+$ with physical quark masses are within reach. However, such studies will require an extension of the existing three-particle formalism~\cite{\ThreeBody} to include the $DD\pi$ system.

Pioneering lattice QCD studies of the $T_{\rm cc}^+$ at heavier-than-physical pion masses have already begun~\cite{Padmanath:2022cvl,Lyu:2023xro,Chen:2022vpo}. In this setting, the $D^*$ meson is stable and the system can be treated as a two-body problem. This allows one to use the L\"uscher finite-volume formalism~\cite{\Luscher}, appropriately generalized~\cite{\RaulSpin}, to determine the elastic $D D^*$ scattering amplitude. By analytically continuing the amplitude from physical scattering energies into the complex plane, the presence of (and, where applicable, the position of) a pole associated with the $T_{\rm cc}^+$ can be investigated. However, a major drawback of this method is that the tentative position of the pole is very close to, or even on top of, the left-hand branch cut generated by one-pion exchange (hereafter referred to as the ``left-hand cut'' or ``lhc''), an issue identified in refs.~\cite{Padmanath:2022cvl,Du:2023hlu}. This cut runs along the negative real axis, from a branch point located at a subthreshold energy below which the two-particle formalism breaks down. Applying the formalism in this regime introduces an uncontrolled theoretical error that raises concerns about conclusions derived from the approach.\footnote{%
These concerns may not apply to the calculations of ref.~\cite{Lyu:2023xro}, which uses the HALQCD
potential method.}

In this work we extend the relativistic field-theoretic (RFT) three-particle formalism~\cite{\HSQCa,\HSQCb} to describe generic $DD\pi$ systems in the sector with charm $C=2$ in isosymmetric QCD. This allows us to provide the formalism necessary both to study the $T_{\rm cc}^+$ at or near physical quark masses, where it decays into three particles, and to avoid the breakdown of the two-particle formalism near or on the left-hand cut in the heavier quark regime, where the $D^*$ is stable and the $T_{\rm cc}^+$ decays into two particles. The key idea for the latter case is that the $D^*$ can be incorporated as a pole in the subthreshold $p$-wave $I=1/2$ $D\pi$ scattering amplitude. In this case, the finite-volume effects from one-pion exchange diagrams are naturally incorporated in the three-particle formalism, and the finite-volume $D D^*$ spectrum can be described with the three-particle quantization condition. By solving the three-particle integral equations and using the Lehmann-Symanzik-Zimmermann (LSZ) reduction formula, following the work of refs.~\cite{Jackura:2020bsk,Dawid:2021fxd,Dawid:2023jrj}, one can extend the range of validity for the relation between finite-volume energies and the $D^*D$ amplitude to include energies below the nearest left-hand branch point.\footnote{
See refs.~\cite{Raposo:2023oru,Meng:2023bmz} for alternative approaches to solving the left-hand cut problem.}

The formalism presented here is derived by a combination of methods used for $2+1$ systems (i.e. with two identical and one distinct particle) from ref.~\cite{\BStwoplusone} with those used for three pions of arbitrary isospin from ref.~\cite{\isospin}. The formalism also applies for generic $KK\pi$ systems in the strangeness $S=-2$ sector, and $BB\pi$ systems with beauty $B=2$. In fact, in the later case there is evidence for tetraquark resonances~\cite{Brown:2012tm,Bicudo:2016ooe,Francis:2016hui,Junnarkar:2018twb,Leskovec:2019ioa,Mohanta:2020eed,Meinel:2022lzo,Hudspith:2023loy}, and a similar left-hand cut problem to that described above is observed if the $B^*$ meson is stable.

This paper is organized as follows. In \Cref{sec:lefthandcut}, we discuss the problem of the left-hand cut in $DD^*$ amplitudes, and explain why the three-body formalism provides a solution. In \Cref{sec:formalism}, we present the formalism for $DD\pi$ systems, beginning with the quantization condition relating finite-volume energies to an intermediate, cut-off-dependent three-body interaction, referred to as the divergence-free three-body K-matrix $\kdf$. We additionally present the integral equations relating $\kdf$ to physical scattering amplitudes, as well as parametrizations of $\kdf$. Then, in \Cref{sec:latticeinputs}, we summarize the lattice input required to study the $T_{\rm cc}^+$ from lattice QCD. We conclude in \Cref{sec:conclu}. Two derivations of the formalism are presented in the appendices. First, a derivation based on time-ordered perturbation theory (TOPT) is presented in \Cref{app:topt_derivation}; second, a derivation based on an intuitive extension of the flavor space is discussed in \Cref{app:flavor_derivation}.

\section{A three-particle solution to the left-hand cut problem}
\label{sec:lefthandcut}

In this section, we discuss how one-pion exchange processes lead to finite-volume effects that are unaccounted for in the two-particle L\"uscher formalism. We then describe qualitatively how the three-particle finite-volume formalism naturally incorporates those effects.

\subsection{The left-hand cut problem in \texorpdfstring{$DD^*$}{D D-star} scattering}

The $D^*$ is a vector resonance in isospin $I=1/2$ $D\pi$ scattering. When the quark masses are slightly larger than physical, the $D^*$ becomes stable, and the $s$-wave $DD^*$ scattering amplitude can be studied. In this scenario, the $T_{\rm cc}^+$, assuming that its mass remains below the $DD^*$ threshold, will appear as a bound-state pole in the ${(I)J^P=(0)1^+}$ $DD^*$ scattering amplitude.

\begin{figure}[!h]
\centering
\begin{subfigure}[b]{0.49\textwidth}
\includegraphics[width=\linewidth]{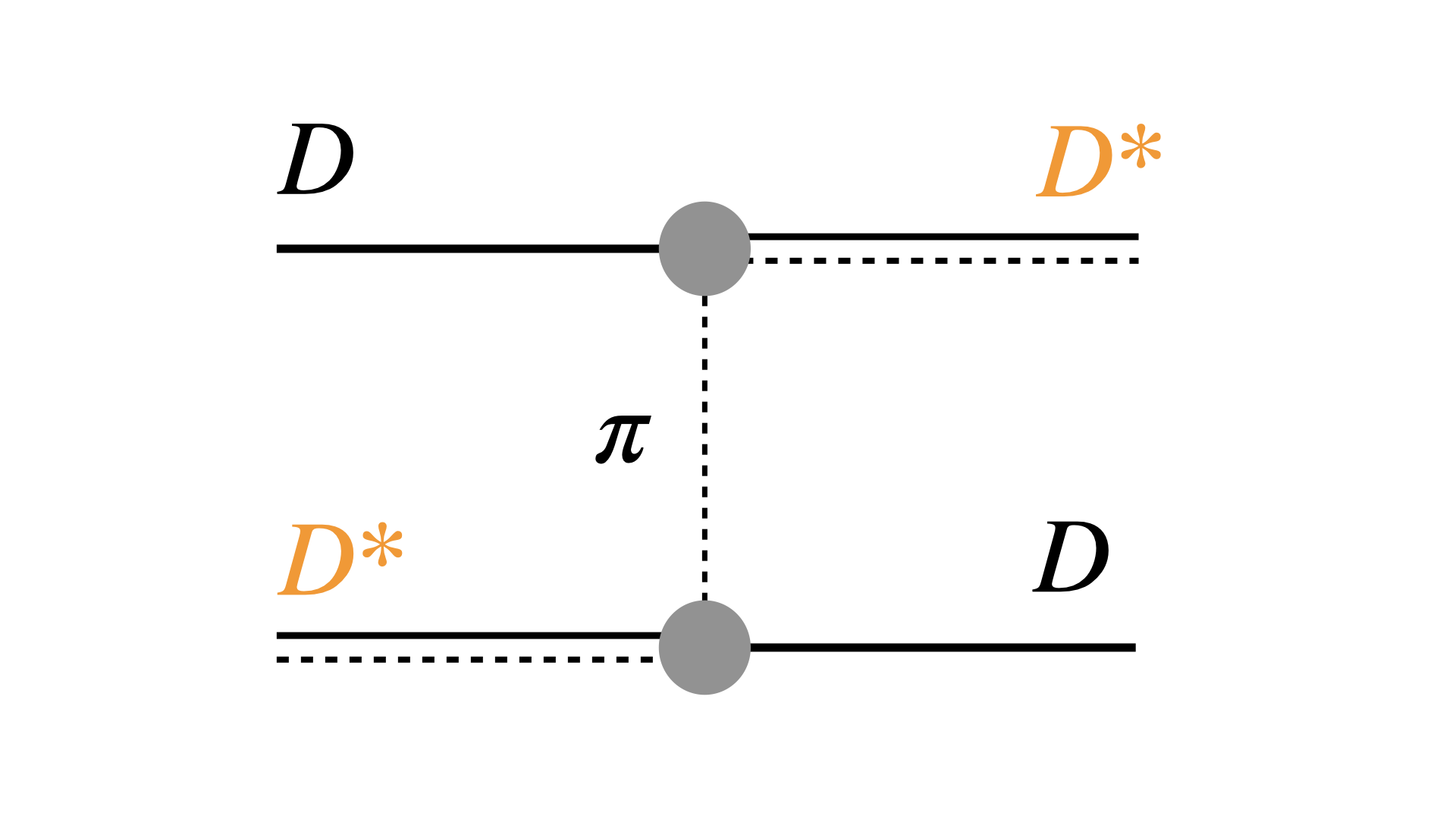}
\caption{}
\label{fig:tchannel}
\end{subfigure}
\hfill
\begin{subfigure}[b]{0.49\textwidth}
\includegraphics[width=\linewidth]{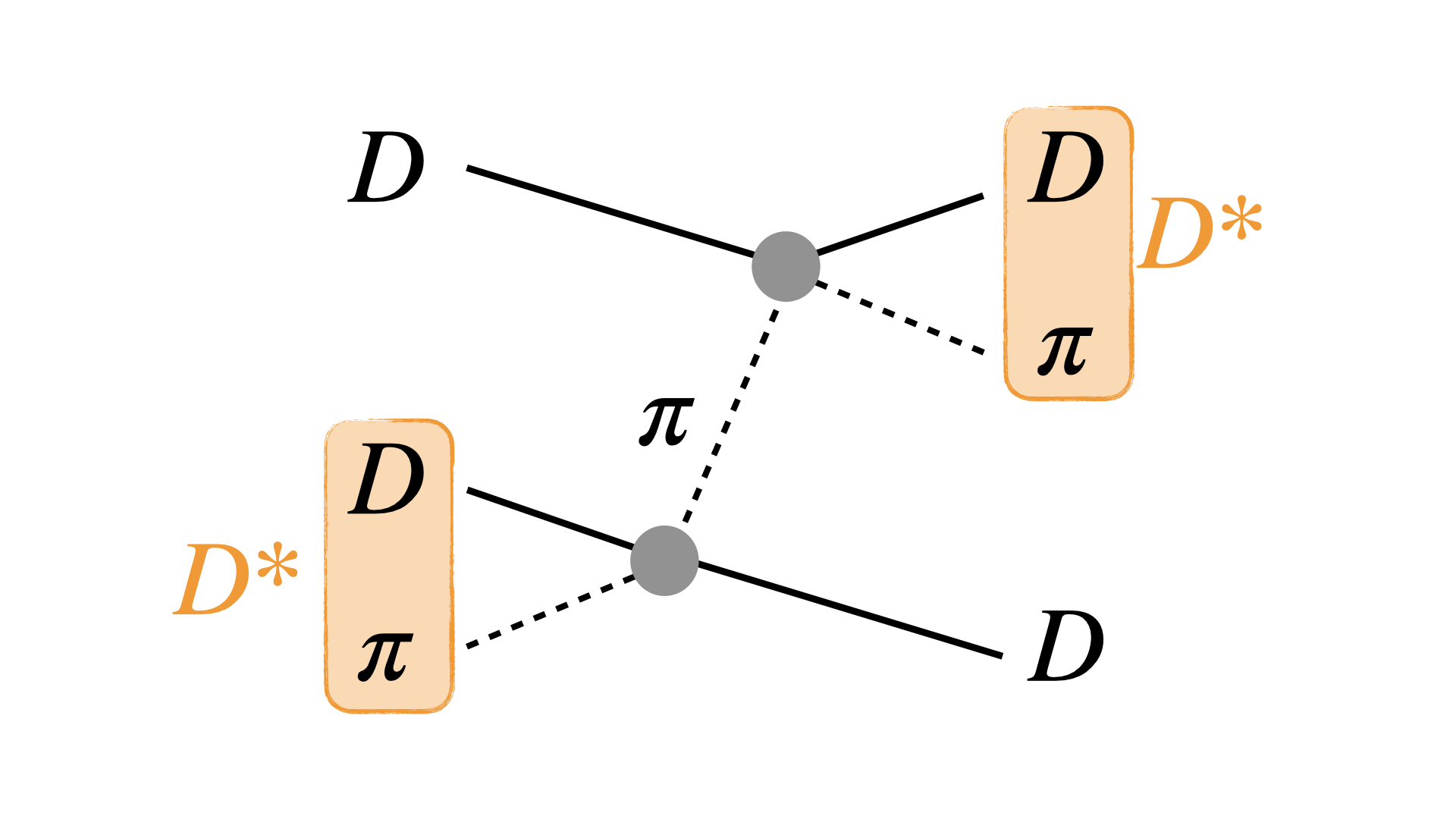}
\caption{}
\label{fig:Gdiag}
\end{subfigure}
\caption{{\bf Left:} $u$-channel one-pion exchange in $DD^*$ scattering giving rise to the left-hand cut that invalidates the L\"uscher formalism for $s \lesssim s_{\rm lhc}$. {\bf Right:} One-pion exchange diagram in $DD\pi$ scattering. The $D^*$ appears as a pole (resonant or bound) in the $D\pi$ scattering amplitude.}
\end{figure}

If the $D^*$ is stable, the $DD^*$ scattering amplitude can be studied with lattice QCD using two-particle methods. In particular, the $DD^*$ spectrum can in principle be mapped to the two-particle scattering amplitude using the L\"uscher formalism, as has been done in refs.~\cite{Padmanath:2022cvl,Chen:2022vpo}. This procedure allows one to constrain the scattering amplitude in the energy region where the scattering is elastic, $E_{\rm th}\equiv M_D+M_{D^*} \le E_{\rm cm} < E_{\rm inel}\equiv 2 M_D + M_\pi$. The method continues to work below the $DD^*$ threshold, until one approaches the left-hand cut, which is caused by one-pion exchange in the $u$ channel, a process shown in \Cref{fig:tchannel}. Projecting onto definite angular momentum turns the resulting pole into a cut, which begins when $u=M_\pi^2$ and $t=0$, and thus at $E_{\rm cm}^2=s_{\rm lhc}$, with
\begin{equation}
s_{\rm lhc} - s_{\rm th} = -M_\pi^2 + (M_{D^*} - M_{D})^2\,,
\end{equation}
where the right-hand side is less than zero, provided $M_{D^*} < M_\pi + M_D$. Here $s_{\rm th} = E_{\rm th}^2$, and we have used $s+t+u= 2 M_D^2+2M_{D^*}^2$.
The cut is present in the two-particle Bethe-Salpeter kernel, and the resultant nonanalyticity is in conflict with the assumptions in the derivation of the L\"uscher formula~\cite{Luscher:1986n2,\KSS}. In particular, it leads to additional finite-volume effects that must be accounted for, such that the L\"uscher formula cannot be applied on the cut.

The conclusion is that the full range of validity of the two-particle formalism is $\sqrt{s_{\rm lhc}} < E_{\rm cm} < E_{\rm inel}$. In practice, however, the range of validity is smaller. This is because as one approaches the end-points, exponentially-suppressed volume effects, which decay with the distance to the endpoint, and which are dropped in the two-particle formalism, can become substantial.

It is helpful to give a numerical example of the range of validity. In ref.~\cite{Padmanath:2022cvl}, one choice of charm quark mass leads to $M_D\approx 1925$MeV, $M_{D^*} \approx 2050$MeV, and $M_\pi \approx 280$MeV, so that
\begin{equation}
\{\sqrt{s_{\rm lhc}},\ E_{\rm th},\ E_{\rm inel} \} = \{3966,\ 3975,\ 4130\} \ {\rm MeV}\,.
\label{eq:numex}
\end{equation}
Thus the left-hand cut lies only 9 MeV below threshold. Perhaps coincidentally, the position of a putative virtual bound state obtained in ref.~\cite{Padmanath:2022cvl}, determined by extrapolating the scattering amplitude from above-threshold energies, lies very close to that of the left-hand cut. However, as stressed in ref.~\cite{Du:2023hlu}, this extrapolation itself is not reliable as it relies on the effective-range expansion, whose convergence fails at the left-hand cut. This is an example of the fact that, even if one only uses lattice results in the allowed energy region, the left-hand cut must still be accounted for. Altogether, the conclusions about the nature of the tetraquark from present lattice studies using the L\"uscher approach are clouded by these issues.

As the pion mass is lowered, the left-hand cut approaches the threshold, reaching it at the point that the $D^*$ becomes unstable. At that point one must use a three-particle formalism.

The finite-volume effects of the left-hand cut can be incorporated into the two-particle formalism using the recently proposed approach of ref.~\cite{Raposo:2023oru}, which explicitly accounts for the pion pole.\footnote{See also ref.~\cite{Meng:2023bmz}, where the effects are included in the context of a particular effective field theory.} The present work provides an alternative method, which not only accounts for the physics of the left-hand cut, but also extends the upper limit of the range of validity to either the $DD\pi\pi$ or $D^* D^*$ threshold, depending on kinematic details of the system.

\subsection{Three-particle formalism in the particle-dimer regime}
\label{subsec:partdimerregime}

The relativistic three-particle formalism~\cite{\HSQCa,\HSQCb} connects the finite-volume spectrum of three particles to the three-particle scattering amplitude in the energy region where only three-particle effects contribute. As already noted, the upper end of the range of validity is increased compared to the two-particle formalism, in the present case to $E_{\rm cm}=2 M_D + 2 M_\pi$ (or $2M_D^*$ if this is lower). The lower end of the range will turn out to be well below the energies of interest, and we return to this below.

Of particular importance here is the fact that the three-particle formalism remains valid in the presence of two-body bound states, as long as one uses an appropriately modified two-particle K matrix. This is discussed in detail in ref.~\cite{\largera}. In particular, one can study the three-particle amplitude below threshold, and by going to the $D^*$ pole in the $D\pi$ subchannel one can extract the $D^* D$ amplitude. This set-up is shown in \Cref{fig:Gdiag}. In other words, the three-particle formalism will correctly predict both levels that are primarily $DD\pi$ states, as well as those that correspond primarily to $D^* D$ states.\footnote{%
In finite volume there will be mixing, in general small, between $DD\pi$ and $D^* D$ states,
so there is no precise one-to-one correspondence.}
While this is a more complicated way of obtaining predictions for $D^* D$ states compared to using the two-particle formalism, it has the advantage that it incorporates the physical effects corresponding to the left-hand cut. This is illustrated in the comparison between \Cref{fig:tchannel} and \Cref{fig:Gdiag}. In the former, pion exchange leads to the left-hand cut in the $DD^*$ K matrix, which invalidates the two-particle formalism, while in the latter, pion exchange is included as it leads to a three-particle intermediate state.\footnote{%
The pion exchange Feynman diagram also includes a time ordering in which the intermediate state has five particles. It is straightforward to see that this time ordering, when considered at the $D^*$ pole, does not contribute to the left-hand cut in the $D^* D$ amplitude, and thus does not need to be included explicitly in the finite-volume formalism. Its contribution is contained in the infinite-volume quantities that appear in the formalism.}
In particular, it is associated with the ``switch-matrix" $G$ in the three-particle formalism.

\begin{figure}[!h]
\centering
\begin{subfigure}[b]{0.49\textwidth}
\includegraphics[width=\linewidth]{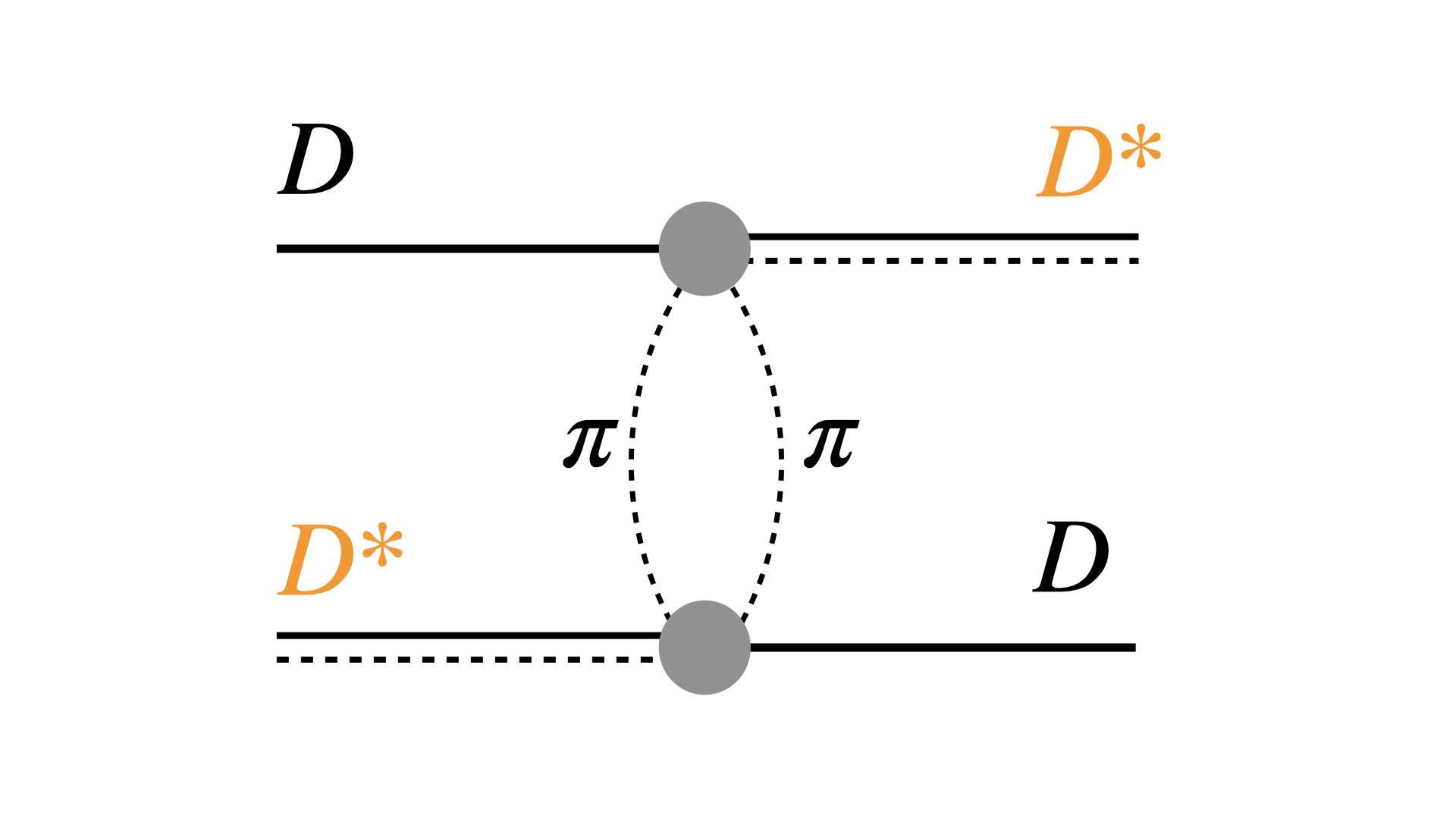}
\caption{}
\label{fig:2pi_u}
\end{subfigure}
\hfill
\begin{subfigure}[b]{0.49\textwidth}
\includegraphics[width=\linewidth]{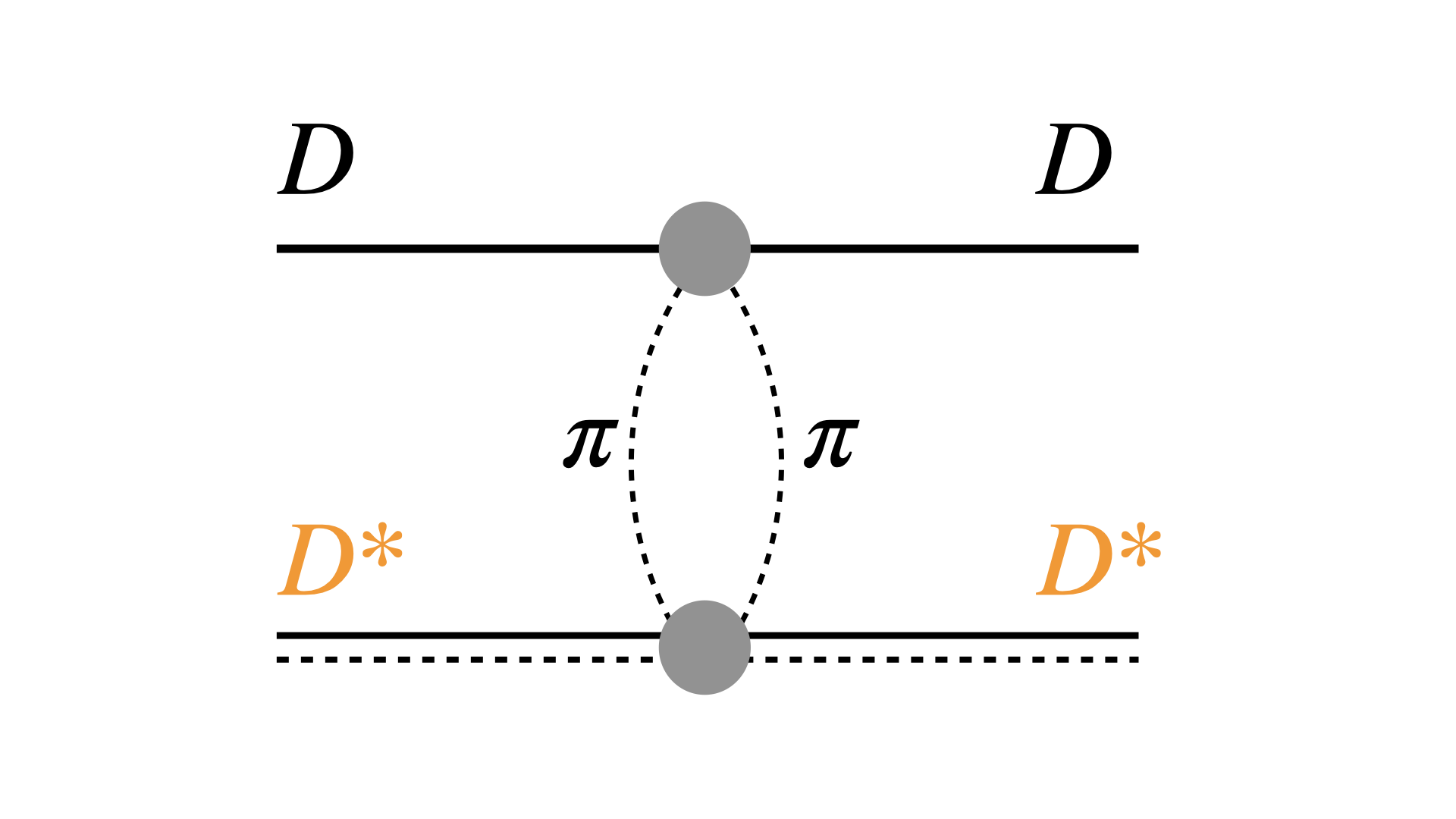}
\caption{}
\label{fig:2pi_t}
\end{subfigure}
\caption{Two-pion exchange processes giving rise to the two-pion left-hand cut that invalidates the three-particle formalism for $s \lesssim s_{2\pi, u, {\rm lhc}}$. {\bf Left:} $u$-channel two-pion exchange. {\bf Right:} $t$-channel two-pion exchange.\label{fig:2picut}}
\end{figure}

In summary, the approach that we are proposing when the $D^*$ is stable is to use the three-particle quantization condition applied to states with $DD^*$ quantum numbers, including $DD\pi$ states, in order to obtain the $DD\pi$ divergence-free three-particle K matrix.\footnote{%
Further details on the input required from lattice QCD are described in \Cref{sec:latticeinputs}.}
One can then solve the integral equations relating this K matrix to the $DD\pi$ scattering amplitude. Continuing this amplitude below threshold, going to the $D^*$ poles in the $D\pi$ channel, and using the LSZ reduction formula, one obtains the $DD^*$ amplitude. This amplitude accounts for pion exchange, and can be studied below threshold to search for (virtual) bound states, irrespective of their position relative to the left-hand cut. This approach will break down only when one is far enough below the $D D^*$ threshold that the nonanalyticity due to two-pion exchange appears, either as t- or u-channel exchanges, as depicted in \Cref{fig:2picut}. In the u-channel, this starts at $u=4 M_\pi^2$ and $t=0$, so that $s= s_{2\pi,\rm lhc}$, with
\begin{equation}
s_{2\pi, u , \rm lhc } - s_{\rm th} = -4M_\pi^2 + (M_{D^*} - M_{D})^2\, ,
\end{equation}
while in the t-channel, the exchange implies $t=4 M_\pi^2$, so that
\begin{equation}
s_{2\pi, t , \rm lhc } - s_{\rm th} = 2\sqrt{M_D^2 - M_\pi^2}\sqrt{M_{D^*}^2 - M_\pi^2} - 2M_D M_{D^*} -2M_\pi^2 \,.
\end{equation}
In the numerical example given in \Cref{eq:numex} above, the cut associated with u-channel two-pion exchange is the one closest to threshold, and it occurs at $E_{\rm cm}= 3937\,$MeV,
i.e.~29 MeV below the first left-hand cut.

This approach has been successfully carried out in a model theory of three identical particles with the two-particle $s$-wave scattering length chosen so that there is a bound state~\cite{Romero-Lopez:2019qrt,Jackura:2020bsk,Dawid:2023jrj,Dawid:2023kxu}. In particular, the integral equations, solved in refs.~\cite{Jackura:2020bsk,Dawid:2023jrj,Dawid:2023kxu}, allow the extraction of the particle $+$ bound-state amplitude below the left-hand cut. The failure of the L\"uscher formalism applied to finite-volume particle $+$ bound-state levels at or below this cut is also clearly seen~\cite{Dawid:2023jrj}. Indeed, it was this work that inspired the present proposal.

\section{Formalism}
\label{sec:formalism}

In this section we describe the formalism, obtained in the RFT approach, needed to study $DD\pi$ systems using lattice QCD. Here we present only the results; the derivation is provided in two appendices. In the first, \Cref{app:topt_derivation}, we give a complete derivation using the TOPT approach of ref.~\cite{\BSQC}, generalizing previous work for nonidentical scalars~\cite{Blanton:2020gmf} and ``2+1'' systems such as $K^+K^+\pi^+$~\cite{Blanton:2021mih}. In the second, \Cref{app:flavor_derivation}, we provide an intuitive derivation based on generalizing the approach used in ref.~\cite{\isospin} for the case of three pions of general isospin. The key generalization compared to previous work is including the effects of having $DD\pi$ systems with different quantum numbers, i.e.~$D^+D^+\pi^-$, $D^+ D^0 \pi^0$, and $D^0D^0\pi^+$.

\subsection{\texorpdfstring{$DD\pi$}{D D pion} states}
\label{sec:DDpi_states}

We first describe the different three-meson states needed for the formalism. We work in isospin-symmetric QCD, which leads to the considerable simplification that all the $DD\pi$ states are degenerate. In particular, the $D^+$ and $D^0$ pair are mass-degenerate and form an isospin doublet, with positive third component in the case of $D^+$. Similarly, the charged and neutral pions are part of an isospin 1 triplet. The $DD\pi$ system thus can have total isospins given by
\begin{equation}
\frac{\mathbf 1}{\mathbf 2} \otimes \frac{\mathbf 1}{\mathbf 2} \otimes {\mathbf 1}
= \mathbf 0 \oplus \mathbf 1 \oplus \mathbf 1 \oplus \mathbf 2\,,
\label{eq:isospindecomp}
\end{equation}
using the standard notation for irreps of SU(2). We will present the formalism for all three choices of total isospin. The appearance of two $I=1$ irreps on the right-hand side of \Cref{eq:isospindecomp} indicates that, for noninteracting particles, there are two physically distinct states with $I=1$, and thus that the corresponding scattering amplitude in the interacting theory will involve two channels and three independent amplitudes (assuming $PT$ symmetry). This means that, in total, there are five independent three-particle scattering amplitudes: one each for isospin 0 and 2, and three for isospin 1.

It will suffice to focus on the sector with one unit of electric charge and charm $C=+2$.
This is sufficient to ensure that all allowed total isospin states are represented.
In the basis of definite meson flavor (flavor basis), we find it useful to introduce four states,
\begin{align}
\begin{split}
& \ket{D^+(k_1) D^0(k_2) \pi^0(k_3)} , \quad \ket{D^0(k_1) D^+(k_2) \pi^0(k_3)}\,, \\
& \ket{D^+(k_1) D^+(k_2) \pi^-(k_3)} , \quad \ket{D^0(k_1) D^0(k_2) \pi^+(k_3)}\,.
\label{eq:ddpi_flavor_states}
\end{split}
\end{align}
Here $k_i$ label the four-momenta of the particles, and in the first line the two states differ only in the momentum assignments. States of definite total isospin can then be constructed as
\begin{align}
\begin{split}
\ket{[D D]_1 \pi}_{2} &= \sqrt{\frac16} \left(\ket{D^0 D^0 \pi^+} + \ket{D^+ D^+ \pi^-} + \sqrt{2}\ket{D^+ D^0 \pi^0} + \sqrt{2}\ket{D^0 D^+ \pi^0} \right), \\
\ket{[D D]_1 \pi}_{1,a} &= \sqrt{\frac12} \left(\ket{D^+ D^+ \pi^-} - \ket{D^0 D^0 \pi^+} \right) ,\\
\ket{[D D]_0 \pi}_{1,b} &= \sqrt{\frac12} \left( \ket{D^+ D^0 \pi^0} - \ket{D^0 D^+ \pi^0} \right) ,\\
\ket{[D D]_1 \pi}_{0} &= \sqrt{\frac13} \left(\ket{D^0 D^0 \pi^+}+ \ket{D^+ D^+ \pi^-}
- \sqrt{\frac12}\ket{D^+ D^0 \pi^0} - \sqrt{\frac12} \ket{D^0 D^+ \pi^0} \right)\,,
\label{eq:ddpi_isospin_states}
\end{split}
\end{align}
where momentum labels [ordered as in \Cref{eq:ddpi_flavor_states}] have been omitted for brevity. The rest of the notation is exemplified by the meaning of $\ket{[D D]_1 \pi}_{1,a}$, which indicates the $DD$ subsystem has isospin $1$, while the three-meson system has isospin $1$. The two total isospin 1 states noted above are labelled as ``$1,a$" and ``$1,b$".

In the isospin decomposition, we could instead first combine a $D$ and a $\pi$, instead of the choice made in \Cref{eq:isospindecomp} of first combining the two $D$s. This leads to linear combinations of the states in \Cref{eq:ddpi_isospin_states}:
\begin{align}
\begin{split}
\ket{D [D\pi]_{3/2}}_{2} &= \ket{(D D)_1 \pi}_{2}, \\
\ket{D [D\pi]_{3/2}}_{1,a'} &= \sqrt{\frac13} \ket{(D D)_1 \pi}_{1,a} + \sqrt{\frac23} \ket{(D D)_0 \pi}_{1,b},\\
\ket{D [D\pi]_{1/2}}_{1,b'} &= -\sqrt{\frac23} \ket{(D D)_1 \pi}_{1,a} +\sqrt{\frac13} \ket{(D D)_0 \pi}_{1,b}, \\
\ket{D [D\pi]_{1/2}}_{0} &= \ket{(D D)_1 \pi}_{0}.
\label{eq:ddpi_isospin_states2}
\end{split}
\end{align}

\subsection{Quantization condition}
\label{sec:QC}

Energy levels in finite volume are given by solutions of the three-particle quantization condition,
which in the RFT approach takes the form
\begin{equation}
\underset{k, \ell, m, i}{\operatorname{det}}
\left[1+ \widehat{\mathcal{K}}_{\mathrm{df}, 3} \ \widehat{F}_3\right]=0\,.
\label{eq:QC3}
\end{equation}
Here the ``hatted'' quantities are matrices in a space labeled by $\{k\ell m i\}$, where $k$ (which is shorthand for $\boldsymbol{k}$) labels the three momentum of the spectator particle (which is drawn from the finite-volume set $(2\pi/L) \mathbb Z^3$), $\ell,m$ are partial-wave indices for the interacting pair, and $i$ is a channel index related to the flavors of the particles and the choice of spectator. The quantity $\widehat{\mathcal{K}}_{\mathrm{df}, 3}$ is the divergence-free three-particle K matrix, while $\widehat{F}_3$ is a matrix that depends on kinematical, volume-dependent functions as well as the two-particle K matrix. A finite-dimensional quantization condition is obtained by (1) neglecting interactions in two-particle partial waves for $\ell > \ell_\text{max}$, and (2) truncating the spectator-momentum using a smooth cutoff function, $H^{(i)}(\boldsymbol k)$.
A suitable form for the cutoff function, applicable to nondegenerate systems,
 is given in eqs.~(2.24) and (2.25) of ref.~\cite{Blanton:2021eyf}.
 The precise forms we use here will be explained below.

We first consider the size of the flavor space (i.e. the values taken by the index $i$). There are three $DD\pi$ systems, $D^+D^+\pi^-$, $D^0D^0\pi^+$ and $D^0D^+\pi^0$. The first two consist of two identical particles and a distinct third (so-called ``$2+1$'' systems), while the last has three distinguishable particles (a ``$1+1+1$'' system). Since the matrices in the quantization condition for $2+1$ systems are of dimension 2~\cite{Blanton:2021mih}, while those for $1+1+1$ systems are of dimension 3~\cite{Blanton:2020gmf}, one would naively expect a $2+2+3=7$-dimensional flavor space. This corresponds to all the different choices for the flavor of the spectator. In fact, it is advantageous to use an $8$-dimensional space, in which the parts of the $D^+D^0\pi^0$ state that are symmetric and antisymmetric under $DD$ exchange are separated. This is because these two parts correspond to different isospins for the $DD$ subsystem ($I=1$ and $0$, respectively). Further details concerning this expansion to $8$ dimensions are given in \Cref{app:7to8}, as well as in \Cref{app:Bspace}.

Since total isospin is conserved, when the quantization condition is expressed in the total-isospin basis (a transformation discussed in \Cref{app:converttoI,app:BfinalQC3}) all the matrices that it contains block diagonalize, and correspondingly the quantization condition factorizes as
\begin{equation}
\prod_{I \in \{0,1,2\}} \underset{k, \ell, m, i}
{\operatorname{det}}\left[1+ \widehat{\mathcal{K}}^{[I]}_{\mathrm{df}, 3}
\ \widehat{F}^{[I]}_3\right]=0,
\label{eq:QC3isospin}
\end{equation}
where the superscript $[I]$ labels the total isospin of the system. The $I=0$ and $2$ blocks have dimension two, while the $I=1$ block has dimension four. These dimensions can be understood as follows. For $I=0,2$ there are two choices of spectator flavor, either the $D$ or the $\pi$. For $I=1$ there are, in addition, two independent channels (labeled $a$ and $b$ above), thus doubling the dimensionality to four. The ordering of indices in the total-isospin basis is given by
\begin{equation}
\begin{pmatrix}
[[D\pi]_{3/2} D]_{I_{\rm tot}=2}\\
[[DD]_1\pi]_{I_{\rm tot}=2} \\
[[D\pi]_{3/2} D]_{I_{\rm tot}=1}\\
[[D\pi]_{1/2} D]_{I_{\rm tot}=1}\\
[[DD]_1\pi]_{I_{\rm tot}=1}\\
[[DD]_0\pi]_{I_{\rm tot}=1}\\
[[D\pi]_{1/2} D]_{I_{\rm tot}=0} \\
[[DD]_1\pi]_{I_{\rm tot}=0}
\end{pmatrix}\,,
\label{eq:totalisospinorder}
\end{equation}
where the third flavor is that of the spectator, while the isospin of the remaining pair is as indicated.

A caveat of the quantization condition in \Cref{eq:QC3isospin} that must be taken into account when describing the finite-volume spectrum involves the mixing between $DD$ and $DD\pi$ (or $DD^*$) states that can occur for non-maximal isospin channels. Thus, strictly speaking, \Cref{eq:QC3isospin} is only applicable for irreducible representations (irreps) of the finite-volume symmetry group in which $DD$ states are absent. In the third item of \Cref{sec:latticeinputs} below, we describe what this implies for the $I=0$ case in which the $T_{\rm cc}^+$ is present.

We now discuss the building blocks of the quantization condition,
$\widehat{F}_3^{[I]}$ and $\widehat{\mathcal{K}}^{[I]}_{\mathrm{df}, 3}$.
The former is given by the standard RFT form
\begin{equation}
\widehat{F}_3^{[I]} \equiv \frac{\widehat{F}^{[I]}}{3}-\widehat{F}^{[I]} \frac{1}{1+\widehat{\mathcal{M}}_{2, L}^{[I]}
\widehat{G}^{[I]}} \widehat{\mathcal{M}}_{2, L}^{[I]} \widehat{F}^{[I]}, \quad \quad \widehat{\mathcal{M}}_{2, L}^{[I]}
\equiv \frac{1}{\widehat{\mathcal{K}}_{2,L}^{[I]-1}+\widehat{F}^{[I]}}\,,
\label{eq:F3def}
\end{equation}
where now each quantity is a matrix both in flavor space (as denoted by the hat),
as well as implicitly in $\{k \ell m \}$ space.
$\widehat F$ is a kinematical quantity, closely related to the L\"uscher zeta function, with matrix structure
\begin{align}
\widehat F^{[I=2]} &= \widehat F^{[I=0]} = \text{diag }\left( \widetilde F^{D}, \widetilde F^{\pi} \right),
\label{eq:FhatI20}
\\
\widehat F^{[I=1]} &=\text{diag }\left( \widetilde F^{D}, \widetilde F^{D}, \widetilde F^{\pi}, \widetilde F^{\pi} \right)\,.
\label{eq:FhatI1}
\end{align}
where~\cite{\BSnondegen,\BStwoplusone}
\begin{multline}
\left[\wt F^{(i)}\right]_{p' \ell' m';p \ell m} =
\delta_{\bm p' \bm p} \frac{H^{(i)}(\bm p)}{2\omega_{p}^{(i)} L^3}
\left[ \frac1{L^3} \sum_{\bm a}^{\rm UV} - \PV \int^{\rm UV} \frac{d^3 a}{(2\pi)^3} \right]
\\
\times \left[
\frac{\cY_{\ell' m'}(\bm a^{*(i,j,p)})}{\big(q_{2,p'}^{*(i)}\big)^{\ell'}}
\frac1{4\omega_{a}^{(j)} \omega_{b}^{(k)}
\big(E\!-\!\omega_{p}^{(i)}\!-\!\omega_{a}^{(j)}\!-\!\omega_{b}^{(k)}\big)}
\frac{\cY^*_{\ell m}(\bm a^{*(i,j,p)})}{\big(q_{2,p}^{*(i)}\big)^{\ell}}
\right]
\,.
\label{eq:Ft}
\end{multline}
The superscripts $i$, $j$, and $k$ label flavors: $i$ indicates that of the spectator
while $j$ and $k$ are those of the pair, ordered as in \Cref{eq:totalisospinorder}.\footnote{%
We stress that if the pair are $D\pi$, and thus nondegenerate, our use of the convention $j=D$ and $k=\pi$ (which is consistent with that of ref.~\cite{\BStwoplusone}) leads to specific positions for the factors of $P^{(\ell)}$ appearing below in \Cref{eq:GhatI20,eq:GhatI1}.}
The superscript UV on the sum and integral indicates that the sum-integral difference must be regularized in some manner---see Appendix A of ref.~\cite{Blanton:2021eyf} for further discussion. On-shell energies are given by
\begin{equation}
\omega_{p}^{(i)} = \sqrt{ \bm p^2 + m_i^2}\,,
\label{eq:omegadef}
\end{equation}
and the momentum needed for $\omega_b^{(k)}$ is $\bm b = \bm P - \bm a -\bm p$. The $\cY_{\ell m}$ are harmonic polynomials normalized as $\cY_{\ell m}(\bm a) = \sqrt{4\pi}Y_{\ell m}(\hat a) |\bm a|^\ell $. The quantity $q_{2,p}^{*(i)}$ is the magnitude of the relative momentum of the nonspectator pair in their center-of-momentum frame (CMF), assuming that all three particles are on shell. Finally, $\bm a^{*(i,j,p)}$ is the spatial part of the four momentum $(\omega_a^{(j)}, \bm a)$ after boosting it with boost velocity $\bm \beta_p^{(i)} = -(\bm P -\bm p)/(E-\omega_a^{(j)})$, i.e., to the CMF of the nonspectator pair. For explicit expressions for the kinematic quantities see, e.g., Appendix A of ref.~\cite{\BStwoplusone}.

The cutoff function is explicitly defined as
\begin{equation}
H^{(i)}(\boldsymbol p) = J \big (z_i(\boldsymbol p) \big), \qquad z_i(\boldsymbol p) = (1+\epsilon_H)\frac{\sigma_i(\boldsymbol p) - \sigma^{\rm min}_{i}}{ \sigma^{\rm th}_{i} - \sigma^{\rm min}_{i}},
\label{eq:cutoff}
\end{equation}
where $J(z)$ is given, for instance, in eq.~(2.23) of ref.~\cite{Blanton:2021eyf}. This function is defined to vanish for $z<0$, to equal unity for $z \geq 1$, and to vary smoothly and monotonically in the region $0 < z < 1$. The argument of $J$ is expressed in terms of $\sigma_i(\bm p) = (E- \omega^{(i)}_p)^2 - (\bm P - \bm p)^2$,\linebreak$\sigma_i^{\rm th} = (m_j+m_k)^2$, and $\sigma_i^{\rm min}$. The last of these, $\sigma_i^{\rm min}$, specifies the value of $\sigma_i$ below which $z_i(\boldsymbol p)$ goes negative and therefore the cutoff function vanishes. This must be chosen so that left-hand cuts in the corresponding two-particle scattering amplitude occur in the region where the cutoff vanishes. If the spectator is the $D$ meson, $\sigma_D^{\rm min} \geq M_D^2 - M_\pi^2$  ensures that the two-pion exchange left-hand cut in $D\pi$ interactions is avoided. If the spectator is a pion, $\sigma_\pi^{\rm min} \geq 4M_D^2 - 4M_\pi^2$ avoids the two-pion left-hand cut in the $DD$ scattering amplitude. Finally, $\epsilon_H$ is a small positive number, included in eq.~\eqref{eq:cutoff} so that the value of $\sigma_i(\boldsymbol p)$ for which $J(z_i(\boldsymbol p))$ first deviates from unity is slightly below $\sigma_i^{\rm  th }$.

The second kinematic quantity is $\widehat{G}$, which accounts for finite-volume effects in diagrams where the spectator is switched. Its matrix form, determined in \Cref{app:symmon,app:7to8,app:converttoI} as well as in \Cref{app:Bblocks,app:BfinalQC3}, is
\begin{align}
\begin{split}
\widehat{G}^{[I=2]} &=
\begin{pmatrix}
\wt G^{DD} & \sqrt2 P^{(\ell)}\, \wt G^{D\pi} \\
\sqrt2\, \wt G^{\pi D} P^{(\ell)} & 0
\end{pmatrix}\,,
\label{eq:GhatI20}
\\[5pt]
\widehat{G}^{[I=0]} & =
\begin{pmatrix}
\wt G^{DD} & -\sqrt2 P^{(\ell)}\, \wt G^{D\pi} \\
-\sqrt2\, \wt G^{\pi D} P^{(\ell)} & 0
\end{pmatrix}\,,
\end{split}
\\[5pt]
\widehat{G}^{[I=1]} &=
\begin{pmatrix}
- \frac13 \wt G^{DD}
& -\sqrt{\frac89} \wt G^{DD} & -\sqrt{\frac23} P^{(\ell)}\, \wt G^{D\pi} & -\sqrt{\frac43} P^{(\ell)}\, \wt G^{D\pi}
\\
-\sqrt{\frac89} \wt G^{DD}
& \frac13 \wt G^{DD} & \sqrt{\frac43} P^{(\ell)}\,\wt G^{D\pi} & -\sqrt{\frac23} P^{(\ell)}\, \wt G^{D\pi} \\
-\sqrt{\frac23} \wt G^{\pi D} P^{(\ell)}& \sqrt{\frac43} \wt G^{\pi D} P^{(\ell)} & 0 & 0
\\
-\sqrt{\frac43} \wt G^{\pi D} P^{(\ell)}& -\sqrt{\frac23} \wt G^{\pi D} P^{(\ell)} & 0 & 0
\end{pmatrix}\,.
\label{eq:GhatI1}
\end{align}

The superscripts on $\wt G$ indicate the flavors of the spectator particles after and before the switch. For instance, $G^{\pi D}$ means that the outgoing spectator is $\pi$ and the incoming spectator is $D$. Each entry is given by~\cite{\BSnondegen,\BStwoplusone}
\begin{equation}
\left[\wt G^{(ij)}\right]_{p \ell' m';r \ell m} =
\frac1{2\omega^{(i)}_{p} L^3}
\frac{\cY_{\ell' m'}(\bm r^{*(i,j,p)})}{\big(q_{2,p}^{*(i)}\big)^{\ell'}}
\frac{H^{(i)}(\bm p) H^{(j)}(\bm r)}{b_{ij}^2-m_k^2}
\frac{\cY^*_{\ell m}(\bm p^{*(j,i,r)})}{\big(q_{2,r}^{*(j)}\big)^{\ell}}
\frac1{2\omega^{(j)}_{r} L^3}\,,
\label{eq:Gt}
\end{equation}
where $k$ labels the flavor of the particle that is neither the initial nor the final spectator, and thus is the flavor of the particle that is switched. Its four-momentum is given by ${b_{ij}=(E\!-\!\omega^{(i)}_{p}\!-\!\omega^{(j)}_{r}, \bm P \!-\! \bm p \!-\! \bm r)}$. The matrix $P^{(\ell)}$ is
\begin{equation}
[P^{(\ell)}]_{p \ell' m';r \ell m} = \delta_{\bm p \bm r}\delta_{\ell' \ell} \delta_{m' m} (-1)^\ell \,.
\label{eq:Pelldef}
\end{equation}
Note that \Cref{eq:Gt} uses the explicitly Lorentz invariant version of $G$, introduced in ref.~\cite{Briceno:2018aml}. This form is achieved by adding a term that accounts for the particle propagating backwards in time to the original RFT $G$ function~\cite{Hansen:2014eka}. This does not introduce additional complications in the formalism, since it is simply a smooth term in the energy region allowed by the cutoff.

The final component of $\widehat F_3$ is the two-particle K matrix, given by
\begin{align}
\widehat{\cK}_{2,L}^{[I=2]} &=
\text{diag}\left({\cK}_{2,L}^{D\pi, I=3/2}, \frac{1}{2} {\cK}_{2,L}^{DD, I=1} \right)\,,
\label{eq:KhatI2}
\\
\widehat{\cK}_{2,L}^{[I=1]} &=
\text{diag}\left({\cK}_{2,L}^{D\pi, I=3/2}, {\cK}_{2,L}^{D\pi, I=1/2},\frac{1}{2} {\cK}_{2,L}^{DD, I=1},
\frac{1}{2}{\cK}_{2,L}^{DD, I=0} \right),
\label{eq:KhatI1}
\\
\widehat{\cK}_{2,L}^{[I=0]} &=
\text{diag}\left( {\cK}_{2,L}^{D\pi, I=1/2}, \frac{1}{2} {\cK}_{2,L}^{DD, I=1} \right),
\label{eq:KhatI0}
\end{align}
where ${\cK}_{2,L}^{jk, I}$ is related to the two-particle K matrix for the $\{jk\}$ system with isospin $I$~\cite{\HSQCa,\BHSnum,\dwave,\BSnondegen},
\begin{align}
\left[{\cK}_{2,L}^{jk, I\,}\right]_{p\ell' m'; r \ell m} &=
\delta_{\bm p \bm r} 2\omega^{(i)}_{r} L^3 \delta_{\ell' \ell} \delta_{m' m} \left[\cK^{jk, I}_2(\bm r) \right]_{\ell}\,,
\label{eq:K2Li}
\\
\left[\cK^{jk,I\,}_2(\bm r)^{-1}\right]_{\ell}
&=
\frac{ \eta_i}{8 \pi \sqrt{\sigma_i}}
\left\{ q_{2,r}^{*(i)} \cot \delta_\ell^{(j,k), I}(q_{2,r}^{*(i)}) + |q_{2,r}^{*(i)}| [1-H^{(i)}(\bm r)] \right\} \,.
\label{eq:K2}
\end{align}
Here $i$ is the flavor of the spectator, and the spherical harmonics are defined relative to particle of flavor $j$, or to the $D^+$ in the case of the $[DD]_0$. The symmetry factor is $\eta_D=1$ and $\eta_\pi=1/2$. We note only even partial waves appear in $\overline{\cK}_{2,L}^{(DD), I=1\,}$, while $\overline{\cK}_{2,L}^{(DD), I=0\,}$ contains only odd waves. Note that it is expected that low-lying resonances or bound states are present in some of these K matrices. This can lead to spurious poles in finite-volume, which can be cured using the modifications presented in ref.~\cite{Romero-Lopez:2019qrt}. Following this approach, only $\cK_2$ and $F$ must be changed with a factor proportional to $I_{PV}$---see eqs. (3.2) and (3.3) of that work.

Finally, we comment on the structure of the matrix $\widehat{\cK}_{\df,3}$. Determining this is nontrivial, and we describe two approaches in \Cref{app:Kdfform,app:Bkdf}, both of which lead to the same result. For both $I=0$ and $2$, this result has the following outer product form
\begin{align}
\widehat{\cK}_\text{df,3}^{[I=0,2]} &=
{\boldsymbol{\cY}^{[I=0,2]}}\circ \cK_{\df,3}^{[I=0,2]}(\{p_i\}; \{k_i\})\circ \boldsymbol{\cY}^{[I=0,2]\dagger}\,,
\label{eq:KdfI02}
\\[5pt]
\begin{split}
{\boldsymbol{\cY}^{[I=2]}} & = \begin{pmatrix}
\YR123 \\
\sqrt{\tfrac12} \YR312
\end{pmatrix}\,,
\quad
{\boldsymbol{\cY}^{[I=2]}}^\dagger = \begin{pmatrix}
\YL123, & \sqrt{\frac12} \YL312
\end{pmatrix}\,. \\
{\boldsymbol{\cY}^{[I=0]}} & = \begin{pmatrix}
-\YR123 \\
\sqrt{\tfrac12} \YR312
\end{pmatrix}\,,
\quad
{\boldsymbol{\cY}^{[I=0]}}^\dagger = \begin{pmatrix}
-\YL123, & \sqrt{\frac12} \YL312
\end{pmatrix}\,.
\label{eq:v0def}
\end{split}
\end{align}
The operators $\boldsymbol{\mathcal Y}^{[kab]}_{\boldsymbol \sigma}$ and $\boldsymbol{\mathcal Y}^{[kab]\dagger}_{\boldsymbol\sigma}$ are defined in \Cref{eq:YRdef} and the subsequent discussion. They convert from the momentum basis to the $\{k\ell m\}$ basis. For each choice of isospin, all entries in the $2\times 2$ matrices $\widehat{\cK}_\text{df,3}^{[I=0,2]}$ contains the same kinematic function, but expressed in different coordinates. The momentum labels in this function are given by those in \Cref{eq:ddpi_flavor_states}, combined into isospin states following \Cref{eq:ddpi_isospin_states}. Note that for both $I=0,2$ the underlying function is symmetric under the separate exchanges $p_1\leftrightarrow p_2$ and $k_1\leftrightarrow k_2$, corresponding to the interchange of the momenta of the two $D$s.

For $I=1$, there are four underlying functions, corresponding to a $2\times 2$ matrix structure relating the $I=1,a$ and $I=1,b$ states in \Cref{eq:ddpi_isospin_states}. These are $\cK_{\df,3}^{[I=1],aa}$, $\cK_{\df,3}^{[I=1],ab}$, $\cK_{\df,3}^{[I=1],ba}$, and $\cK_{\df,3}^{[I=1],bb}$, each of which is a function of the final and initial momenta.\footnote{Note that $a$ and $b$ are used to indicate both momenta and isospin states. The momenta always appear in the package $[kab]$ while the isospin labels always appear as $[I=1]$ followed by either $a$ or $b$.} PT symmetry relates the $ab$ and $ba$ forms, with final and initial momenta interchanged. Those with an $a$ index are symmetric under interchange of the corresponding $D$ momenta, while those with a $b$ index are antisymmetric. The form of $\widehat{\cK}_{\df,3}$ is a sum of outer products,
\begin{equation}
\widehat{\cK}_{\rm df,3}^{[I=1]} = \sum_{x,y \in \{a,b\}}
\boldsymbol {\mathcal Y}^{[I=1],x} \circ \cK_{\df,3}^{[I=1],xy} \circ \boldsymbol {\mathcal Y}^{[I=1],y\dagger}
\,,
\label{eq:Kdf3I1form}
\end{equation}
where we have suppressed the momentum arguments of the K matrices, and the vectors are
\begin{align}
\boldsymbol {\mathcal Y}^{[I=1],a\dagger} &= \left(
-\sqrt{\tfrac13} \YL123, \
\sqrt{\tfrac23} \YL123,\
\sqrt{\tfrac12} \YL312,\
0
\right)\,,
\label{eq:YI1a}
\\
\boldsymbol {\mathcal Y}^{[I=1],b\dagger} &= \left(
-\sqrt{\tfrac23}\YL123, \
-\sqrt{\tfrac13}\YL123,\
0, \
\sqrt{\tfrac12}\YL312
\right)\,.
\label{eq:YI1b}
\end{align}

The formalism for the $I=2$ $DD\pi$ system is mathematically identical to that for $K^+K^+\pi^+$ systems~\cite{Blanton:2021mih}, aside from trivial changes in kinematics due to the difference between the $D$ and $K$ masses. This is as expected, since although we have considered the $m=0$ component of the $I=2$ system, the result must be the same as for the $m=2$ component, i.e.~the $D^+D^+\pi^+$ system. We also observe that the form of equations is similar between the $I=0$ and the $I=2$ $DD\pi$ systems. Two key differences arise: the first is that the signs of the off-diagonal elements of $G$ are opposite between $I=0$ and $I=2$, and the second is that the $I=1/2$ $D \pi$ amplitude is required for total isospin zero while the $I=3/2$ $D \pi$ amplitude is needed for total isospin two. The overall similarities mean that details of the implementaiton of $K^+ K^+ \pi^+$ can be used for both $I=0$ and $I=2$ $DD\pi$ systems with minimal changes.

\subsection{Integral equations relating \texorpdfstring{$\Kdf$}{the divergence-free K-matrix} to \texorpdfstring{$\cM_3$}{the scattering amplitude}}
\label{sec:inteqs}

The relation between the divergence-free three-particle K matrix and the 3-to-3 scattering amplitude is given by integral equations~\cite{Hansen:2015zga}. Here we summarize these for each of the isospin channels. Details of their derivation are given in \Cref{app:inteqs,app:BKtoM}. The expressions below are closely related to those of refs.~\cite{Blanton:2020gmf,Blanton:2021eyf}, and in particular identical to ref.~\cite{Blanton:2021eyf} in the case of $I=2$.

The infinite-volume amplitude in each isospin channel is obtained by first performing an ordered double limit
\begin{equation}
\widehat{\cM}_{3}^{[I]} = \lim_{\epsilon \to 0^+} \lim_{L\to \infty} \widehat{\cM}_{3,L}^{[I]} \,,
\label{eq:M3hatI}
\end{equation}
where $\epsilon$ is first introduced in the denominators of $\wt F$ and $\wt G$,
\Cref{eq:Ft,eq:Gt}.
The $L\to\infty$ limit will convert the matrix equations shown below into the desired integral equations, as explained in ref.~\cite{\HSQCb}. While \Cref{eq:M3hatI} corresponds to the formal definition of the integral equations, it does not provide the most practical procedure to solving them. Solutions can be more efficiently obtained from partial-wave projected integral equations directly in infinite volume, and using contour deformations~\cite{Sadasivan:2020syi,Pang:2023jri,Dawid:2023jrj,Jackura:2023qtp}.

The finite-volume amplitude $\widehat{\cM}_{3,L}^{[I]}$ is constructed from the corresponding unsymmetrized amplitude (denoted by a $(u,u)$ superscript) using
\begin{align}
\begin{split}
\widehat{\cM}_{3,L}^{[I=0,\,2]} &=
\boldsymbol{\mathcal X}_{0,2} \circ %
\widehat{\cM}_{3,L}^{(u,u), [I=0,\,2]}
\circ \boldsymbol{\mathcal X}_{0,2}^\dagger %
\\
\widehat{\cM}_{3,L}^{[I=1]} &= \begin{pmatrix}
\boldsymbol{\mathcal X}_a \circ \widehat{\cM}_{3,L}^{(u,u), [I=1]} \circ \boldsymbol{\mathcal X}_a^\dagger&\
\boldsymbol{\mathcal X}_a \circ \widehat{\cM}_{3,L}^{(u,u), [I=1]} \circ \boldsymbol{\mathcal X}_b^\dagger
\\
\boldsymbol{\mathcal X}_b \circ \widehat{\cM}_{3,L}^{(u,u), [I=1]} \circ \boldsymbol{\mathcal X}_a^\dagger &\
\boldsymbol{\mathcal X}_b \circ \widehat{\cM}_{3,L}^{(u,u), [I=1]} \circ \boldsymbol{\mathcal X}_b^\dagger
\end{pmatrix}\,.
\label{eq:M3L}
\end{split}
\end{align}
Here, the unsymmetrized finite-volume amplitude is given by
\begin{equation}
\widehat{\cM}_{3,L}^{(u,u), [I]} = \widehat{\cD}_L^{(u,u), [I] }+ \widehat{\cM}_{\df,3,L}^{(u,u)\prime, [I]}\,,
\label{eq:MhatuuI}
\end{equation}
and is composed of the ladder amplitude, which contains pairwise rescattering,
\begin{align}
\widehat{\cD}_L^{(u,u), [I]} &= - \widehat{\cM}_{2,L}^{[I]} \widehat G^{[I]} \widehat{\cM}_{2,L}^{[I]}
\frac1{1 + \widehat{G}^{[I]} \widehat{\cM}_{2,L}^{[I]} }\,,
\label{eq:DhatuuI}
\end{align}
and a short-distance piece that depends on the divergence-free three-particle K matrix
\begin{align}
\widehat{\cM}_{\df,3,L}^{(u,u)\prime, [I]} &= \left[ \frac13 - \widehat{\cD}_{23,L}^{(u,u), [I]} \widehat F^{[I]} \right]
\widehat{\cK}_{\df,3}^{[I]} \frac1{1 + \widehat F^{[I]}_3 \widehat{\cK}_{\df,3}^{[I]} }
\left[\frac13 - \widehat F^{[I]} \widehat{\cD}_{23,L}^{(u,u), [I]} \right]\,,
\label{eq:Mhatdf3LI}
\end{align}
where
\begin{equation}
\widehat{\cD}_{23,L}^{(u,u), [I]} = \widehat{\cM}_{2,L}^{[I]} + \widehat{\cD}_{L}^{(u,u), [I]}\,.
\label{eq:Dhat23Ldef}
\end{equation}

The final ingredient is the vectors that project the symmetrized amplitudes from the unsymmetrized ones in \Cref{eq:M3L}, by combining contributions with different spectators. These are given by
\begin{align}
\begin{split}
\boldsymbol{\mathcal X}_0
& = \begin{pmatrix}
-\XR123 - \XR213,& {\sqrt2} \XR312
\end{pmatrix}\,,
\\[5pt]
\boldsymbol{\mathcal X}_2
&= \begin{pmatrix}
\XR123+ \XR213,& {\sqrt2} \XR312
\end{pmatrix}\,,
\label{eq:alphaSvec}
\end{split}
\\[5pt]
\boldsymbol{\mathcal X}_a &= \begin{pmatrix}
-\sqrt{\tfrac13}[\XR123+\XR213], & \sqrt{\tfrac23} [\XR123 +\XR213], & \sqrt2\, \XR312, &0
\end{pmatrix},
\label{eq:alphaavec}
\\[5pt]
\boldsymbol{\mathcal X}_b &= \begin{pmatrix}
-\sqrt{\tfrac23}[\XR123-\XR213], & -\sqrt{\tfrac13} [\XR123-\XR213], &0, & \sqrt2\,\XR312
\end{pmatrix}\,.
\label{eq:alphabvec}
\end{align}
Here the operators $\boldsymbol{\mathcal X}^{\boldsymbol \sigma}_{[kab]}$ and $\boldsymbol{\mathcal X}^{\boldsymbol\sigma \dagger}_{[kab]}$ convert from the $\{k\ell m\}$ basis to that in terms of three on-shell momenta. The superscripts indicate which momenta, and thus which flavor, are associated with the spectator particle. The full definitions are given in \Cref{eq:XRdef,eq:XLdef} and the surrounding discussion.

There is a subtlety in the use of \Cref{eq:M3L} and, in particular, of the operators $\boldsymbol{\mathcal X}^{\boldsymbol \sigma}_{[kab]}$ and $\boldsymbol{\mathcal X}^{\boldsymbol\sigma \dagger}_{[kab]}$, as discussed, for example, in refs.~\cite{\HSQCb,\BSnondegen}. When these operators act in finite volume, the spectator is set to a finite-volume momentum, but those of the nonspectator pair are not, in general, in the finite-volume set. The reason is that (for a given choice of $E$ and $\bm P$) there are no triplets $\{p_i\}$ of finite-volume momenta that satisfy the on-shell condition. This means that the different terms in the sums implied by the matrix multiplications in \Cref{eq:M3L} have different momentum arguments and cannot strictly be combined. The combination does become possible, however, in the $L\to\infty$ limit, and it is only in that limit that we use \Cref{eq:M3L}, for that is when the integral equations are obtained, cf. \Cref{eq:M3hatI}.

The symmetrized short-distance piece, $\widehat{\cM}_{\df,3}^{[I]}$ is defined using projection vectors analogously to $\widehat{\cM}_{3,L}^{[I]}$ in \Cref{eq:M3L}. The normalization of this quantity and $\Kdf$ are such that in the limit of weak two-body interactions
\begin{align}
\begin{split}
\widehat{\cM}_{\df,3}^{[I=0,2]} &= {\cK}_{\df,3}^{[I=0,2]} \left[ 1 + \cO(\cM_{2},\Kdf) \right],
\\[5pt]
\widehat{\cM}_{\df,3}^{[I=1]} &= \begin{pmatrix}
\mathcal{K}_\text{df,3}^{[I=1],a,a} & \mathcal{K}_\text{df,3}^{[I=1],a,b} \\
\mathcal{K}_\text{df,3}^{[I=1],b,a} & \mathcal{K}_\text{df,3}^{[I=1],b,b}
\end{pmatrix} \left[ 1 + \cO(\cM_{2},\Kdf) \right].
\label{eq:M3dflimit}
\end{split}
\end{align}
Similar relations for the case of pions or kaons were used in refs.~\cite{Blanton:2019vdk,Blanton:2021eyf,Baeza-Ballesteros:2023ljl} to invert the integral equations analytically to a given order in the chiral expansion.

\subsection{Parametrization of \texorpdfstring{$\Kdf$}{the divergence-free K-matrix}}
\label{sec:Kdf}

The divergence-free three-particle K-matrix, $\Kdf$, can be parametrized using the symmetries of the theory, in particular $P$ and $T$ invariance, and of the three-meson states. This can be done systematically in an expansion about the three-particle threshold, following ref.~\cite{\dwave}, and also in the vicinity of a three-particle resonance, generalizing the approach of ref.~\cite{\isospin}. In both cases we give expressions for $\Kdf$ written as a function of the three incoming on-shell four-momenta $\{k_i\}$ and the corresponding outgoing momenta $\{p_i\}$. To convert these expressions into the $\{k\ell m\}$ basis is a straightforward exercise that has been worked out explicitly in refs.~\cite{\dwave,\implement}.

We begin with the threshold expansion, which can be written in terms of powers of the variables $t_{ij}=(p_i-k_j)^2$, which vanish at threshold. We use the labels $1$ and $2$ for the $D$ mesons, and $3$ for the pion. For the $I=0$ and $2$ channels, the symmetry under the exchange of the $D$ mesons, which can be seen from \Cref{eq:ddpi_isospin_states}, leads, at linear order, to the form
\begin{equation}
\mathcal{K}^{[I=0,2]}_\text{df,3}(\{p\},\{k\}) = c^I_0 + c^I_1 (t_{11}\!+\! t_{12} \!+\! t_{21} \!+\! t_{22})
+ c^I_2 (t_{13}\!+\!t_{23}\!+\!t_{31}\!+\!t_{31}) + c^I_3 t_{33} + \cO(t^2)\,,
\label{eq:KdfthrI02}
\end{equation}
where the $c_j^I$ are real coefficients. The leading order term $c^I_0$ is independent of the momenta and thus isotropic. The expansion is formally identical to that for the $K^+K^+ \pi^+$ system~\cite{\BStwoplusone,\implement}, although here we use a different set of basis functions.

As discussed around \Cref{eq:Kdf3I1form}, only three independent functions must be parametrized for the $I=1$ case. This can be done using their symmetries. First, $\mathcal{K}_\text{df,3}^{[I=1],aa}$ is symmetric under the exchange of the $D$ mesons, and thus can be expanded as for the $I=0,2$ amplitudes:
\begin{equation}
\mathcal{K}_{\df,3}^{[I=1],aa}(\{p\},\{k\}) = c^{aa}_0 + c^{aa}_1 (t_{11}\!+\! t_{12} \!+\! t_{21} \!+\! t_{22})
+ c^{aa}_2 (t_{13}\!+\!t_{23}\!+\!t_{31}\!+\!t_{31}) + c^{aa}_3 t_{33} + \cO(t^2)\,.
\end{equation}
By contrast, $\mathcal{K}_\text{df,3}^{[I=1],bb}$ is antisymmetric under exchange of either the incoming or outgoing $D$ mesons. This implies that the leading isotropic term is absent, and that there is only a single linear term,
\begin{equation}
\mathcal{\cK}_{\df,3}^{[I=1],bb}(\{p\},\{k\}) = c^{bb}_1 (t_{11} \!+\! t_{22} \!-\! t_{12} \!-\! t_{21})\,.
\end{equation}
Finally, the off-diagonal amplitude $\mathcal{K}_\text{df,3}^{[I=1],a,b}$ is symmetric (antisymmetric) under exchange of outgoing (incoming) $D$ mesons, leading to the form
\begin{equation}
\mathcal{\cK}_{\df,3}^{[I=1],a,b}(\{p\},\{k\}) = c^{ab}_1 (t_{11} - t_{21} + t_{21} - t_{22})
+ c^{ab}_2 (t_{13}-t_{23})\,,
\end{equation}
with an analogous result for the $b,a$ case.

We now turn to the case in which there is a resonance, and focus on the tetraquark channel, i.e. $I=0$ and $J^P=1^+$. Three-particle resonances have been studied in the finite-volume formalism in refs.~\cite{\BHSnum,\largera,Jackura:2020bsk,Mai:2021nul,Garofalo:2022pux,Dawid:2023jrj}. It is not yet clear whether an explicit pole in $\Kdf$ is needed to produce a pole in $\cM_3$ for the tetraquark resonance. Nevertheless, it may be a useful exercise to build a pole term that satisfies the relevant symmetries and has a factorizable residue. For this, it is possible to reuse results from the similar exercise undertaken in ref.~\cite{Hansen:2020zhy} for the $3\pi$ system in the channel of the $J^P=1^+$ resonance $a_1$.

A contribution to $\mathcal{K}^{[I=0]}_\text{df,3}$ can take the form
\begin{equation}
\mathcal{K}^{[I=0]}_\text{df,3} \supset \mathcal K_{cc}\frac{ 1 }{P^2 - M_{cc}^2} V_{12}'^\mu V_{12}^\nu \left( g_{\mu\nu} - \frac{P_\nu P_\mu}{P^2} \right),
\label{eq:KdfpoleI0}
\end{equation}
where the projector $g_{\mu\nu} - {P_\nu P_\mu}/{P^2}$ ensures that only the vector part contributes. The invariance under $1 \leftrightarrow 2$ exchange, separately for the initial and final states, allows only two possibilities for $V_{12}^\nu$ at linear order in momenta,
\begin{equation}
V_{12}^\nu = c_1 (k_1+k_2)^\nu + c_2 P^\nu\,.
\end{equation}
Since the projector removes the second term, we can set $c_2=0$ without loss of generality. There can, however, be contributions to the residue vectors of higher order in the $t_{ij}$.

\section{Required inputs from lattice calculations}
\label{sec:latticeinputs}

In this section we summarize the inputs that will be required from lattice calculations in order to implement the formalism presented in this work, focusing on the $I=0$, $J^P=1^+$ channel in which the doubly-charmed tetraquark lies. We stress that we are considering the case of isosymmetric QCD. We also restrict attention to an $L^3$ spatial geometry and to periodic boundary conditions, though we expect the formalism could be readily generalized to other geometries (of type $L_1 \times L_2 \times L_3$) and to twisted boundary conditions, along the lines discussed in refs.~\cite{\altSetUps}.

Let us assume at first that the light quark masses are chosen such that the $D^*$ is stable, i.e. there is a bound-state pole in the $p$-wave $D\pi$ amplitude. This is the case in all lattice calculations to date~\cite{Padmanath:2022cvl,Chen:2022vpo,Lyu:2023xro}. The inputs that are then required are as follows.
\begin{enumerate}
\item
{\em The finite-volume spectrum in the $I=1/2$ $D\pi$ channel, up to the $D\pi\pi$ threshold.}
Ideally, this should be obtained for several values of the box size $L$, and, for each box size, for several values of the total momentum $\bm P$ and all available irreducible representations of the relevant little group (a comment that applies to all subsequent spectra as well.) The set of operators used in the variational spectrum determination should include both local bilinears with the quantum numbers of the $D^*$ and nonlocal operators of the form $D(p_1) \pi (p_2)$. This spectrum can then be fit using the two-particle (L\"uscher) quantization condition including at least $s$ and $p$ waves, using, for example, a truncated effective-range expansion (ERE) for each channel. Such a parametrization can accommodate the $D^*$ bound state expected in the $p$-wave channel. Studies along these lines have already been done~\cite{Mohler:2012na,Moir:2016srx,Gayer:2021xzv}. In the $s$-wave channel, $D\pi$ interactions are expected to be attractive around threshold, and the $D_0^*$ resonance is found at around $2300$ MeV~\cite{Moir:2016srx,Gayer:2021xzv,Yan:2023gvq}. The latter will not be important for this study, since the $T_{\rm cc}^+$ is located close to the $DD\pi$ threshold while the $D^*_0$ lies above the $D\pi\pi$ threshold.

The two-particle quantization condition has a limited range of validity. One must work with CMF energies below the inelastic $D\pi\pi$ threshold, and above the left-hand cuts. In this case, the nearest left-hand cut involves $t$-channel exchange of two pions, which occurs when $s=M_D^2-M_\pi^2$, $t=4M_\pi^2$ and $u=s$, implying a CMF energy lying about $M_\pi$ below threshold, which is likely far below the position of a $D^*$ bound state. We stress, however, that, in the three-particle formalism, the $D\pi$ amplitude is needed below threshold down to a minimal energy, which, with the standard choice of cutoff function, occurs at the position of the left-hand cut~\cite{\implement}.

\item
{\em The finite-volume spectrum in the $I=1$ $DD$ channel, up to the $DD\pi$ threshold.}
This channel can be studied using operators with $D^+D^+$ quantum numbers. To our knowledge, no studies of this channel have been done to date, but it is likely to be weakly repulsive and not resonant, like other maximal isospin channels such as $K^+ K^+$. Since there are only even partial waves, we expect it to be a good approximation in this kinematic range to keep only the $s$ wave, dropping $d$ waves and above. Thus we imagine fitting the levels with the two-particle quantization condition using an ERE parametrization of the $s$-wave containing only a few parameters. The left-hand cut here also corresponds to $t$-channel exchange of two pions, which occurs when $s=4 M_D^2 - 4 M_\pi^2$. One will need to extrapolate the $DD$ amplitude down to this value when implementing the three-particle quantization condition.

At larger-than-physical quark masses for which the $D^*$ is stable,
the values of the lowest lying $D^*D$ finite-volume energies can lead to CMF energies of the $DD$ subsystem that lie on the two-pion left-hand cut. This implies that the cutoff function  appearing in this subsystem, given by \Cref{eq:cutoff}, vanishes,
and thus that the quantization condition does not have entries in which the pion is the spectator.
This is the case, for example, for the quark masses used in ref.~\cite{Padmanath:2022cvl}.
In such cases, the $DD$ two-particle scattering amplitude simply does not contribute and thus $DD$ finite-volume energies are not needed to apply the formalism.
Physically, this means that power-law finite-volume effects come only from the $D^*$ pole in the
$D\pi$ channel in this energy range.
The effects of $DD$ scattering are included in the infinite-volume quantity $\Kdf$;
they do not produce the poles in the finite-volume correlator that  lead to power-law finite-volume effects.
We stress that this turn-off of the $DD$ subchannel occurs smoothly as the energy is varied and is built into the formalism.

\item
{\em The finite-volume spectrum in the $I=0$ $DD\pi$ channel, up to the $DD\pi\pi$ threshold (or $D^* D^*$ threshold if it is lower).}
In this channel we also include $D^* D$ states---these will be reproduced by the three-particle quantization condition upon inclusion of a two-particle K matrix that leads to a bound state, as discussed in \Cref{sec:lefthandcut}. Thus we imagine a set of operators including single-particle operators (for the putative tetraquark), $D^* D$ operators, and three-particle $DD\pi$ operators.

A subtlety in the spectrum determination is the possible mixing of $DD$ and $DD\pi$ states in some irreducible representations (irreps). In particular, $DD$ states with ${I=0}$ have $J^P=\text{odd}^-$, and can mix with $DD\pi$ states having the same quantum numbers in infinite volume. The formalism presented in this work is then not applicable, since it does not include $2 \leftrightarrow 3$ mixing; to do so, one would need to use an approach along the lines of ref.~\cite{\BHSQC}. Since the $T_{\rm cc}^+$ has $J^P=1^+$, this mixing is not a problem as long as one only considers irreps where $J^P=\text{odd}^-$ is absent. This is exactly true in the rest frame irrep $T_{1g}$, and to good approximation (specifically, if one neglects states with $J \geq 3$) in the $A_2$ irrep in the moving frames with $\bm P$ of the form $[00n]$, $[0nn]$ and $[nnn]$. This can be seen from Table II of ref.~\cite{Dudek:2012gj} and appendix A of ref.~\cite{Woss:2018irj}. Other irreps containing $J^P=1^+$ (e.g. the $E$ irrep for $\bm P \sim [00n]$ frames), finite-volume effects mix with $J^P=1^-$ states, which can be both $DD\pi$ and $DD$. Thus they should be avoided for the study of $T_{\rm cc}^+$.

The $DD\pi$/$DD^*$ spectrum should be fit to the $I=0$ version of the three-particle quantization condition, \Cref{eq:QC3isospin}. What is unclear at this stage is the parametrization to use for $\Kdf$, i.e. whether a threshold expansion as in \Cref{eq:KdfthrI02} will suffice to describe a potential tetraquark bound state, or whether a pole in $\Kdf$, as in \Cref{eq:KdfpoleI0},
will be needed.
\end{enumerate}

We also note that, as in refs.~\cite{Blanton:2019vdk,Blanton:2021llb,Draper:2023boj}, any actual lattice implementation may prefer to do a global simultaneous fit to all three spectra, so as to determine the two- and three-particle K matrices taking into account the full correlations in the spectra.

Finally, we observe that, in the case of an unstable $D^*$, the three requirements listed above are unchanged. The only difference will be in the fitting of levels---the two-particle $D\pi$ $p$-wave K matrix will now need to produce a narrow resonance.

\section{Conclusions}
\label{sec:conclu}

In this paper we have presented the formalism necessary to extract the $DD\pi$ scattering amplitude in isosymmetric QCD, given knowledge of the finite-volume two- and three-particle spectra. Results are given for all choices of allowed isospin, $I=0, 1, 2$, and thus include the $I=0$ channel that contains the doubly-charmed tetraquark candidate, $T_{cc}(3875)^+$. While the formalism for $I=2$ could have been deduced from earlier work on $K^+K^+\pi^+$ systems~\cite{\BStwoplusone,\implement}, that for $I=0$ and $1$ is new. It turns out that the former is almost identical to that for $I=2$. The formalism presented here also applies for the $KK\pi$ and $BB\pi$ systems, as long as one changes the kinematic functions appropriately. The latter system is of particular interest, since lattice QCD studies point towards the existence of a tetraquark with minimal quark content of $b b \bar u \bar d$~\cite{Hudspith:2023loy,Meinel:2022lzo}.

A major motivation for this work is the observation that, if the quarks are heavier than physical, so that the $D^*$ is stable, lattice studies of the $D^*D$ spectrum cannot use the standard L\"uscher two-particle quantization condition to study a possible tetraquark bound state. This is because that state lies very close to the left-hand cut due to single pion exchange, which invalidates the quantization condition. As noted in \Cref{sec:lefthandcut}, the physical origin of this cut arises from the intermediate $DD\pi$ state. The formalism we have presented includes the effect of this state, and thus allows a study of the $D^* D$ amplitude at and below the start of the left-hand cut. It only breaks down when one goes further below threshold and reaches the nonanalyticity due to two-pion exchange.

Alternative methods for incorporating the effects of one-pion exchange in the analysis of finite volume energies have recently been presented in refs.~\cite{Raposo:2023oru,Meng:2023bmz}. Ref.~\cite{Meng:2023bmz} is based on using a plane-wave basis for the quantization condition and using an EFT-based potential to parameterize the scattering amplitude for $D D^*$ as well as the finite-volume energies. In contrast, ref.~\cite{Raposo:2023oru} uses a particular decomposition of the Bethe-Salpeter kernel to treat the effects of single-meson exchange, and is currently only applicable to systems with identical scattering particles such as nucleon-nucleon. There are clear parallels between all approaches, including the need for additional parameters relative to the standard L\"uscher quantization condition, as well as the use of integral equations and an extended matrix space in the quantization condition. At the same time, the approach presented here is clearly different in that it includes the effect of the $D D \pi$ branchcut in the extracted amplitudes and the associated finite-volume effects. A related point is that this method requires the $D \pi$ and $DD$ two-particle scattering line shapes as inputs. Explicitly understanding the relationships between the different methods is a topic for future work. Ultimately, the key question will likely be whether the various effects included or neglected in a given approach are numerically significant.

We have derived the formalism using the RFT approach, first introduced in refs.~\cite{\HSQCa,\HSQCb}. The main issue compared to previous work is to determine the required generalization of the flavor structure. The final results take the by-now standard forms in the RFT formalism, but with an additional flavor index that runs over two values for both $I=0$ and $I=2$, and four values for $I=1$. To derive the flavor structure we have used two independent approaches, described in detail in the two appendices. The first is based on the derivation of the formalism using time-ordered perturbation theory introduced in ref.~\cite{\BSQC}, while the second generalizes the original Feynman-diagram based method of refs.~\cite{\HSQCa,\HSQCb}. In the former, all steps are algebraic and explicit while, in the latter, it is assumed that all the algebra of the derivation for three identical particles holds and only simple operations on the flavor space are needed. The agreement between the two approaches indicates that either method can be used for future generalizations.

The lattice QCD inputs to the quantization conditions necessary to constrain the K matrices describing two- and three-particle interactions are summarized in \Cref{sec:latticeinputs}. In order to study the three-meson scattering amplitude, one additionally needs to carry out the very important second step of solving the integral equations given in \Cref{sec:inteqs}. It is only by solving these equations, which put back in the full analytic structure of the three-particle amplitude, and include, in particular, all initial- and final-state interactions, that one can determine whether there is a three-particle tetraquark bound state, virtual state, or resonance. Techniques for solving these equations have improved apace, and we were particularly inspired by refs.~\cite{Dawid:2021fxd,Dawid:2023jrj} in which virtual states and resonances were studied in similar but simpler problem involving three identical particles with a two-particle bound state and induced three-particle bound states, The present case is more complicated, involving nondegenerate particles and multiple channels, but we hope that results from solving the integral equations will be available based on model interactions in the not-too-distant future.

There are several directions in which the present work can be generalized. The methodology for deriving finite-volume formalism presented in the appendices can be extended to other cases, including particles with spin such as~$N\pi\pi$ systems. Moreover, the approach presented here for avoiding the left-hand cut due to single-particle exchange in two-particle systems is quite general, and should apply as well to the $NN$ system. This would require treating the nucleon as a subthreshold pole in $p$-wave $N\pi$ scattering. In addition, another direction to explore is the inclusion of mixing between two- and three-particle systems necessary for the Roper resonance.

The doubly-charmed tetraquark represents a crucial milestone in understanding the open puzzles in the hadron spectrum. The present work lays out a comprehensive formalism for upcoming lattice QCD studies, and brings us closer to a detailed description of exotic hadrons from first-principles QCD.

\acknowledgments

We are very grateful to Zack Draper for many useful discussions, comments on the text, and initial collaboration on the project. In addition, we thank
Ra\'ul Brice\~no,
Padmanath Madanagopalan,
Maxim Mai,
Sasa Prelovsek,
and
Andr\'e Raposo
for useful discussions.

MTH is supported by UKRI Future Leader Fellowship MR/T019956/1 and in part by UK STFC grant ST/P000630/1. The work of FRL has been supported in part by the U.S.~Department of Energy, Office of Science, Office of Nuclear Physics, under grant Contract Numbers DE-SC0011090 and DE-SC0021006. FRL also acknowledges financial support by the Mauricio and Carlota Botton Fellowship. The work of SRS is supported in part by the U.S. Department of Energy grant No.~DE-SC0011637. This work contributes to the goals of the USDOE ExoHad Topical Collaboration, contract DE-SC0023598.

FRL would like to thank the Physics Department at the University of Washington for its hospitality during a visit in which this work was initiated.

\appendix

\section{Derivation using time-ordered perturbation theory}
\label{app:topt_derivation}

In this appendix we sketch the derivation of the results presented in \Cref{sec:formalism} using the TOPT approach developed in ref.~\cite{\BSQC}, a work referred to as BS1 in the following. We make particular use of the application of this approach to systems with three distinguishable particles~\cite{\BSnondegen} and $2+1$ systems consisting of two identical and a third particle~\cite{\BStwoplusone}. These two papers are referred to hereafter as BS3 and BS2, respectively.

We work in a generic relativistic effective field theory describing the interactions of $D$ mesons and pions, together with any other stable particles that can be produced by the strong interactions (e.g. $K \bar K$ pairs). The vertices can involve arbitrary numbers of particles and derivatives, and we do not need to specify or constrain the associated coupling constants. In the kinematic range of interest (to be discussed below) only $DD\pi$ states can go on shell, which implies that all intermediate states involving more (or fewer) particles are virtual. It follows, as discussed below, that these virtual intermediate states contribute only to infinite-volume Bethe-Salpeter kernels.

As discussed in \Cref{sec:lefthandcut}, the $D^*$ enters the formalism, if it is stable, as a subthreshold pole
in the $D\pi$ $p$-wave scattering amplitude.
In terms of the derivation of the quantization condition,
this approach is valid because a two-particle bound state does not imply that there is
a singularity in either the two-particle Bethe-Salpeter kernel, $\cB_2$, or K matrix, $\mathcal K_2$.
Instead, the singularity appears only in the two-particle scattering amplitude, $\mathcal M_2$,
which does not enter the derivation of the quantization conditions,
and thus does not introduce singularities that lead to explicit finite-volume dependence in
quantization conditions.
This does not imply, however, that finite-volume effects due to on-shell $D^* D$ states are absent.
They are properly incorporated in the three-particle quantization condition due to the contributions
in which the $D$ is a spectator, and its momentum is summed over the finite-volume set.

We restrict to the sector with quantum numbers $C=2$, $U=D=-1$,
with $U$ and $D$ referring to ``upness'' and ``downness'', respectively.
There are three different flavor channels of $DD\pi$ with these quantum numbers,
\begin{equation}
\left\{ D^+(\bm k_1) D^+(\bm k_2) \pi^- (\bm k_3),
\ D^+(\bm k_1) D^0(\bm k_2) \pi^0 (\bm k_3),
\ D^0(\bm k_1) D^0(\bm k_2) \pi^+(\bm k_3)\right\}\,,
\label{eq:flavorchannels}
\end{equation}
which we label with the flavor index $i=1-3$.
(The momentum labels will be useful in the following.)
We note that the first and third are $2+1$ channels, while the second consists of distinguishable particles.
Since we assume exact isospin symmetry, all three channels are degenerate.
We work in Minkowski time, with space restricted to a cubic box of side-length $L$ with periodic boundary
conditions.
We construct the finite-volume Minkowski-space correlation matrix
\begin{equation}
C_L(P)_{jk} \equiv \int d x^0 \!\int_{L^3} d^3 {\boldsymbol x} \,
e^{-i \boldsymbol{P} \cdot {\boldsymbol x}+i E t}
\langle 0 \vert \mathrm{T}\mathcal{O}_j(x) \mathcal{O}_k^{\dagger}(0) \vert 0 \rangle_L \,,
\label{eq:CLdef}
\end{equation}
where $j,k$ run over the flavors, with $\mathcal O_j(x)$ any quasi-local operator with the correct quantum numbers
to destroy states with flavor $j$.
In addition, the operators are assumed to couple to all allowed irreps of the cubic group (the symmetry group
of the cubic box).
The total momentum $\bm P$ is restricted by the boundary conditions
to lie in the finite volume set $2\pi \bm n/L$, where $\bm n \in \mathbb Z^3$.

For a given choice of $\bm P$,
the energies of finite-volume states are given by the poles of $C_L$ as a function of $E$.
We will focus on the kinematic range where only three particles can go on shell.
A conservative choice for this range is $2 M_D < E^*=\sqrt{E^2-\bm P^2} < 2 M_D + 2 M_\pi$,
so that states with one more, or one less, pion cannot go on shell.
For a given choice of finite-volume irrep,
the allowed kinematic range may be larger---this happens if the residual symmetries forbid
the addition or subtraction of a single pion.

For a given choice of $\bm P$,
the energies of finite-volume states are given by the poles of $C_L$ as a function of $E$.
We will focus on the kinematic range where only three particles can go on shell, and above left-hand cuts produced by two-pion exchanges.
This range is $2 M_D + M_\pi - \delta < E^*=\sqrt{E^2-\bm P^2} < 2 M_D + 2 M_\pi$,
where the lower limit is determined the position of the first two-pion lhc parametrized by the positive quantity $\delta$ (see \Cref{subsec:partdimerregime}), and
the upper limit avoids that states with an additional pion go on shell.
For a given choice of finite-volume irrep,
the allowed kinematic range may be larger---this happens if the residual symmetries forbid
the addition or subtraction of a single pion.

\subsection{All orders expression obtained using TOPT}
\label{eq:TOPT1}

By a simple generalization of the work of BS2 and BS3,
in the kinematic range of interest we can write the finite-volume-dependent part of $C_L$ as
a geometric series in terms of TOPT Bethe-Salpeter kernels $B_{2,L}$ and $B_3$ lying between ``cut factors'' $D$
that carry the singularities associated with three-particle states:
\begin{equation}
\Delta C_L \equiv C_L - C_\infty^{(0)} = A' i D \frac1{1 - i (B_{2,L}+B_3) i D} A + \mathcal O(e^{-M_\pi L}) \,.
\label{eq:CLresult}
\end{equation}
Here $A'$ and $A$ are ``endcaps'', composed of all diagrams connecting the operators $\cO_j$
and $\cO^\dagger_k$ to a three-particle intermediate state,
while $C_\infty^{(0)}$ is the contribution that has no three-particle intermediate states.
Almost all quantities in \Cref{eq:CLresult} are matrices, both in flavor space and momentum space.
The only exceptions are the endcaps, for which the external indices run only over the three flavors,
and not the momenta.
The momentum indices are labeled $\{\bm k\}=\{ \bm k_1, \bm k_2, \bm k_3\}$,
where the triplets of finite-volume momenta are constrained by $\bm k_1+\bm k_2 + \bm k_3=\bm P$.
We assign momentum labels using the convention shown in \Cref{eq:flavorchannels}.
The derivation of \Cref{eq:CLresult} makes use of the result that finite-volume momentum sums of nonsingular
summands can be replaced by the corresponding infinite-volume integrals, up to corrections
that are suppressed exponentially in $M_\pi L$~\cite{\KSS}.
Henceforth, such corrections are dropped and not shown explicitly.
This replacement means that $A'$, $A$, and $C_\infty^{(0)}$ are infinite-volume quantities.
Their explicit forms are not needed in the following derivation.

We now describe the remaining quantities contained in \Cref{eq:CLresult}, beginning with the cut factors.
These are diagonal in flavor,
\begin{align}
D &= {\rm diag}(D_0/2, D_0, D_0/2)\,,
\label{eq:Ddef}
\end{align}
where $D_0$ is the usual TOPT energy denominator
\begin{equation}
D_0 = \delta_{\{\bm p\} \{\bm k\}} \frac1{L^6} \frac1{8\omega_1\omega_2\omega_3}
\frac1{E - \omega_1 - \omega_2 -\omega_3}\,,
\end{equation}
with $\omega_1\equiv \omega_{p_1} = \sqrt{\bm p_1^2 + M_D^2}$ etc. and
\begin{equation}
\delta_{\{\bm p\} \{\bm k\}} = \delta_{\bm p_1 \bm k_1} \delta_{\bm p_2 \bm k_2} \delta_{\bm p_3 \bm k_3}
\,.
\end{equation}
Note that for the $2+1$ entries of $D$ the factor of $D_0$ is accompanied by a symmetry factor of $1/2$.

We next consider the matrix $B_{2,L}$, which contains two-particle Bethe-Salpeter kernels, denoted $\cB_2$,
which are given by the sums over all TOPT diagrams that have no two-particle cuts in the $s$ channel.
In these diagrams all momentum sums have been replaced by integrals, so they are infinite-volume quantities.
Extending the analysis in BS2 and BS3, we find the form
(note the use of caligraphic $\cB$ for the individual entries, in contrast to roman $B$ for the overall matrix)
\begin{equation}
{\mathcal B}_{2,L} = \begin{pmatrix}
\mathcal B_{2, L, 11}
&
\mathcal B_{2, L, 12}
&
0
\\
\mathcal B_{2, L, 21}
&
\mathcal B_{2, L, 22}
&
\mathcal B_{2, L, 23}
\\
0
&
\mathcal B_{2, L, 32}
&
\mathcal B_{2, L, 33}
\end{pmatrix}\,,
\label{eq:B2Ldef}
\end{equation}
where we have introduced
\begin{align}
\mathcal B_{2,L,11}& = S^{D} \cB_{2,L}(D^+\pi^-\leftarrow D^+\pi^-) S^{D}
+ \cB_{2,L}(D^+D^+ \leftarrow D^+ D^+) \,,
\\
\mathcal B_{2, L, 12} & =
S^{D} \cB_{2,L}(D^+\pi^- \leftarrow D^0\pi^0) \,,
\\
\mathcal B_{2, L, 21} & =
\cB_{2,L}(D^0\pi^0 \leftarrow D^+\pi^-) S^D \,,
\\
\mathcal B_{2,L,22}& = \cB_{2,L}(D^+ D^0 \leftarrow D^+ D^0)
+ \cB_{2,L}(D^0\pi^0 \leftarrow D^0\pi^0)
+ \cB_{2,L}(D^+\pi^0 \leftarrow D^+\pi^0) \,,
\\
\mathcal B_{2, L, 23} & =\cB_{2,L}(D^+\pi^0 \leftarrow D^0\pi^+) S^{D} \,,
\\
\mathcal B_{2, L, 32} & = S^D \cB_{2,L}(D^0\pi^+ \leftarrow D^+\pi^0) \,,
\\
\mathcal B_{2,L,33} & = S^{D} \cB_{2,L}(D^0\pi^+ \leftarrow D^0\pi^+) S^{D}
+ \cB_{2,L}(D^0D^0 \leftarrow D^0 D^0) \,.
\end{align}

The form of the entries in $B_{2,L}$ is determined by which two-particle interactions can
lead to the appropriate changes in flavor composition.
For example, the top-right and bottom-left entries vanish because there are no two-particle interactions
that connect $D^+D^+ \pi^-$ to $D^0 D^0 \pi^+$, since all three particles need to change.
The form of the quantities $\cB_{2,L}$ is exemplified by
\begin{equation}
\cB_{2,L}(D^0\pi^+\!\leftarrow\! D^0\pi^+)_{\{\bm p\}, \{\bm k\}}
= 2 \omega_2 L^3 \delta_{\bm p_2 \bm k_2}
\cB_2[D^0(\bm p_1) \pi^+(\bm p_3)\leftarrow D^0(\bm k_1) \pi^+(\bm k_3)]\,,
\end{equation}
which appears in the bottom right entry of $B_{2,L}$.
The explicit $L$ dependence arises from keeping track of TOPT propagator factors, as explained in BS1.
The single momentum Kronecker-delta arises because one of the particles (here one of the $D^0$s) spectates.
Note that we have chosen the $D^0$ particles that scatter to be those with initial and final momenta $\bm k_1$
and $\bm p_1$, respectively.
The three other possible choices are included by the factors of $S^D$,
which is a symmetrization operator acting on momentum indices,
\begin{equation}
S^D = 1 + P^D\,,\qquad 1 = \delta_{\bm p_1 \bm k_1} \delta_{\bm p_2 \bm k_2} \delta_{\bm p_3 \bm k_3}\,,\qquad
P^D = \delta_{\bm p_2 \bm k_1} \delta_{\bm p_1 \bm k_2} \delta_{\bm p_3 \bm k_3}\,,
\label{eq:SDdef}
\end{equation}
where $P^D$ interchanges the first two momenta. These operators were introduced in BS2.
All other entries in $B_{2,L}$ have a similar form, with the choice of spectator momentum depending on the
flavors involved in the two-particle scattering.

It remains to define $B_3$. The entries in this flavor matrix are the sum over all TOPT diagrams
connecting initial and final flavors having no three-particle cuts.
Momentum sums are replaced by integrals, so that all entries are infinite-volume quantities.
The only properties of $B_3$ that we will need are that it conserves isospin and is symmetric under
interchanges of identical particles in either the initial or final states.

\subsection{Symmetrization and on-shell projection}
\label{app:symmon}

The next step is to move the symmetrization operators onto the cut factors $D$. This can be achieved following
the same steps as in BS2, making use of the symmetry under $DD$ exchange of quantities containing two identical
$D$ mesons. To formalize this, we introduce
\begin{equation}
\tilde S^D \equiv {\rm diag}(S^D, 1, S^D)\ \ {\rm and}\ \ R = {\rm diag} (\tfrac12, 1, \tfrac12)\,,
\label{eq:tSDdef}
\end{equation}
such that
\begin{equation}
A =\tilde S^D R A\,,\ \
A' = A' R \tilde S^D\,,\ \ {\rm and}\ \
B_3 = \tilde S^D R B_3 R \tilde S^D\,.
\end{equation}
Then, defining
\begin{align}
D_S &= \tilde S^D D \tilde S^D = {\rm diag} (\tfrac12 S^D D_0 S^D, D_0, \tfrac12 S^D D_0 S^D)\,,
\\
\tilde B_3 &= R B_3 R\,, \quad \tilde A' = A' R\,,\quad \tilde A = R A\,,
\label{eq:symm1}
\end{align}
and introducing $\tilde B_{2,L}$, such that
\begin{equation}
B_{2,L} = \tilde S^D \tilde B_{2,L} \tilde S^D\,,
\label{eq:symm2}
\end{equation}
where $\tilde B_{2,L}$ is free of symmetrization factors,
\begin{equation}
\widetilde{B}_{2,L} = \begin{pmatrix}
\widetilde{\mathcal B}_{2, L, 11}
&
\widetilde{\mathcal B}_{2, L, 12}
&
0
\\
\widetilde{\mathcal B}_{2, L, 21}
&
\widetilde{\mathcal B}_{2, L, 22}
&
\widetilde{\mathcal B}_{2, L, 23}
\\
0
&
\widetilde{\mathcal B}_{2, L, 32}
&
\widetilde{\mathcal B}_{2, L, 33}
\end{pmatrix}\,,
\label{eq:symm3}
\end{equation}
with
\begin{align}
\widetilde{\mathcal B}_{2,L,11}& = \cB_{2,L}(D^+\pi^-\leftarrow D^+\pi^-)
+ \cB_{2,L}(D^+D^+ \leftarrow D^+ D^+) / 4 \,,
\\
\widetilde{\mathcal B}_{2, L, 12} & = \cB_{2,L}(D^+\pi^- \leftarrow D^0\pi^0) \,,
\\
\widetilde{\mathcal B}_{2, L, 21} & = \cB_{2,L}(D^0\pi^0 \leftarrow D^+\pi^-) \,,
\\
\widetilde{\mathcal B}_{2,L,22}& = \cB_{2,L}(D^+ D^0 \leftarrow D^+ D^0)
+ \cB_{2,L}(D^0\pi^0 \leftarrow D^0\pi^0)
+ \cB_{2,L}(D^+\pi^0 \leftarrow D^+\pi^0) \,,
\\
\widetilde{\mathcal B}_{2, L, 23} & =\cB_{2,L}(D^+\pi^0 \leftarrow D^0\pi^+) \,,
\\
\widetilde{\mathcal B}_{2, L, 32} & =\cB_{2,L}(D^0\pi^+ \leftarrow D^+\pi^0) \,,
\\
\widetilde{\mathcal B}_{2,L,33} & = \cB_{2,L}(D^0\pi^+ \leftarrow D^0\pi^+)
+ \cB_{2,L}(D^0D^0 \leftarrow D^0 D^0) / 4 \,,
\end{align}
we can rearrange the volume-dependent part of the correlator into the form
\begin{equation}
\Delta C_L = \tilde A' i D_S \frac1{1 - i (\tilde B_{2,L}+ \tilde B_3) i D_S} \tilde A\,.
\label{eq:CLresulttilde}
\end{equation}

\bigskip
Now we introduce additional matrix structure corresponding to the different choices of spectator:
two options each for the first and third flavor indices (following BS2) and three options for the second flavor index
(following BS3). This leads to matrices of flavor dimension $2+3+2=7$.
We denote these larger matrices
(as well as the vectors needed for endcaps)
with carets (``hats''), so as to distinguish them from the $3$-d flavor matrices
discussed above.
Here we follow the prescriptions presented in BS2 and BS3,
generalized slightly to deal with flavor off-diagonal terms. We also apply on-shell projection at this stage,
which converts the factors of $D$ into either $F$ or $G$ cuts, and introduces the usual $\{k \ell m\}$
indices, although now we must specify in addition the flavor of the spectator.
After some algebra, the result is
\begin{equation}
\Delta C_L = \hat A' i \widehat F_G \frac1{1 - i (\widehat{\cK}_{2,L}+ \widehat{\cK}_{\rm df,3}^{(u,u)}) i \widehat F_G}
\widehat A\,.
\label{eq:CLresulthat}
\end{equation}
We now proceed to explain the elements of this result.

First we note that, in writing the 7-d matrices,
we choose the ordering of the pair-spectator decompositions as
\begin{equation}
\left( [D^+\pi^-] D^+, [D^+ D^+]\pi^-,
[D^+\pi^0]D^0, [D^0\pi^0]D^+, [D^+ D^0]\pi^0,
[D^0 \pi^+]D^0, [D^0 D^0]\pi^+ \right)\,,
\label{eq:pairspect}
\end{equation}
with the spectator being the last entry in each case.

Second, $\widehat F_G$ is the result of on-shell projection applied to the factors of $D_0$.
It is a block-diagonal matrix consisting of $2\times 2$, $3\times 3$ and $2\times 2$ blocks:
\begin{align}
\widehat F_G &= \begin{pmatrix} F_G^{2+1} & 0 & 0 \\ 0 & F_G^{1+1+1} & 0 \\ 0 & 0 & F_G^{2+1} \end{pmatrix}\,,
\label{eq:FGa}
\\
F_G^{2+1} &= \begin{pmatrix} \tilde F^{D} + \tilde G^{DD} & \sqrt2 P^{(\ell)} \tilde G^{D\pi} P_e
\\ \sqrt2 P_e \tilde G^{\pi D} P^{(\ell)} & P_e \tilde F^\pi P_e \end{pmatrix}\,,
\label{eq:FG2p1}
\\
F_G^{1+1+1} &=
\begin{pmatrix}
\tilde F^{D} & \tilde G^{DD} & P^{(\ell)} \tilde G^{D\pi} P^{(\ell)}
\\
\tilde G^{DD} & \tilde F^{D} & P^{(\ell)} \tilde G^{D\pi}
\\
P^{(\ell)} \tilde G^{\pi D} P^{(\ell)} & \tilde G^{\pi D} P^{(\ell)} & \tilde F^{\pi}
\end{pmatrix}\,.
\label{eq:FG1p1p1}
\end{align}
Here $\tilde F^{D}$ and $\tilde F^\pi$ are given by \Cref{eq:Ft} [equivalent to eq.~(A20) of BS2],
with the superscript indicating the spectator particle.
This definition also requires specifying which of the remaining pair is used to define the spherical
harmonics [denoted flavor $j$ in \Cref{eq:Ft}].
Our convention is that this particle is the first one in each of the triplets in \Cref{eq:pairspect}.
Note that \Cref{eq:Ft} does not contain the factor of $1/2$ for identical particles---symmetry
factors are all explicit and are given by the factors of $\sqrt2$ here and $1/2$ below.

For the $\tilde G^{ij}$, the superscript indicates the final, $i$, and initial, $j$, spectator flavor and mass.
The explicit expression is given in \Cref{eq:Gt} [equivalent to eq.~(A21) of BS2].
Here, the spherical harmonics are always defined relative to the particle that will become the
spectator. This convention can differ from that used for $\tilde F^i$, leading to the need for
correction factors:
\begin{equation}
P^{(\ell)}_{k' \ell' m'; k \ell m}= \delta_{\bm k' \bm k} \delta_{\ell \ell'} \delta_{mm'} (-1)^\ell \,.
\label{eq:Plm}
\end{equation}
For example
in the upper-right element of $F_G^{2+1}$, the initial-state spectator is a pion,
and thus in $\tilde G^{D\pi}$ the final state pair is decomposed in harmonics relative to
the pion direction, rather than the standard choice of the $D$ meson based on the triplets
in \Cref{eq:pairspect}. Thus, a factor of $P^{(\ell)}$ to the left (final-state) side of $\tilde G^{D\pi}$ is required.

The factors of $P_e=(1+P^{(\ell)})/2$ in $F_G^{2+1}$ project onto even partial waves.
They reflect the fact that $F_G^{2+1}$ is always adjacent to either an endcap or a K matrix,
and these quantities are symmetric when the interacting pairs are made of identical particles.

Third, the two-particle $K$ matrices are packaged as
\begin{equation}
\widehat{\cK}_{2,L} =
\begin{pmatrix}
\cK_{2,L}^{D\pi,a} & 0 & 0 & \cK_{2,L}^{D\pi,b} & 0 & 0 & 0
\\
0 & \tfrac12\cK_{2,L}^{DD,a} & 0 & 0 & 0 & 0 & 0
\\
0 & 0 & \cK_{2,L}^{D\pi,c} & 0 & 0 & \cK_{2,L}^{D\pi,d} & 0
\\
\cK_{2,L}^{D\pi,b,T} & 0 & 0 & \cK_{2,L}^{D\pi,e} & 0 & 0 & 0
\\
0 & 0 & 0 & 0 & \cK_{2,L}^{DD,b} & 0 & 0
\\
0 & 0 & \cK_{2,L}^{D\pi,d,T} & 0 & 0 & \cK_{2,L}^{D\pi,f} & 0
\\
0 & 0 & 0 & 0 & 0 & 0 & \tfrac12 \cK_{2,L}^{DD,c}
\end{pmatrix},
\label{eq:K2Lhat}
\end{equation}
where
\begin{align}
\left[\cK_{2,L}^{D\pi,a}\right]_{k'\ell' m',k\ell m}
&\equiv 2\omega_k^{(D)} L^3 \delta_{\bm k', \bm k} \delta_{\ell' \ell} \delta_{m' m}
\cK_2^{\ell}[D^+\pi^-\leftarrow D^+\pi^-]\,,
\\
\left[\cK_{2,L}^{D\pi,b}\right]_{k'\ell' m',k\ell m}
&\equiv 2\omega_k^{(D)} L^3 \delta_{\bm k', \bm k} \delta_{\ell' \ell} \delta_{m' m}
\cK_2^{\ell}[D^+\pi^- \leftarrow D^0\pi^0 ]\,,
\\
\left[\cK_{2,L}^{D\pi,c}\right]_{k'\ell' m',k\ell m}
&\equiv 2\omega_k^{(D)} L^3 \delta_{\bm k', \bm k} \delta_{\ell' \ell} \delta_{m' m}
\cK_2^{\ell}[D^+\pi^0 \leftarrow D^+\pi^0]\,,
\\
\left[\cK_{2,L}^{D\pi,d}\right]_{k'\ell' m',k\ell m}
&\equiv 2\omega_k^{(D)} L^3 \delta_{\bm k', \bm k} \delta_{\ell' \ell} \delta_{m' m}
\cK_2^{\ell}[D^+\pi^0 \leftarrow D^0\pi^+]\,,
\\
\left[\cK_{2,L}^{D\pi,e}\right]_{k'\ell' m',k\ell m}
&\equiv 2\omega_k^{(D)} L^3 \delta_{\bm k', \bm k} \delta_{\ell' \ell} \delta_{m' m}
\cK_2^{\ell}[D^0\pi^0\leftarrow D^0\pi^0]\,,
\\
\left[\cK_{2,L}^{D\pi,f}\right]_{k'\ell' m',k\ell m}
&\equiv 2\omega_k^{(D)} L^3 \delta_{\bm k', \bm k} \delta_{\ell' \ell} \delta_{m' m}
\cK_2^{\ell}[D^0\pi^+\leftarrow D^0\pi^+]\,,
\\
\left[\cK_{2,L}^{DD,a}\right]_{k'\ell' m',k\ell m}
&\equiv 2\omega_k^{(\pi)} L^3 \delta_{\bm k', \bm k} \delta_{\ell' \ell} \delta_{m' m}
\cK_2^{\ell}[D^+D^+\leftarrow D^+ D^+]\,,
\\
\left[\cK_{2,L}^{DD,b}\right]_{k'\ell' m',k\ell m}
&\equiv 2\omega_k^{(\pi)} L^3 \delta_{\bm k', \bm k} \delta_{\ell' \ell} \delta_{m' m}
\cK_2^{\ell}[D^+ D^0\leftarrow D^+ D^0]\,,
\\
\left[\cK_{2,L}^{DD,c}\right]_{k'\ell' m',k\ell m}
&\equiv 2\omega_k^{(\pi)} L^3 \delta_{\bm k', \bm k} \delta_{\ell' \ell} \delta_{m' m}
\cK_2^{\ell}[D^0 D^0\leftarrow D^0 D^0]\,,
\end{align}
and the superscript $T$ indicates interchanging the initial and final flavors, e.g.
\begin{equation}
\left[\cK_{2,L}^{D\pi,b,T}\right]_{k'\ell' m',k\ell m}
\equiv 2\omega_k^{(D)} L^3 \delta_{\bm k', \bm k} \delta_{\ell' \ell} \delta_{m' m}
\cK_2^{\ell}[D^0\pi^0 \leftarrow D^+\pi^-]\,.
\end{equation}
For the scattering of distinguishable particles, both even and odd partial waves
are allowed, and for the latter we define the spherical harmonics relative to the first
particle listed in the argument of $\cK_2^\ell$, e.g. the $D^+$ for both initial and final
states in $\cK_{2,L}^{D\pi,a}$. On-shell energies are defined as in \Cref{eq:omegadef}.
The partial wave amplitudes $\cK_2^\ell$ have an implicit dependence on the
relative momentum in the CMF of the scattering pair.

The K-matrices can be written in the isospin basis as follows (dropping the superscript $\ell$ for brevity)
\begin{align}
\cK_2[D^+ D^+\leftarrow D^+ D^+] &= \cK_2[D^0 D^0\leftarrow D^0 D^0] = \cK_2^{DD,I=1}\,,
\label{eq:K2iso1}
\\
\cK_2[D^+ D^0\leftarrow D^+ D^0] &= \cK_2[D^0 D^+\leftarrow D^0 D^+]
= \frac12 (\cK_2^{DD,I=1} + \cK_2^{DD,I=0})\,,
\\
\cK_2[D^+ D^0\leftarrow D^0 D^+] &= \cK_2[D^0 D^+\leftarrow D^+ D^0]
= \frac12 (\cK_2^{DD,I=1} - \cK_2^{DD,I=0})\,,
\\
\cK_2[D^0\pi^+\leftarrow D^0\pi^+] &= \frac13 (\cK_2^{D\pi,I=3/2} + 2\cK_2^{D\pi,I=1/2} )\,,
\\
\cK_2[D^+\pi^0\leftarrow D^+\pi^0] &= \frac13 (2\cK_2^{D\pi,I=3/2} + \cK_2^{D\pi,I=1/2} )\,,
\\
\cK_2[D^0\pi^+\leftarrow D^+\pi^0] &=
\cK_2[D^+\pi^0 \leftarrow D^0 \pi^+] = \frac{\sqrt2}3 (\cK_2^{D\pi,I=3/2} - \cK_2^{D\pi,I=1/2} )\,,
\\
\cK_2[D^+\pi^-\leftarrow D^+\pi^-] &= \frac13 (\cK_2^{D\pi,I=3/2} + 2 \cK_2^{D\pi,I=1/2} )\,,
\\
\cK_2[D^0\pi^0\leftarrow D^0\pi^0] &= \frac13 (2\cK_2^{D\pi,I=3/2} + \cK_2^{D\pi,I=1/2} )\,,
\\
\cK_2[D^0\pi^0\leftarrow D^+\pi^-] &=
\cK_2[D^+\pi^-\leftarrow D^0\pi^0] = \frac{\sqrt2}3 (\cK_2^{D\pi,I=3/2} - \cK_2^{D\pi,I=1/2} )\,.
\label{eq:K2ison}
\end{align}

The $7\times 7$ flavor matrix $\widehat{\cK}_{\rm df,3}^{(u,u)}$ is a three-particle K matrix.
It can be related algebraically to the quantities discussed above, including $B_3$, by generalizing results from BS2 and BS3,
but this relation plays no role in the following, so we do not display it.
All we need to know about $\widehat{\cK}_{\rm df,3}^{(u,u)}$ is that it is an infinite-volume quantity, free
from any singularities associated with three-particle intermediate states, and with nonzero entries
in all elements.
The $(u,u)$ superscript indicates that it is an unsymmetrized quantity, in which, if an external interaction
involves two particles, then the momentum $\bm k$ is always that of the spectator.

Finally, the new endcaps $\widehat A'$ and $\widehat A$ are $3\times 7$ matrices in flavor space.
Their relation to the earlier endcaps is known,
but we do not display the results as it does not impact the form of the quantization conditions that we
obtain below. Indeed, in the manipulations that follow we will not explicitly keep track of changes to the endcaps.

\subsection{Converting from the 7-d to the 8-d basis}
\label{app:7to8}

In order to match with the main text, and to allow the full use of isospin symmetry,
we need to decompose the $D^+ D^0$ pair into even and odd partial waves, corresponding to $I=1$ and $I=0$,
respectively. This can be done using the following $8\times 7$ matrix,
\begin{equation}
C_{7\to8}=
\begin{pmatrix}
1 & 0 & 0 & 0 & 0 & 0 & 0 \\
0 & 1 & 0 & 0 & 0 & 0 & 0 \\
0 & 0 & 1 & 0 & 0 & 0 & 0 \\
0 & 0 & 0 & 1 & 0 & 0 & 0 \\
0 & 0 & 0 & 0 & P_e & 0 & 0 \\
0 & 0 & 0 & 0 & P_o & 0 & 0 \\
0 & 0 & 0 & 0 & 0 & 1 & 0\\
0 & 0 & 0 & 0 & 0 & 0 & 1
\end{pmatrix}\,,
\label{eq:C7to8}
\end{equation}
where $P_o = (1-P^{(\ell)})/2$ projects onto odd waves.
Here we have extended the ordering given in \Cref{eq:pairspect} by the change
\begin{equation}
[D^+ D^0] \pi^0 \longrightarrow
[D^+ D^0]_1 \pi^0,\ [D^+ D^0]_0 \pi^0 \,.
\label{eq:extendbasis}
\end{equation}
To use this matrix we insert the $7\times 7$ identity, written as $C_{7\to 8}^\dagger C_{7\to 8}$,
between all matrices in \Cref{eq:CLresulthat}.
This leads to a result of the same form, except that $\widehat F_G$, $\widehat{\cK}_{2,L}$,
and $\widehat{\cK}_{\rm df,3}^{(u,u)}$ are conjugated by $C_{7\to8}$,
for example $\widehat F_G \to C_{7\to8} \widehat F_G C_{7\to 8}^\dagger$.
For simplicity of notation, we use the same names for these quantities after conjugation.

This leads to the following changes. The block-diagonal form of $\widehat F_G$, \Cref{eq:FGa},
is maintained, but the central block changes to the 4-d form
\begin{equation}
F_G^{1+1+1} \to
\begin{pmatrix}
\tilde F^{D} & \tilde G^{DD} & P^{(\ell)} \tilde G^{D\pi} P_e & -P^{(\ell)} \tilde G^{D\pi} P_o
\\
\tilde G^{DD} & \tilde F^{D} & P^{(\ell)} \tilde G^{D\pi}P_e & P^{(\ell)} \tilde G^{D\pi}P_o
\\
P_e \tilde G^{\pi D} P^{(\ell)} & P_e \tilde G^{\pi D} P^{(\ell)} & P_e \tilde F^{\pi} P_e & 0
\\
-P_o \tilde G^{\pi D} P^{(\ell)} & P_o \tilde G^{\pi D} P^{(\ell)} & 0 & P_o \tilde F^{\pi} P_o
\end{pmatrix}\,.
\label{eq:FG1p1p14d}
\end{equation}
To obtain this we have used $P_e P^{(\ell)} = P_e = P^{(\ell)} P_e$, $P_o P^{(\ell)} = - P_o = P^{(\ell)} P_o$,
and $P_e F^\pi P_o = P_o F^\pi P_e = 0$.

For $\widehat{\cK}_{2,L}$, whose 7-d form is given in \Cref{eq:K2Lhat},
the extension to the 8-d form involves changing the $\{5,5\}$ entry,
$\cK_{2,L}^{DD,b}$, into the following 2-d matrix,
\begin{equation}
{\rm diag} \left(
2 \omega_k^\pi L^3 \delta_{k'k}\delta_{\ell'\ell} \delta_{m' m} \tfrac12 \cK_2^{\ell}(DD[I=1]) ,
2 \omega_k^\pi L^3 \delta_{k'k}\delta_{\ell'\ell} \delta_{m' m} \tfrac12 \cK_2^{\ell}(DD[I=0]) \right)\,,
\end{equation}
with all other additional entries being filled with zeros.
We note the presence of the factors of $1/2$ multiplying the isospin amplitudes $\cK_2^\ell$,
which arise from the fact that
\begin{equation}
\cK_2[I=1] = 2 \cK(D^+D^0\to D^+ D^0)_{\ell \ \rm even}, \quad \cK_2[I=0] = 2 \cK(D^+D^0\to D^+ D^0)_{\ell \ \rm odd}\,.
\end{equation}
Finally, $\widehat{\cK}_{\rm df,3}^{(u,u)}$, whose detailed form in terms of the original quantities such as $B_3$
we are not keeping track of, extends to an $8\times 8$ matrix.

\subsection{Converting to the total isospin basis}
\label{app:converttoI}

We now convert to the total isospin basis, using the ordering given in \Cref{eq:totalisospinorder}.
Using the isospin relations given in \Cref{eq:K2iso1} to \Cref{eq:K2ison}
we can write each of these states in terms of the basis \Cref{eq:pairspect}, extended by \Cref{eq:extendbasis}.
This requires the conversion matrix
\begin{equation}
C_{8 \to I} = \frac1{\sqrt6}
\begin{pmatrix}
1 & 0 & \sqrt2 & \sqrt2 & 0 & 0 & 1 & 0
\\
0 & 1 & 0 & 0 & 2 & 0 & 0 & 1
\\
-1 & 0 & \sqrt2 & -\sqrt2 & 0 & 0 & 1 & 0
\\
\sqrt2 & 0 & 1 & -1 & 0 & 0 & -\sqrt2 & 0
\\
0 & \sqrt3 & 0 & 0 & 0 & 0 & 0 & -\sqrt3
\\
0 & 0 & 0 & 0 & 0 & \sqrt6 & 0 & 0
\\
-\sqrt2 & 0 & 1 & 1 & 0 & 0 & -\sqrt2 & 0
\\
0 & \sqrt2 & 0 & 0 & -\sqrt2 & 0 & 0 & \sqrt2
\end{pmatrix}\,.
\label{eq:C8toI}
\end{equation}
We stress that this rotation only combines quantities with the same flavor of spectator
(i.e. $D$ or $\pi$) and the same symmetry of the pair (e.g. symmetric or antisymmetric $DD$ pairs).
This is necessary for rotation to be allowable.
Inserting $\mathbf 1_{8\times 8} = C_{8\to I}^\dagger C_{8\to I}$ between all matrices in the
correlator, one obtains the matrices in the isospin basis by conjugation,
e.g. $\widehat F_G \to C_{8\to I} \widehat F_G C_{8\to I}^\dagger$.
We will again use the same notation for matrices after the basis change.

The results are block diagonal in total isospin.
The three blocks for $\widehat F_G$ are
\begin{align}
\widehat F_G[I=2] &=
\begin{pmatrix}
\tilde F^{D}+ \tilde G^{DD} & \sqrt2 P^{(\ell)} \tilde G^{D\pi} P_e
\\
\sqrt2 P_e \tilde G^{\pi D} P^{(\ell)} & P_e \tilde F^{\pi} P_e \end{pmatrix}\,,
\\
\widehat F_G[I=1] &=
\begin{pmatrix}
\tilde F^{D} - \frac13 \tilde G^{DD} & -\frac{2\sqrt2}3 \tilde G^{DD}
& -\sqrt{\frac23} P^{(\ell)} \tilde G^{D\pi} P_e & -\frac2{\sqrt3} P^{(\ell)} \tilde G^{D\pi} P_o
\\
-\frac{2\sqrt2}3 \tilde G^{DD} & \tilde F^{D} + \frac13 \tilde G^{DD}
& \frac2{\sqrt3} P^{(\ell)} \tilde G^{D\pi} P_e & -\sqrt{\frac23} P^{(\ell)} \tilde G^{D\pi} P_o
\\
-\sqrt{\frac23} P_e \tilde G^{\pi D} P^{(\ell)} & \frac2{\sqrt3} P_e \tilde G^{\pi D} P^{(\ell)}
& P_e \tilde F^{\pi} P_e & 0
\\
-\frac2{\sqrt3} P_o \tilde G^{\pi D} P^{(\ell)} & -\sqrt{\frac23} P_o \tilde G^{\pi D} P^{(\ell)}
& 0 & P_o \tilde F^{\pi} P_o
\end{pmatrix}\,,
\\
\widehat F_G[I=0] &= \begin{pmatrix}
\tilde F^{D}+ \tilde G^{DD} & -\sqrt2 P^{(\ell)} \tilde G^{D\pi} P_e
\\
-\sqrt2 P_e \tilde G^{\pi D} P^{(\ell)} & P_e \tilde F^{\pi} P_e \end{pmatrix}\,.
\end{align}
These results are reproduced in the main text in \Cref{eq:FhatI20,eq:FhatI1,eq:GhatI20,eq:GhatI1},
where the $F$ and $G$ parts of $F_G$ are separated.
Furthermore, the factors of $P_e$ and $P_o$ appearing here are set to unity in the main text,
because the (anti)symmetry that they enforce is automatically built in to the two- and three-particle
K matrices on which $F_G$ acts.

For $\widehat{\cK}_{2,L}$ we obtain the results given in \Cref{eq:KhatI2,eq:KhatI1,eq:KhatI0}.
As for $\widehat{\cK}_{\rm df,3}^{(u,u)}$, we know that this must also block diagonalize,
since isospin is an exact symmetry,
but we hold off from discussing the form of these blocks until later.

\subsection{Asymmetric form of the three-particle quantization condition}

We have arrived at a form for the volume-dependent part of the correlation function,
given in \Cref{eq:CLresulthat}, in which all the quantities (aside from $\widehat A'$ and $\widehat A$)
are $8\times 8$ matrices in the isospin basis of \Cref{eq:totalisospinorder}.
This allows us to simply read off the asymmetric form of the quantization condition
(i.e. that containing an asymmetric three-particle K matrix)
\begin{equation}
\det \left[1 + (\widehat \cK_{2,L} + \widehat{\cK}_{\rm df,3}^{(u,u)}) \widehat F_G \right] = 0\,.
\label{eq:QC3asymm}
\end{equation}
We will not make use of this form, but display it as it is the result that can be connected
to the quantization conditions obtained in the FVU approach~\cite{\MD}, as shown in ref.~\cite{\BSequiv}.

\subsection{Symmetric form of the quantization condition}
\label{app:QCsym}

To obtain the desired symmetric form of the quantization condition, written
in terms of a symmetrized three-particle K matrix, requires use of the
symmetrization identities discussed in BS1, BS2 and BS3. These are most straightforwardly
implemented on a different finite-volume correlator than $C_L$, namely a finite-volume quantity,
$\cM_{23,L,\rm off}$, that is akin to the three-particle scattering amplitude in infinite volume.
Using this quantity also allows
the determination of the integral equations relating the divergence-free three-particle K matrix to $\cM_3$.

$\cM_{23,L,\rm off}$ is a finite-volume $3\to3$ correlation function
involving external single-particle legs at fixed momenta (drawn from the finite-volume set),
which has been amputated, and has the final and initial times sent to $\pm \infty$, respectively.
The subscript ``$23$” indicates that this quantity contains not only fully-connected contributions,
but also those in which one of the three particles spectates while the others interact.
See Fig.~2 of BS3 for an explanation of the definition.
The subscript ``off” indicates that this is an off-shell amplitude
because the energy $E$ does not, in general, equal the sum of the external on-shell energies.
In the following, we will
use the notation that the absence of the subscript ``off'' implies that the quantity is on shell.

In our set-up, $\cM_{23,L,\rm off}$ is a $3\times 3$ matrix connecting the different flavors in \Cref{eq:flavorchannels},
as well as a matrix in $\{\bm k\}$ space.
Generalizing the arguments in BS2 and BS3, it is straightforward to show that
\begin{equation}
\cM_{23,L,\rm off} = (B_{2,L} + B_3)\frac1{1 -i D i(B_{2,L}+B_3)}\,,
\end{equation}
where $D$, $B_{2,L}$ and $B_3$ are the same as above.
As for $C_L$, the symmetrization factors contained in $B_{2,L}$ [see \Cref{eq:B2Ldef}]
can be moved onto the cut-factor $D$, as well as to the outside. This leads to
\begin{equation}
\cM_{23,L,\rm off} = \tilde S^D (\tilde B_{2,L} + \tilde B_3)
\frac1{1 -i D_S i(\tilde B_{2,L}+ \tilde B_3)} \tilde S^D \,,
\end{equation}
where again the quantities are as above [see eqs.~(\ref{eq:tSDdef})---(\ref{eq:symm3})].

Next we introduce additional matrix indices as above to convert to a 7-d matrix form, and then project on shell.
Following the same steps as in BS2 and BS3, this yields
\begin{align}
\cM_{23,L} &= \CR \circ \widehat{\cM}_{23,L}^{(u,u)} \circ \CL\,,
\label{eq:M23Ldef}
\\
\widehat{\cM}_{23,L}^{(u,u)} &= (\widehat{\cK}_{2,L} + \widehat{\cK}_{\df,3}^{(u,u)})
\frac1{1 -i \widehat F_G i(\widehat{\cK}_{2,L} + \widehat{\cK}_{\df,3}^{(u,u)})} \,,
\end{align}
where $\widehat F_G$, $\widehat{\cK}_{2,L}$ and $\widehat{\cK}_{\df,3}^{(u,u)}$ are the
$7\times 7$ matrices appearing above,
and $\CR$ is a $3\times 7$ matrix of operators with block form
\begin{equation}
\CR = \widetilde S^D \begin{pmatrix}
\boldsymbol{\mathcal V}_{\alpha} & 0 & 0
\\
0 & \boldsymbol{\mathcal V}_{1} & 0
\\
0 & 0 & \boldsymbol{\mathcal V}_{\alpha}
\end{pmatrix}
=
\begin{pmatrix}
S^D \boldsymbol{\mathcal V}_{\alpha} & 0 & 0
\\
0 & \boldsymbol{\mathcal V}_{1} & 0
\\
0 & 0 & S^D \boldsymbol{\mathcal V}_{\alpha}
\end{pmatrix}
\,,
\label{eq:Cdef}
\end{equation}
with the vectors are given by
\begin{align}
\boldsymbol{\mathcal V}_{\alpha} &= \left( \XR{1}{2}{3},\ \sqrt{\tfrac12} \XR{3}{1}{2} \right)\,,
\label{eq:braalphadef}
\\
\boldsymbol{\mathcal V}_{1} &= \Big(\XR{2}{1}{3},\ \XR{1}{2}{3},\ \XR{3}{1}{2} \Big)\,.
\label{eq:bra1def}
\end{align}

The vector $\boldsymbol{\mathcal V}_{\alpha}$ is introduced in sec.~IID of BS2,
where it is denoted $\bra{\alpha}$,
while $\boldsymbol{\mathcal V}_{1}$ is introduced in sec.~V of BS3,
where it is denoted $\bra{1}$.
We prefer the present notation since the action of these vectors is not simply
one of forming a matrix product,
but also includes the combination of quantities having $\{k\ell m\}$ indices
with spherical harmonics, so as to express the result in terms of momenta.
For the same reason we use the operator $\circ$ appearing in \Cref{eq:M23Ldef}.
The combination with spherical harmonics is brought about by
the operator $\boldsymbol{\mathcal X}_{[kab]}^{\boldsymbol \sigma}$,
defined by its action on a vector $f$ with index $\{k \ell m\}$:
\begin{align}
\left[\boldsymbol{\mathcal X}_{[kab]}^{\boldsymbol \sigma} \circ f\right] (\{ p_i \})
= \left[\sum_{\ell m} Y^*_{\ell m}(\hat a^*) f_{k\ell m}
\right]_{\bm k\to \bm p_{\sigma_1} , \, \bm a\to \bm p_{\sigma_2}, \, \bm b \to \bm p_{\sigma_3}}
\,,
\label{eq:XRdef}
\end{align}
where $\boldsymbol \sigma$ is a permutation of $\{1,2,3\}$.
In words, the sum over $\ell m$ yields a function of $\bm k$ and $\hat a^*$.
The former is then equated to $\bm p_{\sigma_1}$ (the spectator momentum),
while the latter is set to the direction of $p_{\sigma_2}$ when boosted to the CMF of
the nonspectator pair, as discussed following \Cref{eq:Ft}. For three on-shell momenta,
this completely determines $p_{\sigma_2}$ and also the final momentum $b=P-k-a$,
which is equated with $p_{\sigma_3}$.
The result is a function of (in this example, the final state) on-shell momenta $p_1, p_2, p_3$.
The choice of momentum assignments in \Cref{eq:braalphadef,eq:bra1def}
follows from the ordering of the flavor channels in \Cref{eq:flavorchannels},
and the ordering of choices of the spectator and dominant member of the pair given in \Cref{eq:pairspect}.

The left-acting version $\boldsymbol{\mathcal X}_{[kab]}^{{\boldsymbol \sigma}\dagger}$ is defined analogously.
\begin{align}
\left[f \circ \boldsymbol{\mathcal X}_{[kab]}^{\boldsymbol \sigma \dagger} \right] (\{ k_i\})
= \left[\sum_{\ell m} f_{k\ell m} Y_{\ell m}(\hat a^*)
\right]_{\bm k\to \bm k_{\sigma_1} , \, \bm a\to \bm k_{\sigma_2}, \, \bm b \to \bm k_{\sigma_3}}
\,,
\label{eq:XLdef}
\end{align}
and appears in
\begin{equation}
\CL = \begin{pmatrix}
S^D \boldsymbol{\mathcal V}^\dagger_{\alpha} & 0 & 0
\\
0 & \boldsymbol{\mathcal V}^\dagger_{1} & 0
\\
0 & 0 & S^D \boldsymbol{\mathcal V}^\dagger_{\alpha}
\end{pmatrix}\,.
\end{equation}

Strictly speaking, \Cref{eq:M23Ldef} only makes sense in the $L\to\infty$ limit,
because \Cref{eq:XRdef,eq:XLdef} lead to momenta $\bm p_{\sigma_2}$ and $\bm p_{\sigma_3}$
that do not lie in the finite-volume set, instead taking on continuous values.
This point is discussed further in BS3 and in the main text.
Validity in the $L\to\infty$ limit is sufficient for our purposes,
as that is where we use the results of this subsection to obtain the desired integral equations.

$\cM_{23,L}^{(u,u)}$ contains both two- and three-particle scattering contributions, and
we need to subtract the former to make contact with $\cM_3$. This is achieved by
\begin{align}
\widehat{\cM}_{3,L}^{(u,u)} &= \widehat{\cM}_{23,L}^{(u,u)} - \widehat{\cM}_{2,L}\,,
\end{align}
where $\widehat{\cM}_{2,L}$ is the correlator involving two-particle scattering with a noninteracting spectator,
\begin{align}
\widehat{\cM}_{2,L} &= \widehat{\cK}_{2,L} \frac1{1-i \widehat F i \widehat{\cK}_{2,L}} \,.
\end{align}
Here we have used the decomposition
\begin{equation}
\widehat F_G = \widehat F + \widehat G\,,
\end{equation}
where $\widehat F$ contains only the $F$ terms in \Cref{eq:FGa} and $\widehat G$ the $G$ terms.

Algebraic manipulations described in Appendix C of BS1 then lead to
\begin{equation}
\widehat{\cM}_{3,L}^{(u,u)} = \widehat{\cD}_L^{(u,u)} + \widehat{\cM}_{\df,3,L}^{(u,u)}\,,
\end{equation}
where
\begin{align}
\widehat{\cD}_L^{(u,u)} &= - \widehat{\cM}_{2,L} \widehat G \widehat{\cM}_{2,L}
\frac1{1 + \widehat{G} \widehat{\cM}_{2,L}}\,,
\end{align}
and
\begin{equation}
\widehat{\cM}_{\df,3,L}^{(u,u)} = \left[ 1 - \widehat{\cD}_{23,L}^{(u,u)} \widehat F_G\right] \widehat{\cK}_{\df,3}^{(u,u)}
\frac1{1 + \left[ 1 - \widehat{\cD}_{23,L}^{(u,u)} \widehat F_G\right] \widehat F_G \widehat{\cK}_{\df,3}^{(u,u)}}
\left[ 1 - \widehat F_G \widehat{\cD}_{23,L}^{(u,u)} \right]\,,
\label{eq:Mdf3Luu}
\end{equation}
where
\begin{equation}
\widehat{\cD}_{23,L}^{(u,u)} = \widehat{\cM}_{2,L} + \widehat{\cD}_{L}^{(u,u)}\,.
\end{equation}

One can now obtain the infinite-volume three-particle scattering amplitude, $\cM_3$, which we recall is a 3-d matrix
in flavor space, as well as a matrix in momentum space, by first evaluating
\begin{equation}
\cM_{3,L} = \CR \circ \widehat{\cM}_{3,L}^{(u,u)} \circ \CL
\label{eq:M3Lfinal}
\end{equation}
and then sending $L\to\infty$ with the $i\epsilon$ factors in $F$ and $G$
kept nonzero until after the limit has been taken.
This procedure is explained in ref.~\cite{\HSQCb}, and leads to a set of coupled integral equations that
relate $\cM_3$ to $\widehat{\cK}_{\df,3}^{(u,u)}$.
The drawback of this equation is that $\widehat{\cK}_{\df,3}^{(u,u)}$ is a matrix in which all entries correspond
to sums of different sets of TOPT diagrams, and thus, in general, are different.
Furthermore, these entries are not Lorentz invariant.

These two problems can be resolved by symmetrizing $\widehat{\cK}_{\df,3}^{(u,u)}$.
This can be done by combining the symmetrization relations for three distinct particles
(introduced in sec.~VIIIA of BS3)
with those for $2+1$ systems
(see sec.~IIIA of BS2).
These relations have the same algebraic form, and can be combined into a single symmetrization operator.
Following the algebraic steps in Appendix D of BS3, we then find\footnote{%
Strictly speaking, one should use finite-volume symmetrization operators instead of
$\CR$ and $\CL$ in \Cref{eq:M3Lnew,eq:Mdf3L}, since, as noted above, $\CR$ and $\CL$ are well defined only
in the $L\to\infty$ limit. The finite-volume versions are given, for three distinct particles,
in eq.~(98) of BS3, and, for $2+1$ systems, in eq.~(80) of BS2.
These versions are what is needed to obtain the symmetric form of the quantization condition,
but we do not give them explicitly to avoid introducing further notation.
To obtain the integral equations, the expressions involving $\CR$ and $\CL$ can be used, since
the $L\to\infty$ limit is taken.
In particular, the quantities $S^D \boldsymbol{\mathcal V}_{\alpha}$
and $\boldsymbol{\mathcal V}_{1} $ contained in $\CR$, \Cref{eq:Cdef},
perform the necessary symmetrization of the initial and final states.
[In the $2+1$ case, this is shown explicitly in eqs.~(87) and (88) of BS2.
For three distinct particles, the corresponding result, given in eq.~(120) of BS3, uses an
explicit symmetrization operator, rather than the vector $\boldsymbol{\mathcal V}_1 $ contained in $\CR$.
The form given here is, however, equivalent.]
}
\begin{align}
\cM_{3,L} &= \cM_{\df,3,L} + \boldsymbol {\mathcal C} \circ \widehat{\cD}_L^{(u,u)} \circ \boldsymbol {\mathcal C}^\dagger\,,
\label{eq:M3Lnew}
\\
\cM_{\df,3,L} &= \boldsymbol {\mathcal C} \circ \widehat{\cM}_{\df,3,L}^{(u,u)\prime} \circ \boldsymbol {\mathcal C}^\dagger\,,
\label{eq:Mdf3L}
\\
\widehat{\cM}_{\df,3,L}^{(u,u)\prime} &= \left[ \frac13 - \widehat{\cD}_{23,L}^{(u,u)} \widehat F \right]
\widehat{\cK}_{\df,3} \frac1{1 + \widehat F_3 \widehat{\cK}_{\df,3}}
\left[\frac13 - \widehat F \widehat{\cD}_{23,L}^{(u,u)} \right]\,,
\label{eq:Mhatdf3L}
\\
\widehat F_3 &= \frac{\widehat F}3 - \widehat F \frac1{\widehat{\cK}_{2,L}^{-1} + \widehat F_G} \widehat F\,.
\label{eq:F3app}
\end{align}
We stress that $\widehat{\cM}_{\df,3,L}^{(u,u)\prime}$ differs from
$\widehat{\cM}_{\df,3,L}^{(u,u)}$, given in \Cref{eq:Mdf3Luu}.
However, when sandwiched between the symmetrization operators $\CR$ and $\CL$,
as in \Cref{eq:M3Lnew}, they lead to the same result.
This point is discussed further in BS1.
The new K matrix $\widehat{\cK}_{\df,3}$ is related algebraically
to $\widehat{\cK}_{\df,3}^{(u,u)}$ by an expression that we will not need.
We stress that the symmetrization procedure does not lead to changes in $\widehat F$ or $\widehat G$.

The key features of $\widehat{\cK}_{\df,3}$
are that, within in each block (in the $2+3+2$ decomposition),
the underlying amplitude is the same for each element, and satisfies the same symmetries
as $\cM_3$, including Lorentz invariance.
One subtlety is there are factors of $\sqrt 2$ in the $2+1$ blocks; see eq.~(93) of BS2.

We can convert to the 8-d isospin basis by inserting the $7\times 7$ identity written as
\begin{equation}
\boldsymbol 1_{7\times7} = C_{7\to8}^\dagger C_{8\to I}^\dagger C_{8\to I} C_{7\to8}\,,
\label{eq:7did}
\end{equation}
between all matrices. Then one obtains the same expressions as above
[in particular, \Cref{eq:M3Lnew,eq:Mhatdf3L,eq:F3app}]
but with all matrices in the total-isospin basis,
and with
\begin{equation}
\CR \to \CR C_{7\to8}^\dagger C_{8\to I}^\dagger\,,\qquad
\CL \to
C_{8\to I} C_{7\to8} \CL\,.
\label{eq:CC7toI}
\end{equation}
As before, we will use the same notation for matrices after conversion to the total-isospin
basis as in the 7-d basis.

Since $\cM_{3,L}$ is a finite-volume matrix element, we can use it to obtain another form
for the quantization condition,
\begin{equation}
\det\left[1 + \widehat F_3 \widehat{\cK}_{\df,3} \right] = 0\,.
\label{eq:QC3app}
\end{equation}
We call this the symmetric form, as it contains a symmetric three-particle K matrix.
Since the components of $F_3$ are block diagonal in total isospin, as is $\widehat{\cK}_{\df,3}$
(a point to be discussed further below), this factorizes into the result given
in \Cref{eq:QC3isospin} of the main text.

\subsection{Extracting the isospin amplitudes \texorpdfstring{$\cM_3^{[I]}$}{}}
\label{app:inteqs}

The result from the previous section, $\cM_{3,L}$ in \Cref{eq:M3Lnew},
is a matrix in the 3-d flavor space of \Cref{eq:flavorchannels}.
In this section we project the amplitude onto those of definite isospin, $\cM_{3,L}^{[I]}$.
The $L\to\infty$ limit can then be taken, as explained in \Cref{sec:inteqs},
to obtain the integral equations determining the infinite-volume amplitudes $\cM_{3}^{[I]}$.

We consider first the $I=0$ amplitude. %
The $I=0$ state is [in agreement with the last line of \Cref{eq:ddpi_isospin_states},
and following the momentum assignments of \Cref{eq:flavorchannels}]
\begin{equation}
\frac1{\sqrt3} \left(
D^+(\bm k_1) D^+(\bm k_2) \pi^-(\bm k_3) - [D^+(\bm k_1) D^0(\bm k_2)]_S \pi^0(\bm k_3)
+ D^0(\bm k_1) D^0(\bm k_2) \pi^+(\bm k_3)
\right)\,,
\end{equation}
where
\begin{equation}
[D^+(\bm k_1) D^0(\bm k_2)]_S =
\frac1{\sqrt2} \left( D^+(\bm k_1) D^0(\bm k_2) + D^+(\bm k_2) D^0(\bm k_1)\right)\,.
\end{equation}
The symmetrization here is equivalent to acting with $S^D/\sqrt2$,
where $S^D$ is defined in \Cref{eq:SDdef}.
Thus we have that
\begin{align}
\cM_{3,L}^{[I=0]} &=
\boldsymbol{\mathcal X}^{[I=0]} \circ
\left( \widehat{\cM}_{\df,3,L}^{(u,u)\prime} + \widehat{\cD}_L^{(u,u)} \right)
\circ \boldsymbol{\mathcal X}^{[I=0]\dagger}
\,,
\label{eq:M3LI0}
\\
\boldsymbol{\mathcal X}^{[I=0]} &=
\frac1{\sqrt3} \left(1, - \frac{S^D}{\sqrt2}, 1 \right) \CR C_{7\to8}^\dagger C_{8\to I}^\dagger\,,
\label{eq:I0bra}
\end{align}
where we have included the transformation matrices such that $\widehat{\cM}_{\df,3,L}^{(u,u)\prime}$ and $\widehat{\cD}_L^{(u,u)}$ are in the total isospin basis, \Cref{eq:totalisospinorder}.
We expect that the 8-d vector $\bra{I=0}$ will project onto the $I=0$ block of these quantities,
but to show that this is indeed the case is rather subtle.

We begin by using the expression for $C$, \Cref{eq:Cdef}, to rewrite the projection vector as
\begin{multline}
\boldsymbol{\mathcal X}^{[I=0]} =
\frac{S^D}{\sqrt6} \times \\
\left(
\sqrt2 \XR123, \XR312,
-\XR213, -\XR123, -\XR312,
\sqrt2 \XR123, \XR312
\right)
C_{7\to8}^\dagger C_{8\to I}^\dagger\,.
\label{eq:I0bra2}
\end{multline}
We then observe that the $7\to I$ transformation matrix,
\begin{equation}
C_{7\to8}^\dagger C_{8\to I}^\dagger =\sqrt{\frac16} \begin{pmatrix}
1 & 0 & -1 & \sqrt2 & 0 & 0 & -\sqrt2 & 0
\\
0 & 1 & 0 & 0 & \sqrt3 & 0 & 0 & \sqrt2
\\
\sqrt2 & 0 & \sqrt2 & 1 & 0 & 0 & 1 & 0
\\
\sqrt2 & 0 & -\sqrt2 & -1 & 0 & 0 & 1 & 0
\\
0 & 2P_e & 0 & 0 & 0 & \sqrt6 P_o & 0 & - \sqrt2 P_e
\\
1 & 0 & 1 & -\sqrt2 & 0 & 0 & -\sqrt2 & 0
\\
0 & 1 & 0 & 0 & - \sqrt3 & 0 & 0 & \sqrt2
\end{pmatrix}\,,
\label{eq:7toITr}
\end{equation}
has a block diagonal structure:
the second, fifth and final rows mix only into the second, fifth, sixth and eighth columns,
while the remaining four rows mix only into the first, third, fourth and seventh columns.
This is as expected since these two blocks correspond, respectively, to the
spectator being a pion or a $D$ meson.
This structure implies that terms with $\XR312$ are combined,
and further, using the result
\begin{equation}
P^D \XR123 = \XR213 \ \ \Rightarrow\ \
S^D \XR123 = S^D \XR213 \,,
\label{eq:PDXR}
\end{equation}
that terms with $\XR213$ are similarly combined.
Another simplification is that the factors of $P_e$ and $P_o$ can both be set to unity,
since the second and eighth columns in the total-isospin basis contain the pair $[DD]_1$,
which is automatically symmetric, while the seventh column contains $[DD]_0$,
which is antisymmetric.
The net result is \footnote{%
We use an arrow rather than an equality here and below when we are using results that
follow from the fact that this vector acts on a quantity with definite symmetry properties.
}
\begin{align}
\boldsymbol{\mathcal X}^{[I=0]}
&\to
S^D \left(0,0,0,0,0,-\sqrt{\tfrac16} \XR312, \ -\XR123,\ \sqrt{\tfrac12} \XR312\right) \,.
\label{eq:I0bra3}
\end{align}
We now observe that, since the sixth entry corresponds to a $[DD]_0$ pair, it is annihilated by the
symmetrization operator $S^D$, so that we do recover the expected projection onto the $I=0$ block.
Furthermore, the final entry corresponds to a symmetric $[DD]_1$ pair, so $S^D$ can be replaced by $2$.
Using \Cref{eq:PDXR} for the seventh entry, we come to the final form
\begin{equation}
\boldsymbol{\mathcal X}^{[I=0]} \to \left(0,0,0,0,0,0, -\XR123-\XR213, {\sqrt2} \XR312 \right) \,.
\label{eq:alpI0f}
\end{equation}
One might have expected that the two terms in the seventh entry could be combined,
since they correspond to a state, $[[D\pi]_{1/2} D]_0$, that is symmetric under exchange of the two $D$s.
However, the direct connection to this state is lost when the amplitude is broken up into spectator
channels, so explicit symmetrization is required.

Using this result, and the block-diagonal nature of
$\widehat{\cM}_{\df,3,L}^{(u,u)\prime}$ and $\widehat{\cD}_L^{(u,u)}$,
leads to the integral equations implicitly described in
\Cref{eq:M3L,eq:MhatuuI,eq:DhatuuI,eq:Mhatdf3LI,eq:Dhat23Ldef},
with the vector $\boldsymbol{\mathcal X}^{[I=0]} $ leading to the result for
$\boldsymbol{\mathcal X}_0$
in \Cref{eq:alphaSvec} once only the $I=0$ part is kept.

To obtain the $I=2$ amplitude, we note that the $I=2$ state is
[in agreement with the first line of \Cref{eq:ddpi_isospin_states}]
\begin{equation}
\frac1{\sqrt6} \left(
D^+(\bm k_1) D^+(\bm k_2) \pi^-(\bm k_3) +2 [D^+(\bm k_1) D^0(\bm k_2)]_S \pi^0(\bm k_3)
+ D^0(\bm k_1) D^0(\bm k_2) \pi^+(\bm k_3)
\right)\,.
\end{equation}
The projector in this case is
\begin{align}
\boldsymbol{\mathcal X}^{[I=2]}
&=
\frac1{\sqrt6} \left(1, \sqrt2 {S^D}, 1 \right) C C_{7\to8}^\dagger C_{8\to I}^\dagger\,,
\label{eq:I2bra}
\\
&\to
\left( \XR123 + \XR213,\ {\sqrt2} \XR312, 0,0,0,0,0, 0\right) \,,
\label{eq:alpI2f}
\end{align}
where to obtain the second line we have used the similar arguments as in the $I=0$ case.
This result leads to the integral equation for the $I=2$ amplitude having a similar form to
that for $I=0$, see \Cref{eq:M3L}.

Finally, we consider the $I=1$ case. The two independent states are given in
\Cref{eq:ddpi_isospin_states}.
The $I=1,a$ state is
\begin{equation}
\frac1{\sqrt2} \left( D^+(\bm k_1) D^+(\bm k_2) \pi^- (\bm k_3) - D^0(\bm k_2) D^0(\bm k_1) \pi^+ (\bm k_3) \right)\,.
\end{equation}
The projector onto this state is
\begin{align}
\boldsymbol{\mathcal X}^{[I=1,a]} &= \sqrt{\tfrac12} (1,0, -1) C C_{7\to 8}^\dagger C_{8 \to I}^\dagger
\\
&\to
S^D \left(0,0,-\sqrt{\tfrac13}\XR213,\
\sqrt{\tfrac23}\XR123,\
\sqrt{\tfrac12} \XR312, 0, 0, 0 \right)\,,
\label{eq:I1abra}
\end{align}
where to obtain the second line the same arguments as for $I=0$ have again been used.
As expected, only entries in the $I=1$ block are nonzero.
We can simplify further by noting that the fifth entry corresponds to a symmetric $DD$ pair
\begin{align}
\boldsymbol{\mathcal X}^{[I=1,a]}
&\to \left(0,0,-\sqrt{\tfrac13}[\XR123 \!+\!\XR213],
\sqrt{\tfrac23} [\XR123 \!+\!\XR213] ,
\sqrt2\, \XR312, 0, 0, 0 \right)\,.
\label{eq:alpI1af}
\end{align}
This projector leads to $\boldsymbol{\mathcal X}_a$ in \Cref{eq:alphaavec} when only the $I=1$ part is kept.

The $I=1,b$ state is
\begin{equation}
[D^+(\bm k_1) D^0(\bm k_2)]_A \pi^0(\bm k_3)
\equiv
\frac1{\sqrt2} \left( D^+(\bm k_1) D^0(\bm k_2) \pi^0 (\bm k_3) - D^0(\bm k_2) D^+(\bm k_1) \pi^0 (\bm k_3) \right)\,,
\end{equation}
which exhibits the new feature of antisymmetrization of the two $D$s.
The corresponding projector is
\begin{align}
\boldsymbol{\mathcal X}^{[I=1,b]}&= A^D (0,\ \sqrt{\tfrac12},\ 0) \ C C_{7\to 8}^\dagger C_{8 \to I}^\dagger \,,
\\[5pt]
& \hspace{-0pt}= \sqrt{\tfrac12} A^D \left(0,0,\XR213,\ \XR123,\ \XR312, 0,0 \right)
C_{7\to 8}^\dagger C_{8 \to I}^\dagger\,,
\label{eq:I1bbra}
\\[5pt]
& \hspace{-0pt}= \sqrt{\tfrac12} \left(0,0,\XR213\!-\!\XR123,\ \XR123\!-\!\XR213,\ 2\XR312,\ 0,0 \right)
C_{7\to 8}^\dagger C_{8 \to I}^\dagger\,,
\label{eq:I1bbrab}
\\[5pt]
& \hspace{-0pt}= \left(0,0, \sqrt{\tfrac23}[\XR213\!-\!\XR123],
-\sqrt{\tfrac13}[\XR123\!-\!\XR213], 0, \sqrt2\, \XR312, 0,0 \right) \,,
\label{eq:I1bbrac}
\end{align}
where $A^D = 1 - P^D$ is the $DD$ antisymmetrization operator [cf. \Cref{eq:SDdef}],
and we have used results $A_D P_e\to 0$ and $A_D P_o\to 2 A_D$, as well as \Cref{eq:PDXR}.
This projector agrees with $\boldsymbol{\mathcal X}_b$ in \Cref{eq:alphabvec}.

\subsection{Symmetry factors for \texorpdfstring{$\Kdf$}{the three-body K-matrix}}
\label{app:Kdfform}

Finally, we discuss the form of $\widehat{\cK}_{\df,3}$ within each of the total isospin blocks.
We note that in this section we are working solely with infinite-volume quantities.
For the $I=0$ and $I=2$ blocks, we expect that each entry is given
by a single underlying amplitude, $\mathcal K_{\rm df,3}^{I}$, which is (a) expressed in terms
of different coordinates for each entry, and (b) comes with a potential symmetry factor for each entry.
For the $I=1$ block there will be an underlying $2\times 2$ matrix distributed through the $4\times 4$ block.

To demonstrate that these expectations are correct,
we start from the original $7\times 7$ matrix form of $\widehat{\cK}_{\df,3}$.
From its derivation, we know that this is composed of blocks, all elements of which contain the same
underlying quantity expressed in different coordinates, along with potential symmetry factors.
To express this explicitly, we introduce operators $\boldsymbol{\mathcal Y}^{[kab]}_{\boldsymbol \sigma}$
that have the inverse action of the $\boldsymbol{\mathcal X}_{[kab]}^{\boldsymbol \sigma}$
introduced in \Cref{eq:XRdef}.
This is a new notation compared to BS3 and BS2, which we hope clarifies the rather confusing
index structure.
The $\boldsymbol{\mathcal Y}^{[kab]}_{\boldsymbol \sigma}$ act on functions $g(\{p_i\})$ of three on-shell momenta
and give rise to objects that have $\{k\ell m\}$ indices:
\begin{align}
\left[{\boldsymbol {\mathcal Y}}_{{{\boldsymbol \sigma}}}^{[kab]} \circ g \right]_{k\ell m}
&=
\frac{1}{4\pi} \int d\Omega_{a^*} Y_{\ell m}(\hat a^*)
g(\{ p_i \})\bigg|_{p_{\sigma(1)}\to k,\ p_{\sigma(2)}\to a, \ p_{\sigma(3)}\to b} \,,
\label{eq:YRdef}
\end{align}
where $\boldsymbol \sigma$ is again a permutation of $\{1,2,3\}$.
In words, we choose $p_{\sigma_1}$ to be the spectator momentum,
leaving $p_{\sigma_2}$ and $p_{\sigma_3}$ to be the remaining pair.
We boost to the CMF of this pair, and decompose into spherical harmonics, defining
$\hat a^*$ as the direction of $\bm p_{\sigma_2}$ in this frame.
An analogous definition holds for the conjugate operator
$\boldsymbol{\mathcal Y}^{[kab]\dagger}_{\boldsymbol \sigma}$,
which acts from the right and includes the complex-conjugated spherical harmonics.

Returning to $\widehat{\cK}_{\df,3}$, we know from BS2 and BS3 that it has the schematic form
\begin{equation}
\widehat{\cK}_{\df,3} \sim
\begin{pmatrix}
\cK^a & \frac{\cK^a}{\sqrt2} & \cK^b &\cK^b &\cK^b &\cK^c&\frac{\cK^c}{\sqrt2}
\\
\frac{\cK^a}{\sqrt2} & \frac{\cK^a}2 & \frac{\cK^b}{\sqrt2} &\frac{\cK^b}{\sqrt2}
&\frac{\cK^{b}}{\sqrt2} &\frac{\cK^c}{\sqrt2} &\frac{\cK^c}2
\\
\cK^{b,{\rm T}} & \frac{\cK^{b,{\rm T}}}{\sqrt2} & \cK^d &\cK^d &\cK^d & \cK^e & \frac{\cK^e}{\sqrt2}
\\
\cK^{b,{\rm T}} & \frac{\cK^{b,{\rm T}}}{\sqrt2}& \cK^d &\cK^d &\cK^d & \cK^e & \frac{\cK^e}{\sqrt2}
\\
\cK^{b,{\rm T}} & \frac{\cK^{b,{\rm T}}}{\sqrt2}& \cK^d &\cK^d &\cK^d & \cK^e &\frac{\cK^e}{\sqrt2}
\\
\cK^{c,{\rm T}} & \frac{\cK^{c,{\rm T}}}{\sqrt2}& \cK^{e,{\rm T}} &\cK^{e,{\rm T}} &\cK^{e,{\rm T}} &\cK^f&\frac{\cK^f}{\sqrt2}
\\
\frac{\cK^{c,{\rm T}}}{\sqrt2} & \frac{\cK^{c,{\rm T}}}2 & \frac{\cK^{e,{\rm T}}}{\sqrt2} & \frac{\cK^{e,{\rm T}}}{\sqrt2} & \frac{\cK^{e,{\rm T}}}{\sqrt2}
& \frac{\cK^f}{\sqrt2} &\frac{\cK^f}2
\end{pmatrix}\,,
\label{eq:Kdfgeneral}
\end{equation}
where $\{k\ell m\}$ indices, as well as the flavors of spectator particles, are implicit.
These can be determined from the position in the blocks,
using the ordering of states in the 7-d basis given in \Cref{eq:pairspect}
and the momentum labelings in \Cref{eq:flavorchannels}.
Each block can be represented as an outer product.
For example, the $2 \times 3$ $\cK^b$ block, the $2 \times 2$ $\cK^{c,{\rm T}}$ block,
and the $3\times 2$ $\cK^e$ block are, respectively
\begin{align}
\begin{pmatrix}
\cK^b & \cK^b & \cK^b
\\
\frac{\cK^b}{\sqrt2} & \frac{\cK^b}{\sqrt2} & \frac{\cK^b}{\sqrt2}
\end{pmatrix}
&\sim
\begin{pmatrix}
{\boldsymbol {\mathcal Y}}^{[kab]}_{(123)} \\[5pt]
\frac1{\sqrt2} {\boldsymbol {\mathcal Y}}^{[kab]}_{(312)}
\end{pmatrix}
\circ
\cK^b(\{p_i\} ; \{k_i\})
\circ
\begin{pmatrix}
{\boldsymbol {\mathcal Y}}^{[kab]\dagger}_{(213)}, \ & \ {\boldsymbol {\mathcal Y}}^{[kab]\dagger}_{(123)}, \ & \ {\boldsymbol {\mathcal Y}}^{[kab]\dagger}_{(312)}
\end{pmatrix}\,,
\\[5pt]
\begin{pmatrix}
\cK^{c,\rm T} & \frac{\cK^{c,\rm T}}{\sqrt2}
\\
\frac{\cK^{c,\rm T}}{\sqrt2} & \frac{\cK^{c,\rm T}}{2}
\end{pmatrix}
&\sim
\begin{pmatrix} \,
{\boldsymbol {\mathcal Y}}^{[kab]}_{(123)} \\[5pt]
\frac1{\sqrt2}{\boldsymbol {\mathcal Y}}^{[kab]}_{(312)}
\end{pmatrix}
\circ
\cK^{c, \rm T}(\{p_i\} ; \{k_i\})
\circ
\begin{pmatrix}
{\boldsymbol {\mathcal Y}}^{[kab]\dagger}_{(123)} , \
& \ \frac1{\sqrt2} {\boldsymbol {\mathcal Y}}^{[kab]\dagger}_{(312)}
\end{pmatrix}\,,
\\[5pt]
\begin{pmatrix}
\cK^e & \frac{\cK^{e}}{\sqrt2}
\\
\cK^e & \frac{\cK^{e}}{\sqrt2}
\\
\cK^e & \frac{\cK^{e}}{\sqrt2}
\end{pmatrix}
& \sim
\begin{pmatrix}
{\boldsymbol {\mathcal Y}}^{[kab]}_{(213)} \\[5pt]
{\boldsymbol {\mathcal Y}}^{[kab]}_{(123)} \\[5pt]
{\boldsymbol {\mathcal Y}}^{[kab]}_{(312)}
\end{pmatrix}
\circ
\cK^e(\{p_i\} ; \{k_i\})
\circ
\begin{pmatrix}
{\boldsymbol {\mathcal Y}}^{[kab]\dagger}_{(123)}, \
& \ \frac1{\sqrt2} {\boldsymbol {\mathcal Y}}^{[kab]\dagger}_{(312)}
\end{pmatrix}\,,
\end{align}
which gives examples of all the vectors needed to express the full matrix explicitly.
In each case, given the momentum labelings of \Cref{eq:flavorchannels},
the flavor of the final (initial) spectator can be determined by the first entry in the subscript
of $\boldsymbol{\mathcal Y}^{[kab]}_{\boldsymbol \sigma}$
($\boldsymbol{\mathcal Y}^{[kab]\dagger}_{\boldsymbol \sigma}$).
Finally, we note that using $PT$ symmetry, the underlying amplitude
in the $\cK^{b,{\rm T}}$, $\cK^{c,{\rm T}}$ and $\cK^{e,{\rm T}}$ blocks
are related to those of the $\cK^b$, $\cK^c$ and $\cK^e$ blocks, respectively.
For example
\begin{equation}
\cK^{c, \rm T}(\{p_i\} ; \{k_i\}) = \cK^c(\{k_i\} ; \{p_i\} )\,.
\end{equation}

As seen above, we transform to the isospin basis by the following conjugation
\begin{equation}
\widehat{\cK}_{\rm df,3} \to C_{8\to I} C_{7\to8} \widehat{\cK}_{\rm df,3}
C_{7\to 8}^\dagger C_{8\to I}^\dagger\,.
\end{equation}
The factors of $P_e$ and $P_o$ contained in $C_{7\to8}$ can be set to unity because
the projection onto definite isospin channels automatically enforces the necessary symmetry
of the $D$ pairs.
The resulting 8-d matrix of $\widehat{\cK}_{\df,3}$ must be block-diagonal in isospin, and this leads to relations
between the six underlying amplitudes that appear in \Cref{eq:Kdfgeneral}.
We do not display these, as they are not needed to describe the following results.

For the $I=0$ block (the lower-right $2\times 2$ block in the isospin basis) we find the outer product form
given in \Cref{eq:KdfI02,eq:v0def},
in which the single kinematic function that appears, $\cK_{\df,3}^{I=0}(\{p_i\};\{k_i\})$,
is a known combination of $\cK^a$--$\cK^f$.
The symmetry factors are arrayed in the matrix in the same way as
for the $2+1$ systems considered in BS2.
We choose the overall normalization such that $\Mdf^{[I=0]}=\cK_{\df,3}^{[I=0]}$
in the limit of small $\Kdf$ and $\cK_2$.
Here $\Mdf$ is given by the $L\to\infty$ limit of $\cM_{\df,3,L}$, given in \Cref{eq:Mdf3L,eq:Mhatdf3L}.
This can be seen using \Cref{eq:M3LI0,eq:alpI0f}:
\begin{align}
\begin{split}
\Mdf^{[I=0]}(\{p_i\} ; \{k_i\}) & =
\\
& \hspace{-60pt}
\frac{1}{3^2}
\begin{pmatrix}
-\XR123-\XR213 \,, \ & \ \sqrt{2} \boldsymbol{\mathcal X}^{(312)}_{[kab]}
\end{pmatrix}
\circ
\begin{pmatrix}
{-\boldsymbol {\mathcal Y}}^{[kab]}_{(123)} \\[5pt]
\frac1{\sqrt2} {\boldsymbol {\mathcal Y}}^{[kab]}_{(312)}
\end{pmatrix}
\circ
\cK_{\df,3}^{[I=0]}(\{p_i\} ; \{k_i\})
\\[5pt]
& \hspace{-30pt}
\circ
\begin{pmatrix}
{-\boldsymbol {\mathcal Y}}^{[kab]\dagger}_{(123)} \,, \
& \ \frac1{\sqrt2}{\boldsymbol {\mathcal Y}}^{[kab]\dagger}_{(312)}
\end{pmatrix}
\circ
\begin{pmatrix}
-\XL123 - \XL213 \\[5pt]
\sqrt{2} {\boldsymbol {\mathcal X}}^{(312)\dagger}_{[kab]}
\end{pmatrix} + O(\cK_2, \cK_{\df,3}^2)
\,,
\end{split} \\[10pt]
& = \cK_{\df,3}^{[I=0]}(\{p_i\} ; \{k_i\}) + O(\cK_2, \cK_{\df,3}^2) \,. \label{eq:Kdf3I0form}
\end{align}
To obtain the last line we have used the results that,
first, the $\boldsymbol{ \mathcal X}_{[kab]}^{\boldsymbol \sigma}$ operators are the inverses of the
corresponding $\boldsymbol{ \mathcal Y}^{[kab]}_{\boldsymbol \sigma}$s,
and, second, $ \cK_{\df,3}^{I=0}(\{p_i\} ; \{k_i\})$ is symmetric under the separate exchanges
$p_1\leftrightarrow p_2$ and $k_1\leftrightarrow k_2$,
so that, for example, the right-acting $\YR123$ can be replaced by $\YR213$ as needed.

Analogous manipulations apply for the $I=2$ block. In this case $\cK_{\df,3}^{[I=0]}$ is replaced by $\cK_{\df,3}^{[I=2]}$,
which is a different combination of $\cK^a$--$\cK^f$.

For the $I=1$ block, we find the sum of outer products given in \Cref{eq:Kdf3I1form}
which depend on three independent amplitudes
$\cK_{\df,3}^{[I=1],aa}$, $\cK_{\df,3}^{[I=1],ab}$, and $\cK_{\df,3}^{[I=1],bb}$,
which themselves are linear combinations of the $\cK^a-\cK^f$.
The normalization of the vectors appearing in the outer products,
$\boldsymbol {\mathcal Y}^{[I=1],a}$ and $\boldsymbol {\mathcal Y}^{[I=1],b}$,
defined in \Cref{eq:YI1a,eq:YI1b}, respectively,
is chosen such that,
\begin{equation}
\boldsymbol {\mathcal X}_a \circ \boldsymbol {\mathcal Y}^{[I=1],a} \circ \cK_{\df,3}^{[I=1], ax}
= 3 \cK_{\df,3}^{[I=1], ax} \,, \quad
\boldsymbol {\mathcal X}_b \circ \boldsymbol {\mathcal Y}^{[I=1],b} \circ \cK_{\df,3}^{[I=1], bx} = 3 \cK_{\df,3}^{[I=1], bx},
\label{eq:XYnorm}
\end{equation}
and we also have the orthogonality result
\begin{equation}
\boldsymbol {\mathcal X}_a \circ \boldsymbol {\mathcal Y}^{[I=1],b} \circ \cK_{\df,3}^{[I=1],bx}
= \boldsymbol {\mathcal X}_b\circ \boldsymbol {\mathcal Y}^{[I=1],a} \circ \cK_{\df,3}^{[I=1],ax} = 0 \,.
\label{eq:XYorthog}
\end{equation}
Here $\boldsymbol {\mathcal X}_a$ and $\boldsymbol {\mathcal X}_b$ are defined
in \Cref{eq:alphaavec,eq:alphabvec}, respectively,
and we have made use of the symmetry properties of the amplitudes on which these operators act.
In particular elements of $\kdf$ with $I=1,a$ are symmetric under interchange of the first two momentum arguments,
while those with $I=1,b$ are antisymmetric.
The results \Cref{eq:XYnorm,eq:XYorthog} together imply that
\begin{equation}
\cM_{\df,3}^{[I=1]} \to \begin{pmatrix} \cK^{[I=1],aa} & \cK^{[I=1],ab} \\
\cK^{[I=1],ba} & \cK^{[I=1],bb} \end{pmatrix}
\end{equation}
in the weakly-interacting limit.

\newcommand{\szero}{\ \phantom{x}0\phantom{x} \ }
\newcommand{\fourzeros}{\szero & \szero & \szero & \szero}

\section{Derivation using intuitive flavor space extension}
\label{app:flavor_derivation}

In this appendix, we give an alternative derivation that is more directly based on the original RFT publications \cite{Hansen:2014eka,Hansen:2015zga}. The approach follows closely that of ref.~\cite{Hansen:2020zhy}, in which we have derived the quantization condition for generic systems of three pions.

\subsection{Correlator and index space}
\label{app:Bspace}

We begin by repeating the matrix of finite-volume correlation functions, already introduced in \Cref{eq:CLdef} above
\begin{equation*}
\widehat{C}_L(P)_{jk} \equiv \int d x^0 \int_{L^3} d^3 {\boldsymbol x} \, e^{-i \boldsymbol{P} \cdot {\boldsymbol x}+i E t} \langle 0 \vert \mathrm{T}\mathcal{O}_j(x) \mathcal{O}_k^{\dagger}(0) \vert 0 \rangle_L \,.
\end{equation*}
At this stage, a choice arises concerning the dimensionality of the matrix $\widehat{C}_L(P)$.
One can obtain the same quantization condition for various choices, including using just a single creation and annihilation operator. However, it is convenient to choose a set $\mathcal O_j(x)$ that correspond to the possible three-particle intermediate states arising in generic Feynman diagrams contributing to the correlator. This has the advantage that the matrices representing the external operators and those representing internal cuts are all square and can be treated similarly.

Even with this goal in mind, freedom still remains in the number of operators used. One natural choice is to identify all permutations of $D D \pi$ fields having charm $C=+2$ and electric charge $+1$.
This leads to a vector of 12 operators
\begin{equation}
\widetilde {\mathcal O}^{{\sf AP}}(p_1, p_2, p_3) \equiv \begin{pmatrix}
\widetilde D^+(p_1) & \widetilde D^+(p_2) & \widetilde \pi^-(p_3) \\
\widetilde D^+(p_1) & \widetilde D^0(p_2) & \widetilde \pi^0(p_3) \\
\widetilde D^0(p_1) & \widetilde D^+(p_2) & \widetilde \pi^0(p_3) \\
\widetilde D^0(p_1) & \widetilde D^0(p_2) & \widetilde \pi^+(p_3) \\
\widetilde D^+(p_1) & \widetilde \pi^-(p_2) & \widetilde D^+(p_3) \\
\widetilde \pi^-(p_1) & \widetilde D^+(p_2) & \widetilde D^+(p_3) \\
\widetilde D^0(p_1) & \widetilde \pi^0(p_2) & \widetilde D^+(p_3) \\
\widetilde \pi^0(p_1) & \widetilde D^0(p_2) & \widetilde D^+(p_3) \\
\widetilde D^0(p_1) & \widetilde \pi^+(p_2) & \widetilde D^0(p_3) \\
\widetilde \pi^+(p_1) & \widetilde D^0(p_2) & \widetilde D^0(p_3) \\
\widetilde D^+(p_1) & \widetilde \pi^0(p_2) & \widetilde D^0(p_3) \\
\widetilde \pi^0(p_1) & \widetilde D^+(p_2) & \widetilde D^0(p_3)
\end{pmatrix} \,.
\label{eq:ops_AP}
\end{equation}
The relation to $\mathcal O_j(x)$ then follows from integrating the momentum-space operators with a smooth function
\begin{equation}
\mathcal O_j(x) = \int \! \frac{d^4 p_1}{(2 \pi)^4} \, \int \! \frac{d^4 p_2}{(2 \pi)^4} \, \int \! \frac{d^4 p_3}{(2 \pi)^4} \, f(p_1, p_2, p_3, x) \, \widetilde {\mathcal O}^{{\sf AP}}(p_1, p_2, p_3) \,. \label{eq:Oj}
\end{equation}
Here we have introduced the superscript $\sf AP$, which stands for ``all permutations''. This is to contrast with an alternative flavor space, to which we now turn.

As we have argued in the main text, in particular in the paragraph following \Cref{eq:QC3}, one can also define the quantization condition on an eight-dimensional space. To motivate this, first note that, if we consider states related by a cyclic momentum permutation to be equivalent, then four distinct states arise with charm $C = +2$ and electric charge $+1$: $D^0 D^0 \pi^+$, $D^+ D^+ \pi^-$, $D^+ D^0 \pi^0$, $D^0 D^+ \pi^0$. These directly correspond to the four isospin states with vanishing third-component of isospin, $m_I = 0$. In the isospin basis, the corresponding operators are
\begin{align}
\begin{split}
\mathcal O_{([D D]_1 \pi)_{2}}(p_1, p_2, p_3) &=
\widetilde D^0 \widetilde D^0 \widetilde \pi^+
+
\widetilde D^+ \widetilde D^+ \widetilde \pi^-
+
\sqrt{2} \widetilde D^+ \widetilde D^0 \widetilde \pi^0 + \sqrt{2} \widetilde D^0 \widetilde D^+ \widetilde \pi^0, \\[5pt]
\mathcal O_{([D D]_1 \pi)_{1,a}}(p_1, p_2, p_3) &= \widetilde D^+ \widetilde D^+ \widetilde \pi^- - \widetilde D^0 \widetilde D^0 \widetilde \pi^+ ,\\[5pt]
\mathcal O_{([D D]_0 \pi)_{1,b}}(p_1, p_2, p_3) &= \widetilde D^+ \widetilde D^0 \widetilde \pi^0 - \widetilde D^0 \widetilde D^+ \widetilde \pi^0 ,\\[5pt]
\mathcal O_{([D D]_1 \pi)_{0}}(p_1, p_2, p_3) &= \sqrt{2} \widetilde D^0 \widetilde D^0 \widetilde \pi^+ + \sqrt{2} \widetilde D^+ \widetilde D^+ \widetilde \pi^-
- \widetilde D^+ \widetilde D^0 \widetilde \pi^0 - \widetilde D^0 \widetilde D^+ \widetilde \pi^0 \,,
\end{split}
\label{eq:flavorops}
\end{align}
where the momentum assignments are implicit on the right-hand side and are understood to be ordered $p_1,p_2,p_3$. These operators correspond to the states defined in \Cref{eq:ddpi_isospin_states}, here with the normalization dropped.

Thus, if the quantization condition space is to include two-particle isospin as a label in the definition of the index space, the size must be a multiple of four.
However, a four-dimensional space is insufficient to describe diagrams such as that shown in \Cref{fig:Gdiag}. When assigning momenta to this diagram, a definite choice (say $p_3$) must be given to the spectator particle at the lower right. Since this particle can be either a pion or a $D$ meson, one requires a flavor space in which both can carry the $p_3$ value. It follows that the minimal dimension of the space is eight: the number of distinct isospin multiplets that can be spectators (in this case two) times the number of distinct isospin states
(in this case four).

For concreteness,
we drop the $6^{\text{th}}$, $8^{\text{th}}$, $10^{\text{th}}$, and $12^{\text{th}}$ operators in \Cref{eq:ops_AP} to define a new eight-dimensional set:
\begin{equation}
\widetilde {\mathcal O}^{{\sf MP}}(p_1, p_2, p_3) \equiv \begin{pmatrix}
\widetilde D^+(p_1) & \widetilde D^+(p_2) & \widetilde \pi^-(p_3) \\
\widetilde D^+(p_1) & \widetilde D^0(p_2) & \widetilde \pi^0(p_3) \\
\widetilde D^0(p_1) & \widetilde D^+(p_2) & \widetilde \pi^0(p_3) \\
\widetilde D^0(p_1) & \widetilde D^0(p_2) & \widetilde \pi^+(p_3) \\
\widetilde D^+(p_1) & \widetilde \pi^-(p_2) & \widetilde D^+(p_3) \\
\widetilde D^0(p_1) & \widetilde \pi^0(p_2) & \widetilde D^+(p_3) \\
\widetilde D^0(p_1) & \widetilde \pi^+(p_2) & \widetilde D^0(p_3) \\
\widetilde D^+(p_1) & \widetilde \pi^0(p_2) & \widetilde D^0(p_3)
\end{pmatrix} \,,
\label{eq:orderMP}
\end{equation}
where we have introduced the superscript ${\sf MP}$, which stands for ``minimal permutations''.
This is related to the eight-dimensional space used in \Cref{app:topt_derivation}
(see \Cref{app:7to8}) in a way that is straightforward to determine. We hold off presenting
the relation, however, until we convert to the isospin basis below.

Although it may seem like overkill
to use different bases in the two appendices, it is well motivated by the different origins of the two constructions. The space of \Cref{app:topt_derivation} originates from the quantization conditions of single flavor channels, with a given number of degenerate or non-degenerate particles. For this reason, the three permutations of $D^+ \pi^0 D^0$, for example, are naturally grouped together. By contrast, the derivation of this section is dedicated to identifying the finite-volume cuts for each individual flavor channel. These are then naturally grouped by which of the three particles is the spectator.

The eight-dimensional {\sf MP} basis is sufficient to represent all cuts in all diagrams in the skeleton expansion of the finite-volume correlator. Thus, the final quantization condition can be derived on this space. If one alternatively derives the result on the twelve-dimensional space, then the resulting matrices are rank deficient and one can use a change of basis to identify a four-dimensional null space. In this way the resulting quantization conditions are exactly equivalent.

In both cases, the basic algebra from the derivation for identical spin-zero particles still holds~\cite{\HSQCa,\HSQCb}, and the three-particle correlation function can be written as
\begin{equation}
\widehat{C}_L(P)=\widehat{C}_{\infty}(P)+i\widehat{A}_3^{\prime} \widehat{F}_3 \frac{1}{1+\widehat{\cK}_{\text{df}, 3} \widehat{F}_3} \widehat{A}_3, \label{eq:FVcorr}
\end{equation}
where
\begin{equation}
\widehat{F}_3 \equiv \frac{\widehat{F}}{3}+\widehat{F} \frac{1}{1-\widehat{\cM}_{2, L} \widehat{G}} \widehat{\cM}_{2, L} \widehat{F}, \qquad \qquad \widehat{\cM}_{2, L} \equiv \frac{1}{\widehat{\cK}_2^{-1}-\widehat{F}}. \label{eq:F3full}
\end{equation}
Note that $\widehat{F},\widehat{G}$ and $\widehat{\cK}_2$ are not the same as $\widehat{F}^{[I]},\widehat{G}^{[I]}$ and $\widehat{\cK}^{[I]}_2$ in the main text, but can be related to those by performing a change of basis and then identifying definite-isospin sub-blocks.

\subsection{Quantization condition building blocks}
\label{app:Bblocks}

In this section we describe the building blocks in both the twelve- and eight-dimensional spaces in detail. We begin with the twelve-dimensional $\widehat{F}$, arising from all permutations of the fields,
\begin{equation}
\widehat{F}^{{\sf AP}} = \frac{1}{2}\text{diag }\left( \widetilde F^\pi, \widetilde F^\pi, \widetilde F^\pi, \widetilde F^\pi, \widetilde F^D , \widetilde F^{D^\prime}, \widetilde F^D , \widetilde F^{D^\prime} , \widetilde F^D, \widetilde F^{D^\prime}, \widetilde F^D, \widetilde F^{D^\prime} \right) \,.
\label{eq:F_AP_def}
\end{equation}
$\wt F^\pi$ and $\wt F^D$ are defined in \Cref{eq:Ft}.
The distinction between $\widetilde F^{D}$ and $\widetilde F^{D^\prime}$ arises in which of the two scattering particles is used to define the angular momentum. In particular, the angular momentum is defined with respect to the $D$ meson in $\widetilde F^{D}$ and with respect to the pion in $\widetilde F^{D^\prime}$. In both cases, the construction begins by boosting to the CM frame of the non-spectator ($\pi, D$) system. One then assigns the momentum $\boldsymbol a^*$ to the $D$ meson and $- \boldsymbol a^*$ to the $\pi$ for $F^D$ while swapping the momentum assignments for $\widetilde F^{D^\prime}$. The final quantity is then reached by projecting the directional freedom within $\boldsymbol a^*$ to definite angular momentum. The relation between the two momentum assignments leads to relative signs in odd angular-momentum components, given explicitly by
\begin{equation}
\widetilde F^{D^\prime} = P^{(\ell)} \cdot \widetilde F^D \cdot P^{(\ell)} \,,
\end{equation}
where $P^{(\ell)}$ is defined in \Cref{eq:Plm}.

The factor of $(1/2)$ in \Cref{eq:F_AP_def} is a symmetry factor arising for identical particles. A feature of working on the ${\sf AP}$ space, as already discussed in ref.~\cite{Hansen:2020zhy}, is that this factor is included across all channels, irrespective of whether the two scattering particles within a given $\widetilde F$ are indistinguishable. This works out because two-particle loops with distinct particles are double counted by the inclusion of all permutations, and this double counting compensates the $(1/2)$ to give the correct overall factor for the loop.

More precisely, the choice of factors is a mixture of conventions and requirements following from the expressions defining the underlying Feynman diagrams. For example, if we consider $\widehat C_L(P)$ in the ${\sf AP}$ basis and expand \Cref{eq:FVcorr} as a power series in the K-matrices, then the leading-order (K-matrix-independent) term can be written as
\begin{equation}
\widehat C_L(P) = \widehat C_\infty(P) + \frac13 i \widehat A'^{\sf AP}_3 \cdot \widehat F^{\sf AP} \cdot \widehat A^{\sf AP}_3 + \cdots \,. \label{eq:CL_leading_AP}
\end{equation}
This can always be made correct with a judicious choice of the factors appearing within the endcap functions, $\widehat A'^{\sf AP}_3$ and $\widehat A^{\sf AP}_3$.

The next choice to be made in this construction is that for $\widehat{\cK}_2$, the two-particle K matrix. From the choice made for $\widehat F^{\sf AP}$, it is fixed as follows:
\begin{equation}
\widehat{\cK}^{{\sf AP}}_2
=
\text{diag}\left [
\widetilde{\mathcal K}^{{\sf AP}}_{2,\,DD,\,++}
\,,
\widetilde{\mathcal K}^{{\sf AP}}_{2,\,DD,\,+}
\,,
\widetilde{\mathcal K}^{{\sf AP}}_{2,\,DD,\,0}
\,,
\widetilde{\mathcal K}^{{\sf AP}}_{2,\,D \pi,\,0}
\,,
\widetilde{\mathcal K}^{{\sf AP}}_{2,\,D \pi,\,+}
\right ]\,.
\end{equation}
Here each of the entries is a square matrix, encoding two-to-two scattering in the subspace indicate by the labels: $(DD,\,++)$, $(DD,\,+)$, $(DD,\,0)$, $(D\pi,\,0)$, and $(D\pi,\,+)$.
Here the first entry indicates the flavors of the two scattering particles,
while the second indicates the total charge.
For example
\begin{align}
\widetilde{\mathcal K}^{{\sf AP}}_{2,\,DD,\,++} & = {\cK}_{2,L}^{DD, I=1} \,,
\end{align}
where ${\cK}_{2,L}^{DD, I=1}$ is defined in \Cref{eq:K2Li} of the main text.
We emphasize that, unlike for $\widehat F^{\sf AP}$, no explicit symmetry factors are
needed in $\widehat {\mathcal K}_2^{\sf AP}$.
Thus, the relation between the K-matrix and the scattering phase shift differs by a factor of two, depending on whether identical or non-identical particles are scattering, as shown in \Cref{eq:K2}.

Rather than explicitly give expressions for all entries of all blocks, we indicate the definitions using compact notation
\begin{align}
\widetilde{\mathcal K}^{{\sf AP}}_{2,\,DD,\,++}
& =
\ \begin{bmatrix} D^+D^+ \end{bmatrix}
\ \ \leftarrow \ \
\begin{bmatrix} D^+D^+ \end{bmatrix} \,,
\\[10pt]
\widetilde{\mathcal K}^{{\sf AP}}_{2,\,DD,\,+}
& =
\ \begin{bmatrix} D^+ D^0 \\ D^0 D^+ \end{bmatrix}
\ \ \leftarrow \ \
\begin{bmatrix} D^+ D^0 \,, \ & \ D^0 D^+ \end{bmatrix} \,,
\\[10pt]
\widetilde{\mathcal K}^{{\sf AP}}_{2,\,DD,\,0}
& =
\ \begin{bmatrix} D^0D^0 \end{bmatrix}
\ \ \leftarrow \ \
\begin{bmatrix} D^0D^0 \end{bmatrix} \,,
\\[10pt]
\widetilde{\mathcal K}^{{\sf AP}}_{2,\,D \pi,\,0}
& =
\ \begin{bmatrix} D^+ \pi^- \\ \pi^- D^+ \\ D^0 \pi^0 \\ \pi^0 D^0 \end{bmatrix}
\ \ \leftarrow \ \
\begin{bmatrix} D^+ \pi^- \,, \ & \ \pi^- D^+ \,, \ & \ D^0 \pi^0 \,, \ & \ \pi^0 D^0 \end{bmatrix} \,,
\\[10pt]
\widetilde{\mathcal K}^{{\sf AP}}_{2,\,D \pi,\,+}
& =
\ \begin{bmatrix} D^0\pi^+ \\ \pi^+ D^0 \\ D^+\pi^0 \\ \pi^0 D^+ \end{bmatrix}
\ \ \leftarrow \ \
\begin{bmatrix} D^0 \pi^+ \,, \ & \ \pi^+ D^0 \,, \ & \ D^+ \pi^0 \,, \ & \ \pi^0 D^+ \end{bmatrix} \,.
\end{align}
The $\leftarrow$ indicates that the two particles to the right are defining the incoming state while
those to the left define the outgoing state.
The column vector and the row vector are combined into square matrices by taking the outer product. For example, the $D^+D^0$ channel is given by
\begin{align}
\widetilde{\mathcal K}_{2,\,DD,\,+} & =
\begin{pmatrix}
\mathcal K_2(D^+D^0 \leftarrow D^+D^0) \ \ & \ \ \mathcal K_2(D^+D^0 \leftarrow D^0D^+) \\[10pt]
\mathcal K_2(D^0D^+ \leftarrow D^+D^0) \ \ & \ \ \mathcal K_2(D^0D^+ \leftarrow D^0D^+)
\end{pmatrix} \,.
\end{align}

This definition of $\widehat{\cK}^{{\sf AP}}_2$ then leads to the correct counting of factors in higher orders of $\widehat C_L(P)$ when expanded in powers of the K-matrices. For example, the expansion to linear order gives
\begin{equation}
\widehat C_L(P) = \widehat C_\infty(P) + \frac13 i \widehat A'^{\sf AP}_3 \cdot \widehat F^{\sf AP} \cdot \widehat A^{\sf AP}_3 + i \widehat A'^{\sf AP}_3 \cdot \widehat F^{\sf AP} \cdot \widehat{\cK}^{{\sf AP}}_2 \cdot \widehat F^{\sf AP} \cdot \widehat A^{\sf AP}_3 + \cdots \,.
\end{equation}
In the final term on the right-hand side,
the contribution from entries containing $D^0 D^+$ scattering,
such as $\mathcal K_2(D^0D^+ \leftarrow D^0D^+)$, appears four times,
canceling the two factors of $(1/2)$ arising from the two entries of $F^{\sf AP}$.
Thus the total contribution is included with the correct overall factor.

Having established these choices we now come to the analogous construction in the eight-dimensional ({\sf MP}) space, \Cref{eq:orderMP}. Here we choose to define $\widehat{F}^{{\sf MP}}$ with no symmetry factors,
\begin{equation}
\widehat{F}^{{\sf MP}} = \text{diag} \! \left( \widetilde F^\pi, \widetilde F^\pi, \widetilde F^\pi, \widetilde F^\pi, \widetilde F^{D}, \widetilde F^{D} , \widetilde F^{D}, \widetilde F^{D} \right)\,.
\end{equation}
As with the ${\sf AP}$ basis, any choice for $\widehat{F}^{{\sf MP}}$ is viable, provided corresponding choices are made for $\widehat A'^{\sf MP}_3$, $\widehat A^{\sf MP}_3$, and $\widehat{\cK}^{\sf MP}_2$.

The required factors are determined by again considering the K-matrix-independent and linear terms in $\widehat C_L(P)$
\begin{equation}
\widehat C_L(P) = \widehat C_\infty(P) + \frac13 i \widehat A'^{\sf MP}_3 \cdot \widehat F^{\sf MP} \cdot \widehat A^{\sf MP}_3 + i \widehat A'^{\sf MP}_3 \cdot \widehat F^{\sf MP} \cdot \widehat{\cK}^{{\sf MP}}_2 \cdot \widehat F^{\sf MP} \cdot \widehat A^{\sf MP}_3 + \cdots \,.
\end{equation}
For example, the first four entries match between the ${\sf AP}$ and ${\sf MP}$ spaces and the remaining components cannot compensate these. Thus, the rescaling of $\widehat F^{\sf MP}$ must be matched with a rescaling of the adjacent factors:
$\widehat A_{3,ij}^{\sf MP} = \widehat A_{3,ij}^{\sf AP}/\sqrt{2}$ for $1 \leq i,j \leq 4$. More important for us, however, is the K-matrix, which must satisfy
\begin{equation}
\widehat{\cK}^{{\sf MP}}_2
=
\text{diag}\left [
\widetilde{\mathcal K}^{{\sf MP}}_{2,\,DD,\,++}
\,,
\widetilde{\mathcal K}^{{\sf MP}}_{2,\,DD,\,+}
\,,
\widetilde{\mathcal K}^{{\sf MP}}_{2,\,DD,\,0}
\,,
\widetilde{\mathcal K}^{{\sf MP}}_{2,\,D \pi,\,0}
\,,
\widetilde{\mathcal K}^{{\sf MP}}_{2,\,D \pi,\,+}
\right ]\,,
\end{equation}
where
\begin{align}
\widetilde{\mathcal K}^{{\sf MP}}_{2,\,DD,\,++}
& = \frac12
\widetilde{\mathcal K}^{{\sf AP}}_{2,\,DD,\,++} \,,
\\[5pt]
\widetilde{\mathcal K}^{{\sf MP}}_{2,\,DD,\,+}
& = \frac12
\widetilde{\mathcal K}^{{\sf AP}}_{2,\,DD,\,+} \,,
\\[5pt]
\widetilde{\mathcal K}^{{\sf MP}}_{2,\,DD,\,0}
& = \frac12
\widetilde{\mathcal K}^{{\sf AP}}_{2,\,DD,\,0} \,,
\\[5pt]
\widetilde{\mathcal K}^{{\sf MP}}_{2,\,D \pi,\,0}
& =
\ \begin{bmatrix} D^+ \pi^- \\ D^0 \pi^0 \end{bmatrix}
\ \ \leftarrow \ \
\begin{bmatrix} D^+ \pi^- \,, \ & \ D^0 \pi^0 \end{bmatrix} \,,\\[10pt]
\widetilde{\mathcal K}^{{\sf MP}}_{2,\,D \pi,\,+}
& =
\ \begin{bmatrix} D^0 \pi^+ \\ D^+ \pi^0 \end{bmatrix}
\ \ \leftarrow \ \
\begin{bmatrix} D^0 \pi^+ \,, \ & \ D^+ \pi^0 \end{bmatrix} \,.
\end{align}
This then ensures that the series of insertions of $\widehat{\cK}^{{\sf MP}}_2$ and $\widehat F^{{\sf MP}}$ all come with the correct symmetry factors.

Our results so far can be summarized by noting that the twelve- and eight-dimensional spaces are related by changes of basis and rescaling. In particular, we have
\begin{align}
\widehat{F}^{\overline{\sf MP}} & = R \cdot S \cdot \widehat{F}^{\sf AP} \cdot S^{\rm T} \cdot R \,, \label{eq:F_change_of_basis} \\
\widehat{\mathcal K}^{\overline{\sf MP}}_2 & = R^{-1} \cdot S \cdot
\widehat{\mathcal K}^{\sf AP}_2 \cdot S^{\rm T} \cdot R^{-1} \,,
\\
\widehat{A}^{\overline{\sf MP}} & = R^{-1} \cdot S \cdot \widehat{A}^{\sf AP} \,,
\\
\widehat{A}'^{\overline{\sf MP}} & = \widehat{A}'^{\sf AP} \cdot S^{\rm T} \cdot R^{-1} \,,
\end{align}
where $R$ is a rescaling to address the factors of $2$ and $S$ is a rotation in the flavor space.
The explicit definitions are given by $R = \sqrt{2} \, \, {\mathbb I}_{12 \times 12}$, where the latter factor simply denotes the twelve-dimensional identity, and
\begin{equation}
S = \begin{pmatrix}
\ 1 \ & \fourzeros & \fourzeros & \szero & \szero & \szero \\
\szero & \ 1 \ & \fourzeros & \fourzeros & \szero & \szero \\
\szero & \szero & \ 1 \ & \fourzeros & \fourzeros & \szero \\
\szero & \szero & \szero & \ 1 \ & \fourzeros & \fourzeros \\
\fourzeros & \frac{1}{\sqrt{2}} & \frac{P^{(\ell)}}{\sqrt{2}} & \fourzeros & \szero & \szero \\
\fourzeros & \szero & \szero & \frac{1}{\sqrt{2}} & \frac{P^{(\ell)}}{\sqrt{2}} & \fourzeros \\
\fourzeros & \fourzeros & \frac{1}{\sqrt{2}} & \frac{P^{(\ell)}}{\sqrt{2}} & \szero & \szero \\
\fourzeros & \fourzeros & \szero & \szero & \frac{1}{\sqrt{2}} & \frac{P^{(\ell)}}{\sqrt{2}} \\
\fourzeros & \frac{1}{\sqrt{2}} & -\frac{P^{(\ell)}}{\sqrt{2}} & \fourzeros & \szero & \szero \\
\fourzeros & \szero & \szero & \frac{1}{\sqrt{2}} & - \frac{P^{(\ell)}}{\sqrt{2}} & \fourzeros \\
\fourzeros & \fourzeros & \frac{1}{\sqrt{2}} & - \frac{P^{(\ell)}}{\sqrt{2}} & \szero & \szero \\
\fourzeros & \fourzeros & \szero & \szero & \frac{1}{\sqrt{2}} & - \frac{P^{(\ell)}}{\sqrt{2}} \\
\end{pmatrix}\,.
\end{equation}

We stress that these are square matrices, so that the result of the change of basis and the rescaling is also twelve-dimensional. We have represented this with the labels ${\overline {\sf MP}}$. As we demonstrate below, for both the K-matrix and the endcap factors, the last four rows and columns of the ${\overline {\sf MP}}$ quantities are zero. This means that they do not contribute to $\widehat C_L(P)$, so the quantization condition can be expressed on the eight-dimensional space.

The final step of this section is to present $\widehat G$.
The ${\sf AP}$-space matrix is given by
\begin{equation}
\widehat{G}^{\sf AP} = \begin{pmatrix}
0 & 0 & 0 & 0 & 0 & \widetilde{G}^{\pi D} & 0 & 0 & 0 & 0 & 0 & 0 \\
0 & 0 & 0 & 0 & 0 & 0 & 0 & \widetilde{G}^{\pi D} & 0 & 0 & 0 & 0 \\
0 & 0 & 0 & 0 & 0 & 0 & 0 & 0 & 0 & 0 & 0 & \widetilde{G}^{\pi D} \\
0 & 0 & 0 & 0 & 0 & 0 & 0 & 0 & 0 & \widetilde{G}^{\pi D} & 0 & 0 \\
0 & 0 & 0 & 0 & \widetilde{G}^{DD } & 0 & 0 & 0 & 0 & 0 & 0 & 0 \\
\widetilde{G}^{D\pi} & 0 & 0 & 0 & 0 & 0 & 0 & 0 & 0 & 0 & 0 & 0 \\
0 & 0 & 0 & 0 & 0 & 0 & 0 & 0 & 0 & 0 & \widetilde{G}^{DD } & 0 \\
0 & \widetilde{G}^{D\pi } & 0 & 0 & 0 & 0 & 0 & 0 & 0 & 0 & 0 & 0 \\
0 & 0 & 0 & 0 & 0 & 0 & 0 & 0 & \widetilde{G}^{DD } & 0 & 0 & 0 \\
0 & 0 & 0 & \widetilde{G}^{D\pi } & 0 & 0 & 0 & 0 & 0 & 0 & 0 & 0 \\
0 & 0 & 0 & 0 & 0 & 0 & \widetilde{G}^{DD } & 0 & 0 & 0 & 0 & 0 \\
0 & 0 & \widetilde{G}^{D\pi } & 0 & 0 & 0 & 0 & 0 & 0 & 0 & 0 & 0 \\
\end{pmatrix} \,,
\end{equation}
and the ${\sf MP}$ counterpart is
\begin{equation}
\widehat{G}^{\sf MP} =
\begin{pmatrix}
0 & 0 & 0 & 0 & \!\!\!\! \sqrt{2}\widetilde G^{\pi D} P^{(\ell)} \!\!\!\! & 0 & 0 & 0\\
0 & 0 & 0 & 0 & 0 & \!\!\!\! \sqrt{2} \widetilde G^{\pi D} P^{(\ell)} \!\!\!\! & 0 & 0\\
0 & 0 & 0 & 0 & 0 & 0 & 0 & \!\!\!\! \sqrt{2} \widetilde G^{\pi D} P^{(\ell)} \!\!\! \\
0 & 0 & 0 & 0 & 0 & 0 & \!\!\!\! \sqrt{2} \widetilde G^{\pi D} P^{(\ell)} \!\!\!\! & 0\\
\!\! \sqrt{2} P^{(\ell)} \widetilde G^{D\pi} \!\!\!\! & 0 & 0 & 0 & \widetilde G^{DD} & 0 & 0 & 0\\
0 & \!\!\!\! \sqrt{2} P^{(\ell)} \widetilde G^{D\pi} \!\!\!\! & 0 & 0 & 0 & 0 & 0 & \widetilde G^{DD}\\
0 & 0 & 0 & \!\!\!\! \sqrt{2} P^{(\ell)} \widetilde G^{D\pi} \!\!\!\! & 0 & 0 & \widetilde G^{DD} & 0\\
0 & 0 & \!\!\!\! \sqrt{2} P^{(\ell)} \widetilde G^{D\pi} \!\!\!\! & 0 & 0 & \widetilde G^{DD} & 0 & 0
\end{pmatrix} \,.
\label{eq:G_MP}
\end{equation}
The relation between the two is exactly as for $\widehat F$ in \Cref{eq:F_change_of_basis} above
\begin{equation}
\widehat{G}^{\overline{\sf MP}} = R \cdot S \cdot \widehat{G}^{\sf AP} \cdot S^{\rm T} \cdot R \,.
\end{equation}
As with the other matrices, the ${\overline{\sf MP}}$ superscript indicates that this is a twelve-dimensional matrix for which the upper left eight-dimensional block is given by the ${{\sf MP}}$ matrix. The last four rows and rightmost four columns are irrelevant, since these are annihilated by the zeros in the K-matrices and the endcaps.

\subsection{Result for \texorpdfstring{$\Kdf = 0$}{vanishing K-matrix}}
\label{app:BfinalQC3}

To understand why the quantization condition is valid with both forms, we return to \Cref{eq:FVcorr} and \Cref{eq:F3full}, first setting $\widehat{\cK}_{\text{df}, 3} = 0$. This yields
\begin{align}
\label{eq:CL_Kdf0}
\widehat{C}_L(P) - \widehat{C}_{\infty}(P) & = i\widehat{A}_3^{\prime} \widehat F_3 \widehat A_3
= i\widehat{A}_3^{\prime} \, \bigg [ \, \frac{\widehat{F}}{3}+\widehat{F} \widehat{\cK}_2 \frac{1}{1 - [ \widehat{F} + \widehat{G} ] \widehat{\cK}_2 } \widehat{F} \, \bigg ] \, \widehat{A}_3 \,,
\end{align}
where we have combined the two equations and rearranged slightly.
We observe that $\widehat{C}_L(P)$ is invariant under the transformations
\begin{align}
\widehat F & \ \ \ \longrightarrow \ \ \ R \cdot S \cdot \widehat F \cdot S^{\rm T} \cdot R \,, \\
\widehat G & \ \ \ \longrightarrow \ \ \ R \cdot S \cdot \widehat G \cdot S^{\rm T} \cdot R \,, \\
\widehat{\mathcal K}_2 & \ \ \ \longrightarrow \ \ \ R^{-1} \cdot S \cdot \widehat{\mathcal K}_2 \cdot S^{\rm T} \cdot R^{-1} \,, \\
\widehat{A}_3^{\prime} & \ \ \ \longrightarrow \ \ \ \widehat{A}_3^{\prime} \cdot S^{\rm T} \cdot R^{-1} \,, \\
\widehat{A}_3 & \ \ \ \longrightarrow \ \ \ R^{-1} \cdot S \cdot \widehat{A}_3 \,.
\label{eq:RStrafos}
\end{align}
For this reason the quantization condition holds with both forms of the matrices.

As mentioned above, $\widehat A^{\overline{\sf MP}}_3$, $\widehat A'^{\overline{{\sf MP}}}_3$, and $\widehat {\mathcal K}^{\overline{\sf MP}}_2$ vanish in their last four entries so that all objects appearing in the second term of \Cref{eq:CL_Kdf0} can be truncated to the eight-dimensional ${\sf MP}$ basis. To see why this holds for $\widehat A^{\overline{\sf MP}}_3$ and $\widehat A'^{\overline{{\sf MP}}}_3$, we recall from ref.~\cite{Hansen:2021ofl}
(see, in particular, appendix A of that work) that the endcaps are related to matrix elements of the operators in \Cref{eq:CLdef} with particular three-particle states. To make this explicit we
consider the particular matrix elements
\begin{align}
M^{\sf AP}_{\{1,11\}}(p_1,p_2,p_3) &= \braket{ 0 | \mathcal O_1(0)| D^+(p_1) \pi^0(p_2) D^0(p_3), \text{in}} \,,
\\
M^{\sf AP}_{\{1,12\}}(p_1,p_2,p_3) &= \braket{ 0 | \mathcal O_1(0)| \pi^0(p_1) D^+(p_2) D^0(p_3), \text{in}} \,,
\end{align}
where the indices on the right side ($\{1,11\}$ and $\{1,12\}$) follow from the fact that the right-hand side is defined with the 1st operator and either the 11th or the 12th state of the ${\sf AP}$ basis.

The key relevant observation is that $M^{\sf AP}_{\{1,11\}}(p_1,p_2,p_3) = M^{\sf AP}_{\{1,12\}}(p_2,p_1,p_3)$. This exchange property is preserved in the conversion to the endcaps and, after angular momentum projection, leads to the relation
\begin{equation}
\widehat A'^{\sf AP}_{3,\{1,11\}} = \widehat A'^{\sf AP}_{3,\{1,12\}} \cdot P^{(\ell)} \,.
\end{equation}
These types of relations hold for all entries related by similar exchanges. Combining this with the definition of $S$, one finds that the last four rows of $\widehat A'^{\overline{\sf MP}}_{3}$ (and similarly the rightmost four columns of $\widehat A^{\overline{\sf MP}}_{3}$) vanish, as claimed. In addition, as this is only a property of the exchange symmetry of the first two particles in a given state, it holds in the same way for $\widehat{\mathcal K}_2$ (and for $\widehat{\mathcal K}_{\text{df},3}$, as we will describe in more detail in the next section).

The final step is to rotate from the basis of individual particle flavors to that of definite two- and three-particle isospins. In particular, we require a change of basis from the eight-dimensional ${\sf MP}$ space to the total isospin basis with the ordering established by \Cref{eq:totalisospinorder} of the main text.
This is achieved by the following rotation matrix
\begin{equation}
C_{{\sf MP} \to I}= \frac1{\sqrt6}
\begin{pmatrix}
0 & 0 & 0 & 0 & 1& \sqrt2 & 1 & \sqrt2
\\
1 & \sqrt2 & \sqrt2 &1 & 0 & 0 & 0 & 0
\\
0 & 0 & 0 & 0 & -1& -\sqrt2 & 1 & \sqrt2
\\
0 & 0 & 0 & 0 & \sqrt2& -1& -\sqrt2 & 1
\\
\sqrt3 &0 & 0 & -\sqrt3 & 0 & 0 & 0 & 0
\\
0 & \sqrt3 & -\sqrt3 & 0 & 0 & 0 & 0 & 0
\\
0 & 0 & 0 & 0 & -\sqrt2 & 1& -\sqrt2 & 1
\\
\sqrt2 & -1 & -1 & \sqrt2 & 0 & 0 & 0 & 0
\end{pmatrix}\,.
\label{eq:C8toI_v2}
\end{equation}
Using this, one can obtain isospin-basis versions of all building blocks, including the endcap factors. These transformations block diagonalize all entries and the resulting blocks are those given in the main text. For example
\begin{equation}
C_{{\sf MP} \to I} \cdot \widehat F^{\sf MP} \cdot C_{{\sf MP} \to I}^{\text{T}} = \text{diag}\left [ \widehat F^{[I=2]}, \,\widehat F^{[I=1]}, \,\widehat F^{[I=0]} \right ] \,,
\end{equation}
where $\widehat F^{[I=2]}$ and $\widehat F^{[I=0]}$ are two-dimensional [defined in \Cref{eq:FhatI20}] and $\widehat F^{[I=0]}$ is four-dimensional [defined in \Cref{eq:FhatI1}]. This same pattern holds for $\widehat {\mathcal K}_2$ and $\widehat G$ and thus also for $\widehat F_3$. This leads to the correlator itself block diagonalizing into the three isospin sectors. To express this, we note that so far we have only performed rotations that leave $\widehat{C}_L(P)$ unchanged. To diagonalize this quantity as well we need to apply the change of basis, first from ${\sf AP}$ to $\overline {\sf MP}$, then the reduction to the eight-dimensional ${\sf MP}$ and finally the rotation to definite isospin. We reach
\begin{multline}
-i \textbf{C} \cdot \left [ \widehat{C}_L(P) - \widehat{C}_{\infty}(P) \right ] \cdot
\textbf{C}^{\text{T}}
= \\[5pt]
\text{diag} \left [ \widehat{A}'^{[I=2]}_3 \cdot \widehat{F}^{[I=2]}_3 \cdot \widehat{A}^{[I=2]}_3, \ \ \widehat{A}'^{[I=1]}_3 \cdot \widehat{F}^{[I=1]}_3 \cdot \widehat{A}^{[I=1]}_3,\ \ \widehat{A}'^{[I=0]}_3 \cdot \widehat{F}^{[I=0]}_3 \cdot \widehat{A}^{[I=0]}_3 \right ] \,,
\end{multline}
where we have introduced
\begin{equation}
\textbf{C} = C_{{\sf MP} \to I} \cdot \mathbb{I}_{12 \to 8} \cdot R^{-1} \cdot S \,.
\end{equation}
Here $\mathbb{I}_{12 \to 8}$ is a $8 \times 12$ matrix populated with an $8 \times 8$ identity and zeros to discard the null rows.

Finally we note that, since the isospin basis is the same here and in \Cref{app:topt_derivation},
we can use the combination $C_{{\sf MP} \to I} C_{8 \to I}^{-1}$ to convert between the {\sf MP} basis
and the eight-dimensional basis used in \Cref{app:topt_derivation}.

\subsection{Result for \texorpdfstring{${\cK}_{\text{df}, 3} \neq 0$}{non-zero K-matrix}}
\label{app:Bkdf}

The generalization of the previous discussion to include nonzero $\widehat{\cK}_{\text{df}, 3}$
requires understanding a few important properties of this quantity.
We start in the ${\sf AP}$ basis, in which $\widehat{\cK}_{\text{df}, 3}$ %
is a twelve-dimensional matrix.
A key feature is that subsets of its entries correspond to the same underlying infinite-volume function,
differing only by the assignments of momenta to the $k \ell m$ variables.
An example is
\begin{align}
\begin{split}
[\widehat \cK^{\sf AP}_{\rm df, 3}]_{1,1} &= {\boldsymbol {\mathcal Y}}^{[kab]}_{(312)} \circ \cK_{\text{df}, 3}( D^+(p'_1) D^+(p'_2) \pi^-(p'_3) \leftarrow D^+(p_1) D^+(p_2) \pi^-(p_3)) \circ {\boldsymbol {\mathcal Y}}^{[kab]\dagger}_{(312)}, \\
[\widehat \cK^{\sf AP}_{\rm df, 3}]_{1,5} &= {\boldsymbol {\mathcal Y}}^{[kab]}_{(312)} \circ \cK_{\text{df}, 3}( D^+(p'_1) D^+(p'_2) \pi^-(p'_3) \leftarrow D^+(p_1) D^+(p_2) \pi^-(p_3)) \circ {\boldsymbol {\mathcal Y}}^{[kab]\dagger}_{(123)}, \\
[\widehat \cK^{\sf AP}_{\rm df, 3}]_{1,6} &= {\boldsymbol {\mathcal Y}}^{[kab]}_{(312)} \circ \cK_{\text{df}, 3}( D^+(p'_1) D^+(p'_2) \pi^-(p'_3) \leftarrow D^+(p_1) D^+(p_2) \pi^-(p_3)) \circ {\boldsymbol {\mathcal Y}}^{[kab]\dagger}_{(132)},
\end{split}
\end{align}
where the $\boldsymbol{\mathcal Y}$ operators have been defined in \Cref{eq:YRdef}.
Note that, as with $\widehat{\cK}_2^{\sf AP}$,
no symmetry factors are needed in the {\sf AP} basis.

We now turn to the $\sf MP$ basis. To obtain the expression in the {\sf MP} basis, we first apply the transformation
\begin{equation}
\widehat{\cK}_{\rm df, 3}^{\overline{\sf MP}} = R^{-1} \cdot S \cdot \widehat{\cK}_{\rm df, 3}^{\sf AP} \cdot S^{\rm T} \cdot R^{-1} \,.
\label{eq:KdfAPtoMP}
\end{equation}
Next, to reduce from the twelve-dimensional $\overline{\sf MP}$ to the eight-dimensional $\sf MP$ case, we use properties of the entries of $\widehat \cK^{\sf AP}_{\rm df, 3}$ to simplify the result. One example are the entries $\{1,5\}$ and $\{1,6\}$, where the only difference is that the pair angular momentum is defined with respect
to the $D$ meson or the pion, respectively.
Thus, they are same up to a factor of $P^{(\ell)}$,
\begin{equation}
[\widehat \cK^{\sf AP}_{\rm df, 3}]_{\{1,5\}} = [\widehat \cK^{\sf AP}_{\rm df, 3}]_{\{1,6\}} P^{(\ell)}\,,
\end{equation}
with similar expressions for other entries of $\widehat \cK^{\sf AP}_{\rm df, 3}$. Utilizing these relations, $ \widehat{\cK}_{\rm df, 3}^{\overline{\sf MP}}$ can be simplified to a block-diagonal structure, with an upper-left eight-dimensional block corresponding to $\widehat \cK^{\sf MP}_{\rm df, 3}$, and a lower-right 4-dimensional block with
all entries vanishing. This is the same block structure found above for $\widehat \cK_2^{\overline{\sf MP}}$. In order to give the expression for $\widehat \cK^{\sf MP}_{\rm df, 3}$, we use the same schematic notation as in \Cref{eq:Kdfgeneral},
\begin{equation}
\widehat \cK^{\sf MP}_{\rm df, 3} \sim
\begin{pmatrix}
\frac{\cK^a}{2} & \frac{\cK^b}{2} & \frac{\cK^b}{2} & \frac{\cK^c}{2} & \frac{\cK^a}{\sqrt{2}} & \frac{\cK^b}{\sqrt{2}} & \frac{\cK^c}{\sqrt{2}} & \frac{\cK^b}{\sqrt{2}} \\
\frac{\cK^{b,{\rm T}}}{2} & \frac{\cK^d}{2} & \frac{\cK^d}{2} & \frac{\cK^e}{2} & \frac{\cK^b}{\sqrt{2}} & \frac{\cK^d}{\sqrt{2}} & \frac{\cK^e}{\sqrt{2}} & \frac{\cK^d}{\sqrt{2}} \\
\frac{\cK^{b,{\rm T}}}{2} & \frac{\cK^d}{2} & \frac{\cK^d}{2} & \frac{\cK^e}{2} &\frac{ \cK^b}{\sqrt{2}} &\frac{ \cK^d}{\sqrt{2}} & \frac{\cK^e}{\sqrt{2}} & \frac{\cK^d}{\sqrt{2}} \\
\frac{\cK^{c,{\rm T}}}{2} & \frac{\cK^{e,{\rm T}}}{2} & \frac{\cK^{e,{\rm T}}}{2} & \frac{\cK^f}{2} &\frac{ \cK^c}{\sqrt{2}} & \frac{\cK^e}{\sqrt{2}} & \frac{\cK^f}{\sqrt{2}} & \frac{\cK^e}{\sqrt{2}} \\
\frac{\cK^a}{\sqrt{2}} & \frac{\cK^{b,{\rm T}}}{\sqrt{2}} & \frac{\cK^{b,{\rm T}}}{\sqrt{2}} & \frac{\cK^{c,{\rm T}}}{\sqrt{2}} & \cK^a & \cK^b & \cK^c & \cK^b \\
\frac{\cK^{b,{\rm T}}}{\sqrt{2}} & \frac{\cK^d}{\sqrt{2}} & \frac{\cK^d}{\sqrt{2}} & \frac{\cK^{e,{\rm T}}}{\sqrt{2}} & \cK^{b,{\rm T}} & \cK^d & \cK^e & \cK^d \\
\frac{\cK^{c,{\rm T}}}{\sqrt{2}} & \frac{\cK^{e,{\rm T}}}{\sqrt{2}} & \frac{\cK^{e,{\rm T}}}{\sqrt{2}} & \frac{\cK^f}{\sqrt{2}} & \cK^{c,{\rm T}} & \cK^{e,{\rm T}} & \cK^f & \cK^e \\
\frac{\cK^{b,{\rm T}}}{\sqrt{2}} & \frac{\cK^d}{\sqrt{2}} & \frac{\cK^d}{\sqrt{2}} & \frac{\cK^{e,{\rm T}}}{\sqrt{2}} & \cK^{b,{\rm T}} & \cK^d & \cK^{e,{\rm T}} & \cK^d
\end{pmatrix}\,, \label{eq:kdfflavMP2}
\end{equation}
where $\{k\ell m\}$ indices and the spectator flavor are implicit and can be determined from the position in the matrix, using the ordering of \Cref{eq:orderMP}.
There are isospin relations between the underlying quantities $\cK^a-\cK^f$ which we do not describe explicitly.\footnote{Note that the entries $\cK^a$, $\cK^b$, etc., used here, do not exactly correspond to those of appendix~\ref{app:topt_derivation}. }

To see how both the $\sf MP$ and $\sf AP$ bases lead to the same result, we can apply the transformations in \Cref{eq:RStrafos} together with
\begin{align}
\widehat \cK_{\rm df, 3} & \ \ \ \longrightarrow \ \ \ R^{-1} \cdot S \cdot \widehat \cK_{\rm df, 3} \cdot S^{\rm T} \cdot R^{-1} \,.
\end{align}
It is then possible to see that the finite-volume correlator at nonzero $\Kdf$ in \Cref{eq:FVcorr} remains invariant, and so $\sf AP$ and $\overline{\sf MP}$ forms are equivalent.
Moreover, the contribution from the lower 4-dimensional block exactly vanishes. This follows from the block structure of $\widehat F_3$, the vanishing entries in the endcaps, and crucially from the fact that $ \widehat{\cK}_{\rm df, 3}^{\overline{\sf MP}}$ also has a block-diagonal structure with zero entries. Therefore, the finite-volume correlators in $\overline{\sf MP}$ and ${\sf MP}$ describe the same finite-volume spectrum.

The next step is to rotate $\Kdf$ to the definite three-particle isospin.
The result is block diagonal,
\begin{equation}
C_{{\sf MP} \to I} \cdot \widehat{\cK}_{\text{df}, 3}^{\sf MP} \cdot C_{{\sf MP} \to I}^{\rm T} =
\begin{pmatrix}
{\widehat \cK}^{[I=2]}_{\text{df}, 3} & 0 & 0 \\
0 & \widehat{\cK}^{[I=1]}_{\text{df}, 3} & 0 \\
0& 0 & \widehat{\cK}^{[I=0]}_{\text{df}, 3}
\end{pmatrix} \,,
\end{equation}
where the matrices ${\widehat \cK}^{[I=0,1,2]}_{\text{df}, 3}$ have been defined in the main text in \Cref{eq:KdfI02,eq:Kdf3I1form}. Note that, in order to obtain this result, the above-mentioned isospin relations between the $\cK^a -\cK^f$ must be used.

\subsection{Relation to the scattering amplitude}
\label{app:BKtoM}

Following the formalism for identical particles, one can construct a finite-volume analog of the unsymmetrized scattering amplitude, given by
\begin{equation}
\widehat{\cM}_{3,L}^{(u,u)} = \widehat{\cD}_L^{(u,u)} + \widehat{\cM}_{{\rm df},3,L}^{(u,u)\prime} \,,
\end{equation}
where the two terms on the right-hand side can be expressed using the same building blocks that enter the quantization condition. The first term, sometimes referred to as the ladder amplitude, takes the form
\begin{align}
\widehat{\cD}_L^{(u,u)} &= - \widehat{\cM}_{2,L} \widehat G \widehat{\cM}_{2,L}
\frac1{1 + \widehat{G} \widehat{\cM}_{2,L} }\,,
\label{eq:DhatuuappB}
\end{align}
and the second, with $\Kdf$ dependence, is given by
\begin{align}
\widehat{\cM}_{\df,3,L}^{(u,u)\prime} &= \left[ \frac13 - \widehat{\cD}_{23,L}^{(u,u)} \widehat F \right]
\widehat{\cK}_{\df,3} \frac1{1 + \widehat F_3 \widehat{\cK}_{\df,3} }
\left[\frac13 - \widehat F \widehat{\cD}_{23,L}^{(u,u)} \right]\,,
\label{eq:Mhatdf3LappB}
\end{align}
where $\widehat{\cD}_{23,L}^{(u,u)}$ is defined similarly to \Cref{eq:Dhat23Ldef}.
In order to construct the full finite-volume amplitude, two additional steps are needed. First, the incoming momenta in the $k \ell m$ basis need to be assigned to the corresponding $p_1, p_2, p_3$ using the $\boldsymbol{\mathcal X}$ operator of \Cref{eq:XRdef} (and similar for outgoing). Second, all possible momentum assignments in cyclic permutations need to be added together.

We begin the discussion in the twelve-dimensional $\sf AP$ basis. The physical amplitude is only four dimensional, since there are only four states whether characterized in the flavor basis of \Cref{eq:ddpi_flavor_states} or the isospin basis of \Cref{eq:ddpi_isospin_states}. Thus, each entry in the physical amplitude must come from 3 different entries in the unsymmetrized amplitude in the $\sf AP$ basis, corresponding to three different cyclic permutations acting on the flavors and the assignments of the momenta.
Focusing directly on the isospin case, the finite-volume amplitude can be defined from the $\sf AP$ matrices as
\begin{equation}
\widehat{\cM}^{[I]}_{3,L} = \sum_{\sigma, \sigma' \, \in \, \text{cyclic} }
\boldsymbol{\mathcal V}_{[I]}^{\sf AP}(\sigma) \circ \widehat{\cM}_{3,L}^{(u,u), {\sf AP}} \circ \boldsymbol{\mathcal V}^{{\sf AP}\dagger}_{[I]}(\sigma') , \label{eq:M3sum1}
\end{equation}
where the operators are defined as
\begin{equation}
\boldsymbol{\mathcal V}_{[I]}^{\sf AP}(\sigma) = \boldsymbol{\mathcal X}^{\sigma}_{[kab]} \cdot v_{[I]} \cdot P^{[312]}_{\sigma} \,.
\end{equation}
Here, $P^{[312]}_\sigma$ is a matrix representing the $\sigma$ permutation, such that particle 3 is assigned to $\sigma_1$, particle 1 is assigned to $\sigma_2$ and particle 2 is assigned to $\sigma_3$. Note that this ordering is chosen to match that of $\boldsymbol{\mathcal X}^{\sigma}_{[kab]}$. The identity matrix therefore corresponds to $\sigma=(3,1,2)$, while
\begin{equation}
P^{[312]}_{(123)} = \begin{pmatrix}
0 & 0 & 0 & 0 & 0 & 1 & 0 & 0 & 0 & 0 & 0 & 0 \\
0 & 0 & 0 & 0 & 0 & 0 & 0 & 0 & 0 & 0 & 0 & 1 \\
0 & 0 & 0 & 0 & 0 & 0 & 0 & 1 & 0 & 0 & 0 & 0 \\
0 & 0 & 0 & 0 & 0 & 0 & 0 & 0 & 0 & 1 & 0 & 0 \\
1 & 0 & 0 & 0 & 0 & 0 & 0 & 0 & 0 & 0 & 0 & 0 \\
0 & 0 & 0 & 0 & 1 & 0 & 0 & 0 & 0 & 0 & 0 & 0 \\
0 & 1 & 0 & 0 & 0 & 0 & 0 & 0 & 0 & 0 & 0 & 0 \\
0 & 0 & 0 & 0 & 0 & 0 & 0 & 0 & 0 & 0 & 1 & 0 \\
0 & 0 & 0 & 1 & 0 & 0 & 0 & 0 & 0 & 0 & 0 & 0 \\
0 & 0 & 0 & 0 & 0 & 0 & 0 & 0 & 1 & 0 & 0 & 0 \\
0 & 0 & 1 & 0 & 0 & 0 & 0 & 0 & 0 & 0 & 0 & 0 \\
0 & 0 & 0 & 0 & 0 & 0 & 1 & 0 & 0 & 0 & 0 & 0
\end{pmatrix}, \quad
P^{[312]}_{(231)} = \left( P^{[312]}_{(123)} \right)^2,
\end{equation}
while the following vectors describing the isospin states,
\begin{align}
\begin{split}
v_2 &= \frac{1}{\sqrt{6}} \bigg( 1, \sqrt{2}, \sqrt{2},1 ,0,0,0,0,0,0,0,0 \bigg)^{\rm T}, \\
v_{1,a} &= \frac{1}{\sqrt{2}} \bigg( 1, 0,0,-1 ,0,0,0,0,0,0,0,0 \bigg)^{\rm T}, \\
v_{1,b} &= \frac{1}{\sqrt{2}} \bigg(0, 1, -1,0 ,0,0,0,0,0,0,0,0 \bigg)^{\rm T}, \\
v_0 &= \frac{1}{\sqrt{6}} \bigg( \sqrt{2},-1,-1,\sqrt{2}, ,0,0,0,0,0,0,0,0 \bigg)^{\rm T} .
\label{eq:isostates12}
\end{split}
\end{align}

Because the sums in \Cref{eq:M3sum1} are applied independently to the incoming and outgoing states, the operators factorize and the equation be written as
\begin{equation}
\widehat{\cM}^{[I]}_{3,L} =
\boldsymbol{\mathcal V}_{[I]}^{\sf AP} \circ \widehat{\cM}_{3,L}^{(u,u), {\sf AP}} \circ \boldsymbol{\mathcal V}^{{\sf AP}\dagger}_{[I]} , \label{eq:M3sum1_factorize}
\end{equation}
where we have introduced
\begin{equation}
\boldsymbol{\mathcal V}_{[I]}^{\sf AP} = \sum_{\sigma \, \in \, \text{cyclic} }\boldsymbol{\mathcal V}_{[I]}^{\sf AP}(\sigma) \,.
\end{equation}

The first step to reach the form in main text is to change to the $\sf MP$ basis on the right-hand side of \Cref{eq:M3sum1}. This results in the following expression
\begin{equation}
\widehat{\cM}^{[I]}_{3,L} =
\boldsymbol{\mathcal V}_{[I]}^{\sf MP} \circ \widehat{\cM}_{3,L}^{(u,u), {\sf MP}} \circ \boldsymbol{\mathcal V}^{{\sf MP}\dagger}_{[I]} , \label{eq:M3sum2}
\end{equation}
where both $\widehat{\cM}_{3,L}^{(u,u), {\sf MP}}$ and $\boldsymbol{\mathcal V}_{[I]}^{\sf MP}(\sigma)$ are obtained from their $\overline{{\sf MP}}$ analogs by keeping only the upper eight-dimensional blocks and the latter are defined by
\begin{gather}
\widehat{\cM}_{3,L}^{(u,u), \overline{\sf MP}} = R^{-1} \cdot S \cdot \widehat{\cM}_{3,L}^{(u,u), {\sf AP}} \cdot S^{\rm T} \cdot R^{-1} \,, \\[5pt]
\boldsymbol{\mathcal V}_{[I]}^{ \overline{\sf MP}} = \boldsymbol{\mathcal V}_{[I]}^{\sf AP} \cdot S^{\rm T} \cdot R, \quad
\boldsymbol{\mathcal V}_{[I]}^{ \overline{\sf MP} \dagger} = R \cdot S \cdot \boldsymbol{\mathcal V}_{[I]}^{{\sf AP}\dagger}.
\end{gather}
The vanishing lower four-dimensional block in $ \widehat{\cM}_{3,L}^{(u,u), \overline {\sf MP}} $ results from having either $\widehat \cK_\text{df,3}$ or $\widehat{\cK}_{L}$ as the rightmost and leftmost factor in every term defining the finite-volume amplitude.

Next, we rotate to the isospin basis. After rearranging the sums, this leads to
\begin{equation}
\widehat{\cM}^{[I]}_{3,L} =
\boldsymbol{\mathcal V}_{[I]}^{\sf MP} \cdot C_{{\sf MP} \rightarrow I}^{\rm T} \circ \widehat{\cM}_{3,L}^{(u,u), {\sf iso}} \circ
C_{{\sf MP} \rightarrow I} \cdot \boldsymbol{\mathcal V}^{{\sf MP}\dagger}_{[I]} , \label{eq:M3sum3}
\end{equation}
where $\widehat{\cM}_{3,L}^{(u,u), {\sf iso}} = C_{{\sf MP} \rightarrow I} \cdot \widehat{\cM}_{3,L}^{(u,u), {\sf MP}} \cdot C_{{\sf MP} \rightarrow I}^{\rm T} $ is block diagonal:
\begin{equation}
\widehat{\cM}_{3,L}^{(u,u), {\sf iso}} =
\begin{pmatrix}
\widehat{\cM}_{3,L}^{(u,u), [I=2]} & 0 & 0 \\
0 & \widehat{\cM}_{3,L}^{(u,u), [I=1]} & 0 \\
0& 0 & \widehat{\cM}_{3,L}^{(u,u), [I=0]}
\end{pmatrix}.
\end{equation}
Finally, we can compute the sums in brackets in \Cref{eq:M3sum3} which leads to $\boldsymbol{\mathcal X}_0$, $\boldsymbol{\mathcal X}_2$, $\boldsymbol{\mathcal X}_a$ and $\boldsymbol{\mathcal X}_b$ defined in the main text in \Cref{eq:alphaSvec,eq:alphaavec,eq:alphabvec}. In particular, working out the sums explicitly, one obtains
\begin{align}
\boldsymbol{\mathcal V}_{[I=2]}^{\sf MP} \cdot C_{{\sf MP} \to I} &= \left( \boldsymbol{\mathcal X}^{(123)}_{[kab]} + P^{(\ell)} \boldsymbol{\mathcal X}^{(231)}_{[kab]}, \sqrt{2} \boldsymbol{\mathcal X}^{(312)}_{[kab]} ,0,0,0,0,0,0\right), \\
\begin{split}
\boldsymbol{\mathcal V}_{[I=1,a]}^{\sf MP} \cdot C_{{\sf MP} \to I} &= \bigg(0,0, -\sqrt{\frac{1}{3}}\left[\boldsymbol{\mathcal X}^{(123)}_{[kab]}+P^{(\ell)} \boldsymbol{\mathcal X}^{(231)}_{[kab]} \right],\\ &\sqrt{\frac{2}{3}}\left[\boldsymbol{\mathcal X}^{(123)}_{[kab]}+P^{(\ell)} \boldsymbol{\mathcal X}^{(231)}_{[kab]} \right],\sqrt{2} \boldsymbol{\mathcal X}^{(312)}_{[kab]},0,0,0\bigg),
\end{split}\\
\begin{split}
\boldsymbol{\mathcal V}_{[I=1,b]}^{\sf MP} \cdot C_{{\sf MP} \to I} &= \bigg(0,0, -\sqrt{\frac{2}{3}}\left[\boldsymbol{\mathcal X}^{(123)}_{[kab]}-P^{(\ell)} \boldsymbol{\mathcal X}^{(231)}_{[kab]} \right],\\ &-\sqrt{\frac{1}{3}}\left[\boldsymbol{\mathcal X}^{(123)}_{[kab]}-P^{(\ell)} \boldsymbol{\mathcal X}^{(231)}_{[kab]} \right],0,\sqrt{2} \boldsymbol{\mathcal X}^{(312)}_{[kab]},0,0 \bigg),
\end{split} \\
\boldsymbol{\mathcal V}_{[I=0]}^{\sf MP} \cdot C_{{\sf MP} \to I} &= \left(0,0,0,0,0,0, -\boldsymbol{\mathcal X}^{(123)}_{[kab]} - P^{(\ell)} \boldsymbol{\mathcal X}^{(231)}_{[kab]}, \sqrt{2} \boldsymbol{\mathcal X}^{(312)}_{[kab]}\right).
\end{align}
Keeping only the entries in the given isospin block, and using that $P^{(\ell)}$ is simply a permutation of the interacting pair, i.e., $ \boldsymbol{\mathcal X}^{(213)}_{[kab]} = P^{(\ell)} \boldsymbol{\mathcal X}^{(231)}_{[kab]}$, one obtains the results for $\boldsymbol{\mathcal X}_0$, $\boldsymbol{\mathcal X}_2$, $\boldsymbol{\mathcal X}_a$, and $\boldsymbol{\mathcal X}_b$ in the main text. An analogous calculation applies for $\boldsymbol{\mathcal X}^\dagger_0$, $\boldsymbol{\mathcal X}^\dagger_2$, $\boldsymbol{\mathcal X}^\dagger_a$, and $\boldsymbol{\mathcal X}^\dagger_b$.

\bibliographystyle{JHEP}
\bibliography{ref.bib}

\providecommand{\href}[2]{#2}\begingroup\raggedright\begin{thebibliography}{10}

\bibitem{Ali:2017jda}
A.~Ali, J.~S. Lange and S.~Stone, \emph{{Exotics: Heavy Pentaquarks and
  Tetraquarks}}, \href{https://doi.org/10.1016/j.ppnp.2017.08.003}{\emph{Prog.
  Part. Nucl. Phys.} {\bfseries 97} (2017) 123}
  [\href{https://arxiv.org/abs/1706.00610}{{\ttfamily 1706.00610}}].

\bibitem{Olsen:2017bmm}
S.~L. Olsen, T.~Skwarnicki and D.~Zieminska, \emph{{Nonstandard heavy mesons
  and baryons: Experimental evidence}},
  \href{https://doi.org/10.1103/RevModPhys.90.015003}{\emph{Rev. Mod. Phys.}
  {\bfseries 90} (2018) 015003}
  [\href{https://arxiv.org/abs/1708.04012}{{\ttfamily 1708.04012}}].

\bibitem{Karliner:2017qhf}
M.~Karliner, J.~L. Rosner and T.~Skwarnicki, \emph{{Multiquark States}},
  \href{https://doi.org/10.1146/annurev-nucl-101917-020902}{\emph{Ann. Rev.
  Nucl. Part. Sci.} {\bfseries 68} (2018) 17}
  [\href{https://arxiv.org/abs/1711.10626}{{\ttfamily 1711.10626}}].

\bibitem{Guo:2017jvc}
F.-K. Guo, C.~Hanhart, U.-G. Mei\ss{}ner, Q.~Wang, Q.~Zhao and B.-S. Zou,
  \emph{{Hadronic molecules}},
  \href{https://doi.org/10.1103/RevModPhys.90.015004}{\emph{Rev. Mod. Phys.}
  {\bfseries 90} (2018) 015004}
  [\href{https://arxiv.org/abs/1705.00141}{{\ttfamily 1705.00141}}].

\bibitem{Liu:2019zoy}
Y.-R. Liu, H.-X. Chen, W.~Chen, X.~Liu and S.-L. Zhu, \emph{{Pentaquark and
  Tetraquark states}},
  \href{https://doi.org/10.1016/j.ppnp.2019.04.003}{\emph{Prog. Part. Nucl.
  Phys.} {\bfseries 107} (2019) 237}
  [\href{https://arxiv.org/abs/1903.11976}{{\ttfamily 1903.11976}}].

\bibitem{Brambilla:2019esw}
N.~Brambilla, S.~Eidelman, C.~Hanhart, A.~Nefediev, C.-P. Shen, C.~E. Thomas
  et~al., \emph{{The $XYZ$ states: experimental and theoretical status and
  perspectives}},
  \href{https://doi.org/10.1016/j.physrep.2020.05.001}{\emph{Phys. Rept.}
  {\bfseries 873} (2020) 1} [\href{https://arxiv.org/abs/1907.07583}{{\ttfamily
  1907.07583}}].

\bibitem{Gershon:2022xnn}
{\scshape LHCb} collaboration, T.~Gershon, \emph{{Exotic hadron naming
  convention}},  \href{https://arxiv.org/abs/2206.15233}{{\ttfamily
  2206.15233}}.

\bibitem{Lebed:2022vfu}
R.~F. Lebed et~al., \emph{{Summary of Topical Group on Hadron Spectroscopy
  (RF07) Rare Processes and Precision Frontier of Snowmass 2021}},  in
  \emph{{Snowmass 2021}} (R.~F. Lebed and T.~Skwarnicki, eds.), 7, 2022,
  \href{https://arxiv.org/abs/2207.14594}{{\ttfamily 2207.14594}}.

\bibitem{LHCb:2021vvq}
{\scshape LHCb} collaboration, R.~Aaij et~al., \emph{{Observation of an exotic
  narrow doubly charmed tetraquark}},
  \href{https://doi.org/10.1038/s41567-022-01614-y}{\emph{Nature Phys.}
  {\bfseries 18} (2022) 751}
  [\href{https://arxiv.org/abs/2109.01038}{{\ttfamily 2109.01038}}].

\bibitem{LHCb:2021auc}
{\scshape LHCb} collaboration, R.~Aaij et~al., \emph{{Study of the doubly
  charmed tetraquark $T_{cc}^{+}$}},
  \href{https://doi.org/10.1038/s41467-022-30206-w}{\emph{Nature Commun.}
  {\bfseries 13} (2022) 3351}
  [\href{https://arxiv.org/abs/2109.01056}{{\ttfamily 2109.01056}}].

\bibitem{Detmold:2011kw}
W.~Detmold and B.~Smigielski, \emph{{Lattice QCD study of mixed systems of
  pions and kaons}},
  \href{https://doi.org/10.1103/PhysRevD.84.014508}{\emph{Phys. Rev. D}
  {\bfseries 84} (2011) 014508}
  [\href{https://arxiv.org/abs/1103.4362}{{\ttfamily 1103.4362}}].

\bibitem{Mai:2018djl}
M.~Mai and M.~D{\"{o}}ring, \emph{{Finite-Volume Spectrum of $\pi^+\pi^+$ and
  $\pi^+\pi^+\pi^+$ Systems}},
  \href{https://doi.org/10.1103/PhysRevLett.122.062503}{\emph{Phys. Rev. Lett.}
  {\bfseries 122} (2019) 062503}
  [\href{https://arxiv.org/abs/1807.04746}{{\ttfamily 1807.04746}}].

\bibitem{Horz:2019rrn}
B.~H\"{o}rz and A.~Hanlon, \emph{{Two- and three-pion finite-volume spectra at
  maximal isospin from lattice QCD}},
  \href{https://doi.org/10.1103/PhysRevLett.123.142002}{\emph{Phys. Rev. Lett.}
  {\bfseries 123} (2019) 142002}
  [\href{https://arxiv.org/abs/1905.04277}{{\ttfamily 1905.04277}}].

\bibitem{Blanton:2019vdk}
T.~D. Blanton, F.~Romero-L\'opez and S.~R. Sharpe, \emph{{$I = 3$ three-pion
  scattering amplitude from lattice QCD}},
  \href{https://doi.org/10.1103/PhysRevLett.124.032001}{\emph{Phys. Rev. Lett.}
  {\bfseries 124} (2020) 032001}
  [\href{https://arxiv.org/abs/1909.02973}{{\ttfamily 1909.02973}}].

\bibitem{Mai:2019fba}
M.~Mai, M.~D\"{o}ring, C.~Culver and A.~Alexandru, \emph{{Three-body unitarity
  versus finite-volume $\pi^+\pi^+\pi^+$ spectrum from lattice QCD}},
  \href{https://doi.org/10.1103/PhysRevD.101.054510}{\emph{Phys.\ Rev.\ D}
  {\bfseries 101} (2020) 054510}
  [\href{https://arxiv.org/abs/1909.05749}{{\ttfamily 1909.05749}}].

\bibitem{Culver:2019vvu}
C.~Culver, M.~Mai, R.~Brett, A.~Alexandru and M.~D\"{o}ring, \emph{{Three pion
  spectrum in the $I=3$ channel from lattice QCD}},
  \href{https://doi.org/10.1103/PhysRevD.101.114507}{\emph{Phys. Rev. D}
  {\bfseries 101} (2020) 114507}
  [\href{https://arxiv.org/abs/1911.09047}{{\ttfamily 1911.09047}}].

\bibitem{Fischer:2020jzp}
M.~Fischer, B.~Kostrzewa, L.~Liu, F.~Romero-L\'opez, M.~Ueding and C.~Urbach,
  \emph{{Scattering of two and three physical pions at maximal isospin from
  lattice QCD}},
  \href{https://doi.org/10.1140/epjc/s10052-021-09206-5}{\emph{Eur. Phys. J. C}
  {\bfseries 81} (2021) 436}
  [\href{https://arxiv.org/abs/2008.03035}{{\ttfamily 2008.03035}}].

\bibitem{Hansen:2020otl}
{\scshape Hadron Spectrum} collaboration, M.~T. Hansen, R.~A. Brice\~no, R.~G.
  Edwards, C.~E. Thomas and D.~J. Wilson, \emph{{Energy-Dependent $\pi^+ \pi^+
  \pi^+$ Scattering Amplitude from QCD}},
  \href{https://doi.org/10.1103/PhysRevLett.126.012001}{\emph{Phys. Rev. Lett.}
  {\bfseries 126} (2021) 012001}
  [\href{https://arxiv.org/abs/2009.04931}{{\ttfamily 2009.04931}}].

\bibitem{NPLQCD:2020ozd}
{\scshape NPLQCD, QCDSF} collaboration, S.~R. Beane et~al., \emph{{Charged
  multihadron systems in lattice QCD+QED}},
  \href{https://doi.org/10.1103/PhysRevD.103.054504}{\emph{Phys. Rev. D}
  {\bfseries 103} (2021) 054504}
  [\href{https://arxiv.org/abs/2003.12130}{{\ttfamily 2003.12130}}].

\bibitem{Alexandru:2020xqf}
A.~Alexandru, R.~Brett, C.~Culver, M.~D\"{o}ring, D.~Guo, F.~X. Lee et~al.,
  \emph{{Finite-volume energy spectrum of the $K^-K^-K^-$ system}},
  \href{https://doi.org/10.1103/PhysRevD.102.114523}{\emph{Phys. Rev. D}
  {\bfseries 102} (2020) 114523}
  [\href{https://arxiv.org/abs/2009.12358}{{\ttfamily 2009.12358}}].

\bibitem{Brett:2021wyd}
R.~Brett, C.~Culver, M.~Mai, A.~Alexandru, M.~D\"oring and F.~X. Lee,
  \emph{{Three-body interactions from the finite-volume QCD spectrum}},
  \href{https://doi.org/10.1103/PhysRevD.104.014501}{\emph{Phys. Rev. D}
  {\bfseries 104} (2021) 014501}
  [\href{https://arxiv.org/abs/2101.06144}{{\ttfamily 2101.06144}}].

\bibitem{Blanton:2021llb}
T.~D. Blanton, A.~D. Hanlon, B.~H\"orz, C.~Morningstar, F.~Romero-L\'opez and
  S.~R. Sharpe, \emph{{Interactions of two and three mesons including higher
  partial waves from lattice QCD}},
  \href{https://doi.org/10.1007/JHEP10(2021)023}{\emph{JHEP} {\bfseries 10}
  (2021) 023} [\href{https://arxiv.org/abs/2106.05590}{{\ttfamily
  2106.05590}}].

\bibitem{Mai:2021nul}
{\scshape GWQCD} collaboration, M.~Mai, A.~Alexandru, R.~Brett, C.~Culver,
  M.~D\"oring, F.~X. Lee et~al., \emph{{Three-Body Dynamics of the a1(1260)
  Resonance from Lattice QCD}},
  \href{https://doi.org/10.1103/PhysRevLett.127.222001}{\emph{Phys. Rev. Lett.}
  {\bfseries 127} (2021) 222001}
  [\href{https://arxiv.org/abs/2107.03973}{{\ttfamily 2107.03973}}].

\bibitem{Garofalo:2022pux}
M.~Garofalo, M.~Mai, F.~Romero-L\'opez, A.~Rusetsky and C.~Urbach,
  \emph{{Three-body resonances in the \ensuremath{\varphi}$^{4}$ theory}},
  \href{https://doi.org/10.1007/JHEP02(2023)252}{\emph{JHEP} {\bfseries 02}
  (2023) 252} [\href{https://arxiv.org/abs/2211.05605}{{\ttfamily
  2211.05605}}].

\bibitem{Draper:2023boj}
Z.~T. Draper, A.~D. Hanlon, B.~H\"orz, C.~Morningstar, F.~Romero-L\'opez and
  S.~R. Sharpe, \emph{{Interactions of $\pi K$, $\pi \pi K$ and $KK\pi$ systems
  at maximal isospin from lattice QCD}},
  \href{https://arxiv.org/abs/2302.13587}{{\ttfamily 2302.13587}}.

\bibitem{Detmold:2008gh}
W.~Detmold and M.~J. Savage, \emph{{The Energy of $n$ Identical Bosons in a
  Finite Volume at $O(L^{-7})$}},
  \href{https://doi.org/10.1103/PhysRevD.77.057502}{\emph{Phys. Rev.}
  {\bfseries D77} (2008) 057502}
  [\href{https://arxiv.org/abs/0801.0763}{{\ttfamily 0801.0763}}].

\bibitem{Beane:2007qr}
S.~R. Beane, W.~Detmold and M.~J. Savage, \emph{{n-Boson Energies at Finite
  Volume and Three-Boson Interactions}},
  \href{https://doi.org/10.1103/PhysRevD.76.074507}{\emph{Phys. Rev.}
  {\bfseries D76} (2007) 074507}
  [\href{https://arxiv.org/abs/0707.1670}{{\ttfamily 0707.1670}}].

\bibitem{Briceno:2012rv}
R.~A. Brice\~no and Z.~Davoudi, \emph{{Three-particle scattering amplitudes
  from a finite volume formalism}},
  \href{https://doi.org/10.1103/PhysRevD.87.094507}{\emph{Phys. Rev.}
  {\bfseries D87} (2013) 094507}
  [\href{https://arxiv.org/abs/1212.3398}{{\ttfamily 1212.3398}}].

\bibitem{Polejaeva:2012ut}
K.~Polejaeva and A.~Rusetsky, \emph{{Three particles in a finite volume}},
  \href{https://doi.org/10.1140/epja/i2012-12067-8}{\emph{Eur.\ Phys.\ J.\ A}
  {\bfseries 48} (2012) 67} [\href{https://arxiv.org/abs/1203.1241}{{\ttfamily
  1203.1241}}].

\bibitem{Hansen:2014eka}
M.~T. Hansen and S.~R. Sharpe, \emph{{Relativistic, model-independent,
  three-particle quantization condition}},
  \href{https://doi.org/10.1103/PhysRevD.90.116003}{\emph{Phys. Rev.}
  {\bfseries D90} (2014) 116003}
  [\href{https://arxiv.org/abs/1408.5933}{{\ttfamily 1408.5933}}].

\bibitem{Hansen:2015zga}
M.~T. Hansen and S.~R. Sharpe, \emph{{Expressing the three-particle
  finite-volume spectrum in terms of the three-to-three scattering amplitude}},
  \href{https://doi.org/10.1103/PhysRevD.92.114509}{\emph{Phys. Rev.}
  {\bfseries D92} (2015) 114509}
  [\href{https://arxiv.org/abs/1504.04248}{{\ttfamily 1504.04248}}].

\bibitem{Briceno:2017tce}
R.~A. Brice\~no, M.~T. Hansen and S.~R. Sharpe, \emph{{Relating the
  finite-volume spectrum and the two-and-three-particle $S$ matrix for
  relativistic systems of identical scalar particles}},
  \href{https://doi.org/10.1103/PhysRevD.95.074510}{\emph{Phys. Rev.}
  {\bfseries D95} (2017) 074510}
  [\href{https://arxiv.org/abs/1701.07465}{{\ttfamily 1701.07465}}].

\bibitem{Hammer:2017uqm}
H.-W. Hammer, J.-Y. Pang and A.~Rusetsky, \emph{{Three-particle quantization
  condition in a finite volume: 1. The role of the three-particle force}},
  \href{https://doi.org/10.1007/JHEP09(2017)109}{\emph{JHEP} {\bfseries 09}
  (2017) 109} [\href{https://arxiv.org/abs/1706.07700}{{\ttfamily
  1706.07700}}].

\bibitem{Konig:2017krd}
S.~K\"onig and D.~Lee, \emph{{Volume Dependence of N-Body Bound States}},
  \href{https://doi.org/10.1016/j.physletb.2018.01.060}{\emph{Phys. Lett. B}
  {\bfseries 779} (2018) 9} [\href{https://arxiv.org/abs/1701.00279}{{\ttfamily
  1701.00279}}].

\bibitem{Hammer:2017kms}
H.~W. Hammer, J.~Y. Pang and A.~Rusetsky, \emph{{Three particle quantization
  condition in a finite volume: 2. General formalism and the analysis of
  data}}, \href{https://doi.org/10.1007/JHEP10(2017)115}{\emph{JHEP} {\bfseries
  10} (2017) 115} [\href{https://arxiv.org/abs/1707.02176}{{\ttfamily
  1707.02176}}].

\bibitem{Mai:2017bge}
M.~Mai and M.~{D\"{o}ring}, \emph{{Three-body Unitarity in the Finite Volume}},
  \href{https://doi.org/10.1140/epja/i2017-12440-1}{\emph{Eur. Phys. J.}
  {\bfseries A53} (2017) 240}
  [\href{https://arxiv.org/abs/1709.08222}{{\ttfamily 1709.08222}}].

\bibitem{Briceno:2018mlh}
R.~A. Brice\~no, M.~T. Hansen and S.~R. Sharpe, \emph{{Numerical study of the
  relativistic three-body quantization condition in the isotropic
  approximation}},
  \href{https://doi.org/10.1103/PhysRevD.98.014506}{\emph{Phys. Rev.}
  {\bfseries D98} (2018) 014506}
  [\href{https://arxiv.org/abs/1803.04169}{{\ttfamily 1803.04169}}].

\bibitem{Briceno:2018aml}
R.~A. Brice\~no, M.~T. Hansen and S.~R. Sharpe, \emph{{Three-particle systems
  with resonant subprocesses in a finite volume}},
  \href{https://doi.org/10.1103/PhysRevD.99.014516}{\emph{Phys. Rev.}
  {\bfseries D99} (2019) 014516}
  [\href{https://arxiv.org/abs/1810.01429}{{\ttfamily 1810.01429}}].

\bibitem{Blanton:2019igq}
T.~D. Blanton, F.~Romero-L\'opez and S.~R. Sharpe, \emph{{Implementing the
  three-particle quantization condition including higher partial waves}},
  \href{https://doi.org/10.1007/JHEP03(2019)106}{\emph{JHEP} {\bfseries 03}
  (2019) 106} [\href{https://arxiv.org/abs/1901.07095}{{\ttfamily
  1901.07095}}].

\bibitem{Pang:2019dfe}
J.-Y. Pang, J.-J. Wu, H.~W. Hammer, U.-G. Mei{$\ss$}ner and A.~Rusetsky,
  \emph{{Energy shift of the three-particle system in a finite volume}},
  \href{https://doi.org/10.1103/PhysRevD.99.074513}{\emph{Phys. Rev.}
  {\bfseries D99} (2019) 074513}
  [\href{https://arxiv.org/abs/1902.01111}{{\ttfamily 1902.01111}}].

\bibitem{Jackura:2019bmu}
A.~W. Jackura, S.~M. Dawid, C.~Fern\'andez-Ram\'\i{}rez, V.~Mathieu,
  M.~Mikhasenko, A.~Pilloni et~al., \emph{{Equivalence of three-particle
  scattering formalisms}},
  \href{https://doi.org/10.1103/PhysRevD.100.034508}{\emph{Phys. Rev. D}
  {\bfseries 100} (2019) 034508}
  [\href{https://arxiv.org/abs/1905.12007}{{\ttfamily 1905.12007}}].

\bibitem{Briceno:2019muc}
R.~A. Brice\~no, M.~T. Hansen, S.~R. Sharpe and A.~P. Szczepaniak,
  \emph{{Unitarity of the infinite-volume three-particle scattering amplitude
  arising from a finite-volume formalism}},
  \href{https://doi.org/10.1103/PhysRevD.100.054508}{\emph{Phys. Rev.}
  {\bfseries D100} (2019) 054508}
  [\href{https://arxiv.org/abs/1905.11188}{{\ttfamily 1905.11188}}].

\bibitem{Romero-Lopez:2019qrt}
F.~Romero-L\'opez, S.~R. Sharpe, T.~D. Blanton, R.~A. Brice\~no and M.~T.
  Hansen, \emph{{Numerical exploration of three relativistic particles in a
  finite volume including two-particle resonances and bound states}},
  \href{https://doi.org/10.1007/JHEP10(2019)007}{\emph{JHEP} {\bfseries 10}
  (2019) 007} [\href{https://arxiv.org/abs/1908.02411}{{\ttfamily
  1908.02411}}].

\bibitem{Hansen:2020zhy}
M.~T. Hansen, F.~Romero-L\'opez and S.~R. Sharpe, \emph{{Generalizing the
  relativistic quantization condition to include all three-pion isospin
  channels}}, \href{https://doi.org/10.1007/JHEP07(2020)047}{\emph{JHEP}
  {\bfseries 07} (2020) 047}
  [\href{https://arxiv.org/abs/2003.10974}{{\ttfamily 2003.10974}}].

\bibitem{Blanton:2020gha}
T.~D. Blanton and S.~R. Sharpe, \emph{{Alternative derivation of the
  relativistic three-particle quantization condition}},
  \href{https://doi.org/10.1103/PhysRevD.102.054520}{\emph{Phys. Rev. D}
  {\bfseries 102} (2020) 054520}
  [\href{https://arxiv.org/abs/2007.16188}{{\ttfamily 2007.16188}}].

\bibitem{Blanton:2020jnm}
T.~D. Blanton and S.~R. Sharpe, \emph{{Equivalence of relativistic
  three-particle quantization conditions}},
  \href{https://doi.org/10.1103/PhysRevD.102.054515}{\emph{Phys. Rev. D}
  {\bfseries 102} (2020) 054515}
  [\href{https://arxiv.org/abs/2007.16190}{{\ttfamily 2007.16190}}].

\bibitem{Pang:2020pkl}
J.-Y. Pang, J.-J. Wu and L.-S. Geng, \emph{{$DDK$ system in finite volume}},
  \href{https://doi.org/10.1103/PhysRevD.102.114515}{\emph{Phys. Rev. D}
  {\bfseries 102} (2020) 114515}
  [\href{https://arxiv.org/abs/2008.13014}{{\ttfamily 2008.13014}}].

\bibitem{Romero-Lopez:2020rdq}
F.~Romero-L\'opez, A.~Rusetsky, N.~Schlage and C.~Urbach, \emph{{Relativistic
  $N$-particle energy shift in finite volume}},
  \href{https://doi.org/10.1007/JHEP02(2021)060}{\emph{JHEP} {\bfseries 02}
  (2021) 060} [\href{https://arxiv.org/abs/2010.11715}{{\ttfamily
  2010.11715}}].

\bibitem{Blanton:2020gmf}
T.~D. Blanton and S.~R. Sharpe, \emph{{Relativistic three-particle quantization
  condition for nondegenerate scalars}},
  \href{https://doi.org/10.1103/PhysRevD.103.054503}{\emph{Phys. Rev. D}
  {\bfseries 103} (2021) 054503}
  [\href{https://arxiv.org/abs/2011.05520}{{\ttfamily 2011.05520}}].

\bibitem{Muller:2020vtt}
F.~M\"{u}ller, A.~Rusetsky and T.~Yu, \emph{{Finite-volume energy shift of the
  three-pion ground state}},
  \href{https://doi.org/10.1103/PhysRevD.103.054506}{\emph{Phys. Rev. D}
  {\bfseries 103} (2021) 054506}
  [\href{https://arxiv.org/abs/2011.14178}{{\ttfamily 2011.14178}}].

\bibitem{Blanton:2021mih}
T.~D. Blanton and S.~R. Sharpe, \emph{{Three-particle finite-volume formalism
  for $\pi^+\pi^+K^+$ and related systems}},
  \href{https://doi.org/10.1103/PhysRevD.104.034509}{\emph{Phys. Rev. D}
  {\bfseries 104} (2021) 034509}
  [\href{https://arxiv.org/abs/2105.12094}{{\ttfamily 2105.12094}}].

\bibitem{Muller:2021uur}
F.~M\"uller, J.-Y. Pang, A.~Rusetsky and J.-J. Wu,
  \emph{{Relativistic-invariant formulation of the three-particle quantization
  condition}},  \href{https://arxiv.org/abs/2110.09351}{{\ttfamily
  2110.09351}}.

\bibitem{Blanton:2021eyf}
T.~D. Blanton, F.~Romero-L\'opez and S.~R. Sharpe, \emph{{Implementing the
  three-particle quantization condition for
  \ensuremath{\pi}$^{+}$\ensuremath{\pi}$^{+}$K$^{+}$ and related systems}},
  \href{https://doi.org/10.1007/JHEP02(2022)098}{\emph{JHEP} {\bfseries 02}
  (2022) 098} [\href{https://arxiv.org/abs/2111.12734}{{\ttfamily
  2111.12734}}].

\bibitem{Jackura:2022gib}
A.~W. Jackura, \emph{{Three-body scattering and quantization conditions from
  $S$ matrix unitarity}},  \href{https://arxiv.org/abs/2208.10587}{{\ttfamily
  2208.10587}}.

\bibitem{Padmanath:2022cvl}
M.~Padmanath and S.~Prelovsek, \emph{{Signature of a Doubly Charm Tetraquark
  Pole in DD* Scattering on the Lattice}},
  \href{https://doi.org/10.1103/PhysRevLett.129.032002}{\emph{Phys. Rev. Lett.}
  {\bfseries 129} (2022) 032002}
  [\href{https://arxiv.org/abs/2202.10110}{{\ttfamily 2202.10110}}].

\bibitem{Lyu:2023xro}
Y.~Lyu, S.~Aoki, T.~Doi, T.~Hatsuda, Y.~Ikeda and J.~Meng, \emph{{Doubly
  charmed tetraquark $T^+_{cc}$ from Lattice QCD near Physical Point}},
  \href{https://arxiv.org/abs/2302.04505}{{\ttfamily 2302.04505}}.

\bibitem{Chen:2022vpo}
S.~Chen, C.~Shi, Y.~Chen, M.~Gong, Z.~Liu, W.~Sun et~al.,
  \emph{{$T_{cc}^+(3875)$ relevant $DD^*$ scattering from $N_f=2$ lattice
  QCD}}, \href{https://doi.org/10.1016/j.physletb.2022.137391}{\emph{Phys.
  Lett. B} {\bfseries 833} (2022) 137391}
  [\href{https://arxiv.org/abs/2206.06185}{{\ttfamily 2206.06185}}].

\bibitem{Luscher:1986n2}
M.~L\"{u}scher, \emph{{Volume Dependence of the Energy Spectrum in Massive
  Quantum Field Theories. 2. Scattering States}},
  \href{https://doi.org/10.1007/BF01211097}{\emph{Commun.Math.Phys.} {\bfseries
  105} (1986) 153}.

\bibitem{Luscher:1991n1}
M.~L\"{u}scher, \emph{{Two particle states on a torus and their relation to the
  scattering matrix}},
  \href{https://doi.org/10.1016/0550-3213(91)90366-6}{\emph{Nucl.Phys.}
  {\bfseries B354} (1991) 531}.

\bibitem{Briceno:2015csa}
R.~A. Brice\~no and M.~T. Hansen, \emph{{Multichannel 0 $\to$ 2 and 1 $\to$ 2
  transition amplitudes for arbitrary spin particles in a finite volume}},
  \href{https://doi.org/10.1103/PhysRevD.92.074509}{\emph{Phys. Rev.}
  {\bfseries D92} (2015) 074509}
  [\href{https://arxiv.org/abs/1502.04314}{{\ttfamily 1502.04314}}].

\bibitem{Du:2023hlu}
M.-L. Du, A.~Filin, V.~Baru, X.-K. Dong, E.~Epelbaum, F.-K. Guo et~al.,
  \emph{{Role of left-hand cut contributions on pole extractions from lattice
  data: Case study for $T_{cc}(3875)^+$}},
  \href{https://arxiv.org/abs/2303.09441}{{\ttfamily 2303.09441}}.

\bibitem{Jackura:2020bsk}
A.~W. Jackura, R.~A. Brice\~no, S.~M. Dawid, M.~H.~E. Islam and C.~McCarty,
  \emph{{Solving relativistic three-body integral equations in the presence of
  bound states}},
  \href{https://doi.org/10.1103/PhysRevD.104.014507}{\emph{Phys. Rev. D}
  {\bfseries 104} (2021) 014507}
  [\href{https://arxiv.org/abs/2010.09820}{{\ttfamily 2010.09820}}].

\bibitem{Dawid:2021fxd}
S.~M. Dawid, \emph{{Infinite volume, three-body scattering formalisms in the
  presence of bound states}},
  \href{https://doi.org/10.22323/1.396.0520}{\emph{PoS} {\bfseries LATTICE2021}
  (2022) 520} [\href{https://arxiv.org/abs/2111.05418}{{\ttfamily
  2111.05418}}].

\bibitem{Dawid:2023jrj}
S.~M. Dawid, M.~H.~E. Islam and R.~A. Brice\~no, \emph{{Analytic continuation
  of the relativistic three-particle scattering amplitudes}},
  \href{https://doi.org/10.1103/PhysRevD.108.034016}{\emph{Phys. Rev. D}
  {\bfseries 108} (2023) 034016}
  [\href{https://arxiv.org/abs/2303.04394}{{\ttfamily 2303.04394}}].

\bibitem{Raposo:2023oru}
A.~B.~a. Raposo and M.~T. Hansen, \emph{{Finite-volume scattering on the
  left-hand cut}},  \href{https://arxiv.org/abs/2311.18793}{{\ttfamily
  2311.18793}}.

\bibitem{Meng:2023bmz}
L.~Meng, V.~Baru, E.~Epelbaum, A.~A. Filin and A.~M. Gasparyan, \emph{{Solving
  the left-hand cut problem in lattice QCD: $T_{cc}(3875)^+$ from finite volume
  energy levels}},  \href{https://arxiv.org/abs/2312.01930}{{\ttfamily
  2312.01930}}.

\bibitem{Brown:2012tm}
Z.~S. Brown and K.~Orginos, \emph{{Tetraquark bound states in the heavy-light
  heavy-light system}},
  \href{https://doi.org/10.1103/PhysRevD.86.114506}{\emph{Phys. Rev. D}
  {\bfseries 86} (2012) 114506}
  [\href{https://arxiv.org/abs/1210.1953}{{\ttfamily 1210.1953}}].

\bibitem{Bicudo:2016ooe}
P.~Bicudo, J.~Scheunert and M.~Wagner, \emph{{Including heavy spin effects in
  the prediction of a $\bar{b} \bar{b} u d$ tetraquark with lattice QCD
  potentials}}, \href{https://doi.org/10.1103/PhysRevD.95.034502}{\emph{Phys.
  Rev. D} {\bfseries 95} (2017) 034502}
  [\href{https://arxiv.org/abs/1612.02758}{{\ttfamily 1612.02758}}].

\bibitem{Francis:2016hui}
A.~Francis, R.~J. Hudspith, R.~Lewis and K.~Maltman, \emph{{Lattice Prediction
  for Deeply Bound Doubly Heavy Tetraquarks}},
  \href{https://doi.org/10.1103/PhysRevLett.118.142001}{\emph{Phys. Rev. Lett.}
  {\bfseries 118} (2017) 142001}
  [\href{https://arxiv.org/abs/1607.05214}{{\ttfamily 1607.05214}}].

\bibitem{Junnarkar:2018twb}
P.~Junnarkar, N.~Mathur and M.~Padmanath, \emph{{Study of doubly heavy
  tetraquarks in Lattice QCD}},
  \href{https://doi.org/10.1103/PhysRevD.99.034507}{\emph{Phys. Rev. D}
  {\bfseries 99} (2019) 034507}
  [\href{https://arxiv.org/abs/1810.12285}{{\ttfamily 1810.12285}}].

\bibitem{Leskovec:2019ioa}
L.~Leskovec, S.~Meinel, M.~Pflaumer and M.~Wagner, \emph{{Lattice QCD
  investigation of a doubly-bottom $\bar{b} \bar{b} u d$ tetraquark with
  quantum numbers $I(J^P) = 0(1^+)$}},
  \href{https://doi.org/10.1103/PhysRevD.100.014503}{\emph{Phys. Rev. D}
  {\bfseries 100} (2019) 014503}
  [\href{https://arxiv.org/abs/1904.04197}{{\ttfamily 1904.04197}}].

\bibitem{Mohanta:2020eed}
P.~Mohanta and S.~Basak, \emph{{Construction of $bb\bar{u}\bar{d}$ tetraquark
  states on lattice with NRQCD bottom and HISQ up and down quarks}},
  \href{https://doi.org/10.1103/PhysRevD.102.094516}{\emph{Phys. Rev. D}
  {\bfseries 102} (2020) 094516}
  [\href{https://arxiv.org/abs/2008.11146}{{\ttfamily 2008.11146}}].

\bibitem{Meinel:2022lzo}
S.~Meinel, M.~Pflaumer and M.~Wagner, \emph{{Search for
  b\textasciimacron{}b\textasciimacron{}us and
  b\textasciimacron{}c\textasciimacron{}ud tetraquark bound states using
  lattice QCD}}, \href{https://doi.org/10.1103/PhysRevD.106.034507}{\emph{Phys.
  Rev. D} {\bfseries 106} (2022) 034507}
  [\href{https://arxiv.org/abs/2205.13982}{{\ttfamily 2205.13982}}].

\bibitem{Hudspith:2023loy}
R.~J. Hudspith and D.~Mohler, \emph{{Exotic tetraquark states with two
  b\textasciimacron{} quarks and JP=0+ and 1+ Bs states in a nonperturbatively
  tuned lattice NRQCD setup}},
  \href{https://doi.org/10.1103/PhysRevD.107.114510}{\emph{Phys. Rev. D}
  {\bfseries 107} (2023) 114510}
  [\href{https://arxiv.org/abs/2303.17295}{{\ttfamily 2303.17295}}].

\bibitem{Kim:2005gf}
C.~h. Kim, C.~T. Sachrajda and S.~R. Sharpe, \emph{{Finite-volume effects for
  two-hadron states in moving frames}},
  \href{https://doi.org/10.1016/j.nuclphysb.2005.08.029}{\emph{Nucl. Phys.}
  {\bfseries B727} (2005) 218}
  [\href{https://arxiv.org/abs/hep-lat/0507006}{{\ttfamily hep-lat/0507006}}].

\bibitem{Dawid:2023kxu}
S.~M. Dawid, M.~H.~E. Islam, R.~A. Brice\~no and A.~W. Jackura,
  \emph{{Evolution of Efimov States}},
  \href{https://arxiv.org/abs/2309.01732}{{\ttfamily 2309.01732}}.

\bibitem{Sadasivan:2020syi}
D.~Sadasivan, M.~Mai, H.~Akdag and M.~D\"oring, \emph{{Dalitz plots and
  lineshape of $a_1(1260)$ from a relativistic three-body unitary approach}},
  \href{https://doi.org/10.1103/PhysRevD.101.094018}{\emph{Phys. Rev. D}
  {\bfseries 101} (2020) 094018}
  [\href{https://arxiv.org/abs/2002.12431}{{\ttfamily 2002.12431}}].

\bibitem{Pang:2023jri}
J.-Y. Pang, R.~Bubna, F.~M\"uller, A.~Rusetsky and J.-J. Wu,
  \emph{{Lellouch-L\"uscher factor for the $K\to 3\pi$ decays}},
  \href{https://arxiv.org/abs/2312.04391}{{\ttfamily 2312.04391}}.

\bibitem{Jackura:2023qtp}
A.~W. Jackura and R.~A. Brice\~no, \emph{{Partial-wave projection of the
  one-particle exchange in three-body scattering amplitudes}},
  \href{https://arxiv.org/abs/2312.00625}{{\ttfamily 2312.00625}}.

\bibitem{Baeza-Ballesteros:2023ljl}
J.~Baeza-Ballesteros, J.~Bijnens, T.~Husek, F.~Romero-L\'opez, S.~R. Sharpe and
  M.~Sj\"o, \emph{{The isospin-3 three-particle K-matrix at NLO in ChPT}},
  \href{https://doi.org/10.1007/JHEP05(2023)187}{\emph{JHEP} {\bfseries 05}
  (2023) 187} [\href{https://arxiv.org/abs/2303.13206}{{\ttfamily
  2303.13206}}].

\bibitem{Briceno:2014oea}
R.~A. Brice\~no, \emph{{Two-particle multichannel systems in a finite volume
  with arbitrary spin}},
  \href{https://doi.org/10.1103/PhysRevD.89.074507}{\emph{Phys. Rev.}
  {\bfseries D89} (2014) 074507}
  [\href{https://arxiv.org/abs/1401.3312}{{\ttfamily 1401.3312}}].

\bibitem{Lee:2017igf}
F.~X. Lee and A.~Alexandru, \emph{{Scattering phase-shift formulas for mesons
  and baryons in elongated boxes}},
  \href{https://doi.org/10.1103/PhysRevD.96.054508}{\emph{Phys. Rev. D}
  {\bfseries 96} (2017) 054508}
  [\href{https://arxiv.org/abs/1706.00262}{{\ttfamily 1706.00262}}].

\bibitem{Mohler:2012na}
D.~Mohler, S.~Prelovsek and R.~M. Woloshyn, \emph{{$D \pi$ scattering and $D$
  meson resonances from lattice QCD}},
  \href{https://doi.org/10.1103/PhysRevD.87.034501}{\emph{Phys. Rev. D}
  {\bfseries 87} (2013) 034501}
  [\href{https://arxiv.org/abs/1208.4059}{{\ttfamily 1208.4059}}].

\bibitem{Moir:2016srx}
G.~Moir, M.~Peardon, S.~M. Ryan, C.~E. Thomas and D.~J. Wilson,
  \emph{{Coupled-Channel $D\pi$, $D\eta$ and $D_{s}\bar{K}$ Scattering from
  Lattice QCD}}, \href{https://doi.org/10.1007/JHEP10(2016)011}{\emph{JHEP}
  {\bfseries 10} (2016) 011}
  [\href{https://arxiv.org/abs/1607.07093}{{\ttfamily 1607.07093}}].

\bibitem{Gayer:2021xzv}
{\scshape Hadron Spectrum} collaboration, L.~Gayer, N.~Lang, S.~M. Ryan,
  D.~Tims, C.~E. Thomas and D.~J. Wilson, \emph{{Isospin-1/2 D\ensuremath{\pi}
  scattering and the lightest $ {D}_0^{\ast } $ resonance from lattice QCD}},
  \href{https://doi.org/10.1007/JHEP07(2021)123}{\emph{JHEP} {\bfseries 07}
  (2021) 123} [\href{https://arxiv.org/abs/2102.04973}{{\ttfamily
  2102.04973}}].

\bibitem{Yan:2023gvq}
H.~Yan, C.~Liu, L.~Liu, Y.~Meng and H.~Xing, \emph{{Isospin-$\frac{1}{2}$$D\pi$
  scattering and the $D_0^*$ resonance}},  in \emph{{40th International
  Symposium on Lattice Field Theory}}, 12, 2023,
  \href{https://arxiv.org/abs/2312.01078}{{\ttfamily 2312.01078}}.

\bibitem{Dudek:2012gj}
J.~J. Dudek, R.~G. Edwards and C.~E. Thomas, \emph{{S and D-wave phase shifts
  in isospin-2 $\pi \pi$ scattering from lattice QCD}},
  \href{https://doi.org/10.1103/PhysRevD.86.034031}{\emph{Phys. Rev.}
  {\bfseries D86} (2012) 034031}
  [\href{https://arxiv.org/abs/1203.6041}{{\ttfamily 1203.6041}}].

\bibitem{Woss:2018irj}
A.~Woss, C.~E. Thomas, J.~J. Dudek, R.~G. Edwards and D.~J. Wilson,
  \emph{{Dynamically-coupled partial-waves in $\rho\pi$ isospin-2 scattering
  from lattice QCD}},
  \href{https://doi.org/10.1007/JHEP07(2018)043}{\emph{JHEP} {\bfseries 07}
  (2018) 043} [\href{https://arxiv.org/abs/1802.05580}{{\ttfamily
  1802.05580}}].

\bibitem{Hansen:2021ofl}
M.~T. Hansen, F.~Romero-L\'opez and S.~R. Sharpe, \emph{{Decay amplitudes to
  three hadrons from finite-volume matrix elements}},
  \href{https://doi.org/10.1007/JHEP04(2021)113}{\emph{JHEP} {\bfseries 04}
  (2021) 113} [\href{https://arxiv.org/abs/2101.10246}{{\ttfamily
  2101.10246}}].

\end{thebibliography}\endgroup

\end{document}